\documentclass[longauth]{aa} 

\usepackage{graphicx}
\usepackage{subcaption}
\usepackage{mwe}
\usepackage{booktabs}
\usepackage{tabularx}
\usepackage{multirow}
\usepackage{natbib}
\usepackage{xcolor}

\bibpunct{(}{)}{;}{a}{}{,}
\usepackage{gensymb}
\usepackage{graphicx}
\usepackage{txfonts}

\begin{document}

   \title{The eROSITA Final Equatorial-Depth Survey (eFEDS):} 
   
   \subtitle{Identification and characterization of the counterparts to the point-like sources\footnote{The two catalogs associated to this  paper are available in electronic form
at the CDS via anonymous ftp to cdsarc.u-strasbg.fr or via http://cdsweb.u-strasbg.fr/cgi-bin/qcat?J/A+A/}}

   \author{M. Salvato\fnmsep\thanks{mara@mpe.mpg.de}
          \inst{1,2},
J. Wolf\inst{1,2},
T. Dwelly\inst{1},
A. Georgakakis\inst{3},
M. Brusa\inst{4,5},
A. Merloni\inst{1},
T. Liu\inst{1},
Y. Toba\inst{6,7,8},
K. Nandra\inst{1},
G. Lamer\inst{9},
J. Buchner\inst{1},
C. Schneider\inst{10},
S. Freund\inst{10},
A. Rau\inst{1},
A. Schwope\inst{9},
A. Nishizawa\inst{11},
M. Klein\inst{12},
R. Arcodia\inst{1},
J. Comparat\inst{1},
B. Musiimenta\inst{4,5},
T. Nagao\inst{8},
H. Brunner\inst{1},
A. Malyali\inst{1},
A. Finoguenov\inst{1},
S. Anderson\inst{13},
Y. Shen\inst{14},
H. Ibarra-Medel\inst{14},
J. Trump\inst{15},
W.N., Brandt\inst{16,17,18},
C.M., Urry\inst{19},
C. Rivera\inst{9},
M. Krumpe\inst{9},
T. Urrutia\inst{9},
T. Miyaji\inst{20},
K. Ichikawa\inst{1,21,22},
D.P., Schneider\inst{23,24},
A. Fresco\inst{1},
T. Boller\inst{1},
J. Haase\inst{1},
J. Brownstein\inst{24},
R.R., Lane\inst{25},
D. Bizyaev\inst{27},
C. Nitschelm\inst{28}
          }
  \titlerunning{Point-like sources in eFEDS}
    \authorrunning{Salvato et al.}
\institute{Max-Planck-Institut f\"ur extraterrestrische Physik, Giessenbachstr. 1, 85748 Garching, Germany
\and Exzellenzcluster ORIGINS, Boltzmannstr. 2, D-85748 Garching, Germany 
\and Institute for Astronomy and Astrophysics, National Observatory of Athens, V. Paulou and I. Metaxa, 11532, Greece
\and Dipartimento di Fisica e Astronomia "Augusto Righi", Universit\`a di Bologna,  via Gobetti 93/2,  40129 Bologna, Italy 
\and INAF - Osservatorio di Astrofisica e Scienza dello Spazio di Bologna, via Gobetti 93/3,  40129 Bologna, Italy 
\and Department of Astronomy, Kyoto University, Kitashirakawa-Oiwake-cho, Sakyo-ku, Kyoto 606-8502, Japan
\and Academia Sinica Institute of Astronomy and Astrophysics, Taipei 10617, Taiwan 
\and Research Center for Space and Cosmic Evolution, Ehime University, 2-5 Bunkyo-cho, Matsuyama, Ehime 790-8577, Japan 
\and Leibniz-Institut f\"ur Astrophysik Potsdam (AIP). An der Sternwarte 16. 14482 Potsdam, Germany
\and Universit\"at Hamburg, Hamburger Sternwarte, Gojenbergsweg 112, D-21029 Hamburg, Germany
\and Institute for Advanced Research, Nagoya University Furocho, Chikusa-ku, Nagoya, 464-8602 Japan
\and Faculty of Physics, Ludwig-Maximilians-Universit\"at, Scheinerstr 1, D-81679 Munich, Germany
\and Department of Astronomy, University of Washington, Box 351580, Seattle, WA 98195, USA	True
\and Department of Astronomy, University of Illinois at Urbana-Champaign, Urbana, IL 61801, USA
\and Department of Physics, University of Connecticut, 2152 Hillside Road, Unit 3046, Storrs, CT 06269, USA
\and 1Department of Astronomy and Astrophysics, 525 Davey Lab, The Pennsylvania State University, University Park, PA 16802, USA
\and 2Institute for Gravitation and the Cosmos, The Pennsylvania State University, University Park, PA 16802, USA
\and Department of Physics, 104 Davey Laboratory, The Pennsylvania State University, University Park, PA 16802, USA
\and Department of Physics and Yale Center for Astronomy and Astrophysics, Yale University, PO Box 208120, New Haven, CT 06520-8120, USA
\and Instituto de Astronomía sede Ensenada, Universidad Nacional Autónoma de Me\'xico, AP 106, Ensenada, 22800,  Me\'xico
\and Frontier Research Institute for Interdisciplinary Sciences, Tohoku University, Sendai 980-8578, Japan
\and Astronomical Institute, Tohoku University, Aramaki, Aoba-ku, Sendai, Miyagi 980-8578, Japan
\and Department of Astronomy and Astrophysics, The Pennsylvania State University,University Park, PA 16802
\and Institute for Gravitation and the Cosmos, The Pennsylvania State University,  University Park, PA 16802
\and Department of Physics and Astronomy, University of Utah, 115 S. 1400 E., Salt Lake City, UT 84112, USA
 \and Centro de Investigaci\'on en Astronom\'ia, Universidad Bernardo O'Higgins, Avenida Viel 1497, Santiago, Chile
 \and Apache Point Observatory and New Mexico State University, P.O. Box 59, Sunspot, NM 88349 
  \and Centro de Astronomía, Universidad de Antofagasta, Avenida Angamos 601, Antofagasta 1270300, Chile
 }
\date{Received September XX, 2020; accepted XX, 2020}

  \abstract
   {In November 2019, eROSITA on board of SRG observatory started to map the entire sky in X-rays. After the 4-year survey program, it will reach flux limits about 25 times deeper than ROSAT. During the SRG Performance Verification phase, eROSITA observed a contiguous 140 deg$^2$ area of the sky down to the final depth of the eROSITA all-sky survey ("eROSITA Final Equatorial-Depth Survey": eFEDS), with the goal of obtaining a census of the X-ray emitting populations (stars, compact objects, galaxies, clusters of galaxies, AGN) that will be discovered over the entire sky.}
   {This paper presents the identification of the counterparts to the point-sources detected in eFEDS in the Main and Hard samples described in \citet{Brunner2021}, and their multi-wavelength properties, including redshift.}
   {For the identification of the counterparts  we combined the results from two independent methods ({\sc{nway}} and {\sc{astromatch}}), trained on the multi-wavelength properties of a sample of 23k {\it XMM-Newton} sources detected in the DESI Legacy Imaging Survey DR8. Then spectroscopic redshifts and photometry from ancillary surveys are collated for the computation of photometric redshifts.}
   {The eFEDS sources with a reliable counterparts are 24774/27369 (90.5\%) in the Main sample and 231/246 (93.9\%) in the Hard sample, including 2514 (3) sources for which a second counterpart is equally likely. By means of reliable spectra, Gaia parallaxes, and/or multi-wavelength properties we have classified the reliable counterparts in both samples as 'Galactic'  (2695) and 'extragalactic' (22079). For about 340 of the extragalactic sources  we cannot rule out the possibility that they are unresolved clusters or belong to clusters. Inspection of the distributions of the X-ray sources in various optical/IR colour-magnitude spaces reveal a rich variety of diverse classes of objects. The photometric redshifts are most reliable within the KiDS/VIKING area, where also deep near-infrared data are available.}
   {This paper is accompanying the eROSITA early data release of all the observations performed during the performance and verification phase. Together with the catalogs of primary and secondary counterparts to the Main and Hard samples of the eFEDS survey this paper releases their multi-wavelength properties and redshifts.}


   \maketitle
%

\section{Introduction}\label{intro}

Across the electromagnetic spectrum, sensitive wide-area surveys serve multiple purposes. First and foremost, they help astronomers draw a map of our cosmic neighbourhood, and in doing so they reveal the inner workings of the Milky Way, the local group, and the filamentary large scale structure underpinning the distribution of matter. Secondly, by observing and cataloguing large numbers of stars, galaxies, groups, clusters and super-clusters of galaxies that are the main visible tracers of this large-scale structure, wide area surveys also provide new statistical tools for the study of classes and populations of astronomical objects, thus helping astronomers to better understand their life-cycles, their interactions and, ultimately, their physical properties.

X-ray surveys, in particular, reveal fundamental physical processes invisible at other wavelengths. 
The hot, diffuse plasma that virialises and thermalises within massive dark matter knots; accretion of matter onto compact objects, both Galactic and extra-galactic; the magnetic coronae of mostly young, fast rotating stars are all phenomena accessible by X-ray sensitive instruments.

eROSITA \citep[extended ROentgen Survey with an Imaging Telescope Array;][]{Predehl2021}, onboard the  Spektrum-R\"Ontgen-Gamma (SRG) mission \citep[][]{Sunyaev2021},
was designed to provide sensitive X-ray imaging and spectroscopy over a large field of view, thus unlocking unprecedented capabilities for surveying large areas of the sky to deep flux levels. Moreover, the SRG mission plan includes a long (4 years), uninterrupted all-sky survey program (the eROSITA All-Sky Survey: eRASS; \citealt{Predehl2021}) capable of detecting, for the first time, millions of X-ray sources.

\begin{figure*}[!ht]
\centering
\includegraphics[width=\textwidth]{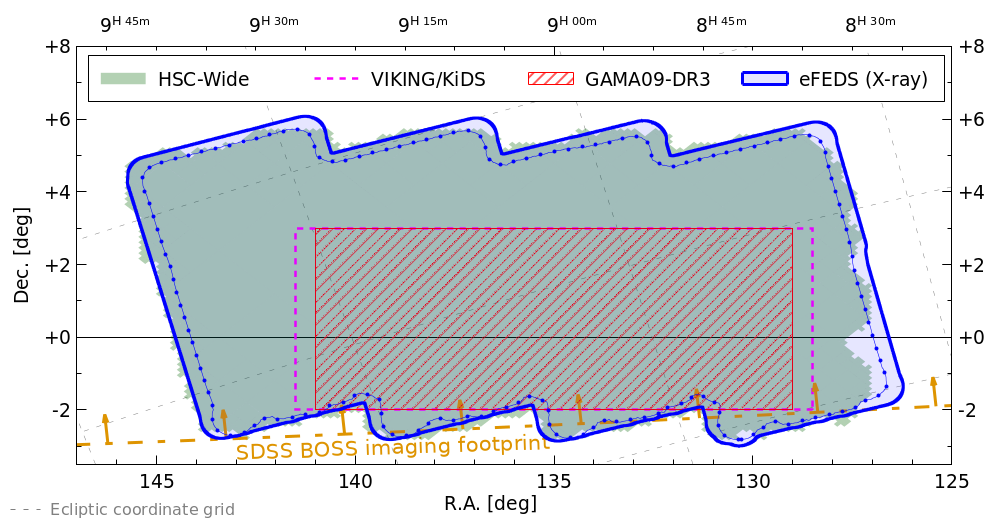}
\caption{eFEDS X-ray and multi-wavelength coverage. The thick blue line shows the outer bound 
of the region that was searched for X-ray sources. The thin blue beaded line shows the region 
having at least 500 seconds of  effective X-ray exposure depth. We indicate the approximate coverage 
of several selected surveys that are particularly important for this work; 
Subaru HSC-Wide (green shaded region), KiDS/VIKING (magenta dashed box), GAMA09-DR3 (red hatched box). 
The eFEDS field is also immersed in several other important surveys that completely (or almost-completely) enclose the displayed region: 
e.g. Galex all-sky surveys (in UV), Gaia (in optical), Legacy Survey DR8 (optical combined 
with Gaia and WISE), VHS and UKIDSS (in Near-Infrared), WISE/NEOWISE-R, and SDSS (optical imaging and spectroscopy).}
\label{fig:AncillaryCoverage}
\centering
\end{figure*}


In order to demonstrate these ground-breaking survey capabilities, and prepare for the science exploitation of the upcoming all-sky survey,
the contiguous 140 square degrees of the eROSITA Final Equatorial-Depth survey \citep[eFEDS;][]{Brunner2021} was observed during the SRG Calibration and Performance Verification phase, between the 3$^{\rm rd}$ and the 7$^{\rm th}$ of November 2019. The entire field,  centreed at R.A. 136 and Dec. +2 (see Figure~\ref{fig:AncillaryCoverage}), was observed to an approximate depth of $\sim$2.2~ks ($\sim$1.2~ks after correcting for telescope vignetting), corresponding to a limiting flux of $ F_{\rm 0.5-2 \, keV} \sim 6.5 \times 10^{-15} \, \mathrm{erg\, s^{-1}\, cm^{-2}}$.
The eFEDS field was chosen from among the extragalactic areas with the richest multi-wavelength coverage visible by eROSITA in November 2019. The observations are just about 50\% deeper than anticipated for eRASS:8 at the end of the planned 4-year' program in the ecliptic equatorial region ($\sim1.1\times10^{-14}$ erg cm$^{-2}$ s$^{-1}$, \citealt{Predehl2021}). Said otherwise, the eFEDS exposure corresponds roughly to the 80th percentile of the expected eRASS:8 exposure distribution over the whole sky and, as such, eFEDS is a fair representation of what the final eROSITA all-sky survey will be, enabling scientists to face and solve the challenges that  will accompany their work for the duration of the survey.

As discussed in detail in \citet{Brunner2021}, the X-ray catalogs generated by the analysis of the eFEDS eROSITA data comprise a {\it Main} one, with 27910 sources detected above a detection likelihood of 6 in the most sensitive 0.2-2.3 keV band, and a {\it Hard} one, containing 246 sources detected above a detection likelihood of 10 in the less-sensitive 2.3-5 keV band.

In this paper, we focus our attention on the point-like (i.e. with an extension likelihood \texttt{EXT\_LIKE$=$0}\footnote{this parameter is obtained from the task
srctool of the eSASS software; \citep[][]{Brunner2021}}) X-ray sources contained in these catalogs (27369 and 246 for the {\it Main} and {\it Hard} sample, respectively),
and describe in detail the procedure to (i) identify reliably multi-wavelength counterparts to the eROSITA sources, (ii) classify and characterise their properties and (iii) provide reliable  redshift measurements (spectroscopic when available and photometric otherwise). 
The identification and determination of the reliability of the counterparts, the computation of the photometric redshifts (photo-z), and the characterisation of the sample follow the same procedure for both {\it Main} and {\it Hard} samples, and for simplicity  we discuss here specifically only the {\it Main} sample, given the large overlap between the two catalogs (226/246 hard sources are in common). While we provide here the catalog of counterparts for all the sources in both samples, the properties of the sources in the {\it Hard} sample are presented and discussed in Nandra et al. (in prep.).
The papers on  X-ray spectral analysis \citep[][]{TLiu2021}, variability \citep[][]{Boller2021, Buchner2021}, X-ray Luminosity Function (Buchner et al., in prep; Wolf et al, in prep) and host properties of eFEDS AGN (Li et al., in prep.) are all based on the catalog and/or the methodology presented in this work. 
Based on this work are also the papers presenting interesting single objects \citep[][]{Brusa2021,Toba2021,Wolf21}, the X-ray properties of WISE sources in eFEDS \citep[][]{Toba2021} , radio properties of unresolved clusters \citep[][]{Bulbul2021} and photo-z computed via machine learning (Nishizawa et al., in prep),

The structure of the paper is as follows: In section~\ref{sec:supporting} we summarise the availability of ancillary data that will be used for the identification of the X-ray counterparts and photo-z  estimates.
Sections~\ref{section:counterparts}  and ~\ref{section:nway_astromatch_validation} describe in detail the methods used in this paper for identifying the counterparts, while in section~\ref{section:comparison}  the counterparts are finally assigned. In section~\ref{section:properties} the counterparts are separated in Galactic and extragalactic, using morphological, photometric and proper motion information.

\begin{figure*}[!ht]
\centering

\includegraphics[width=0.45\textwidth]{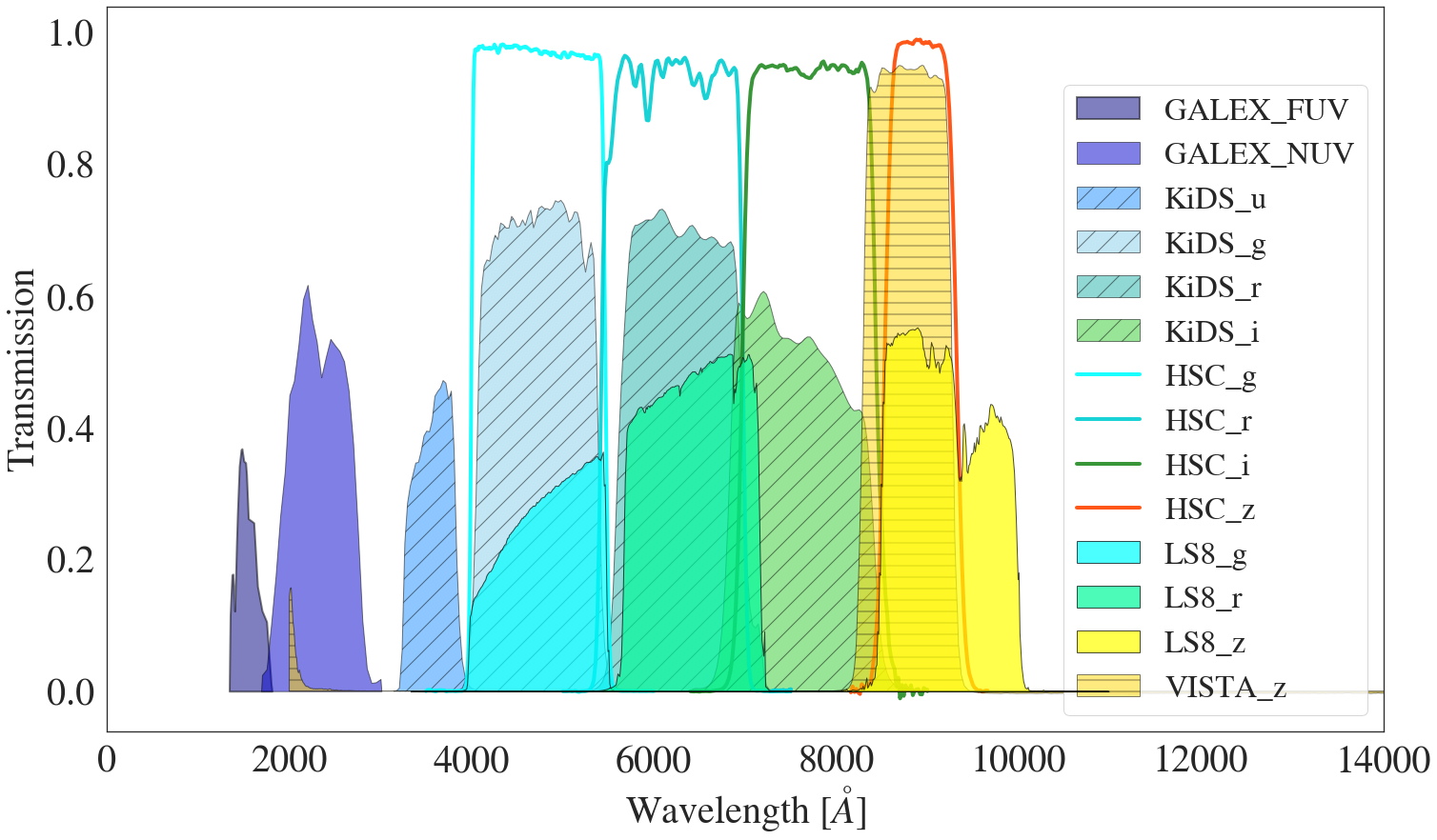}
\includegraphics[width=0.45\textwidth]{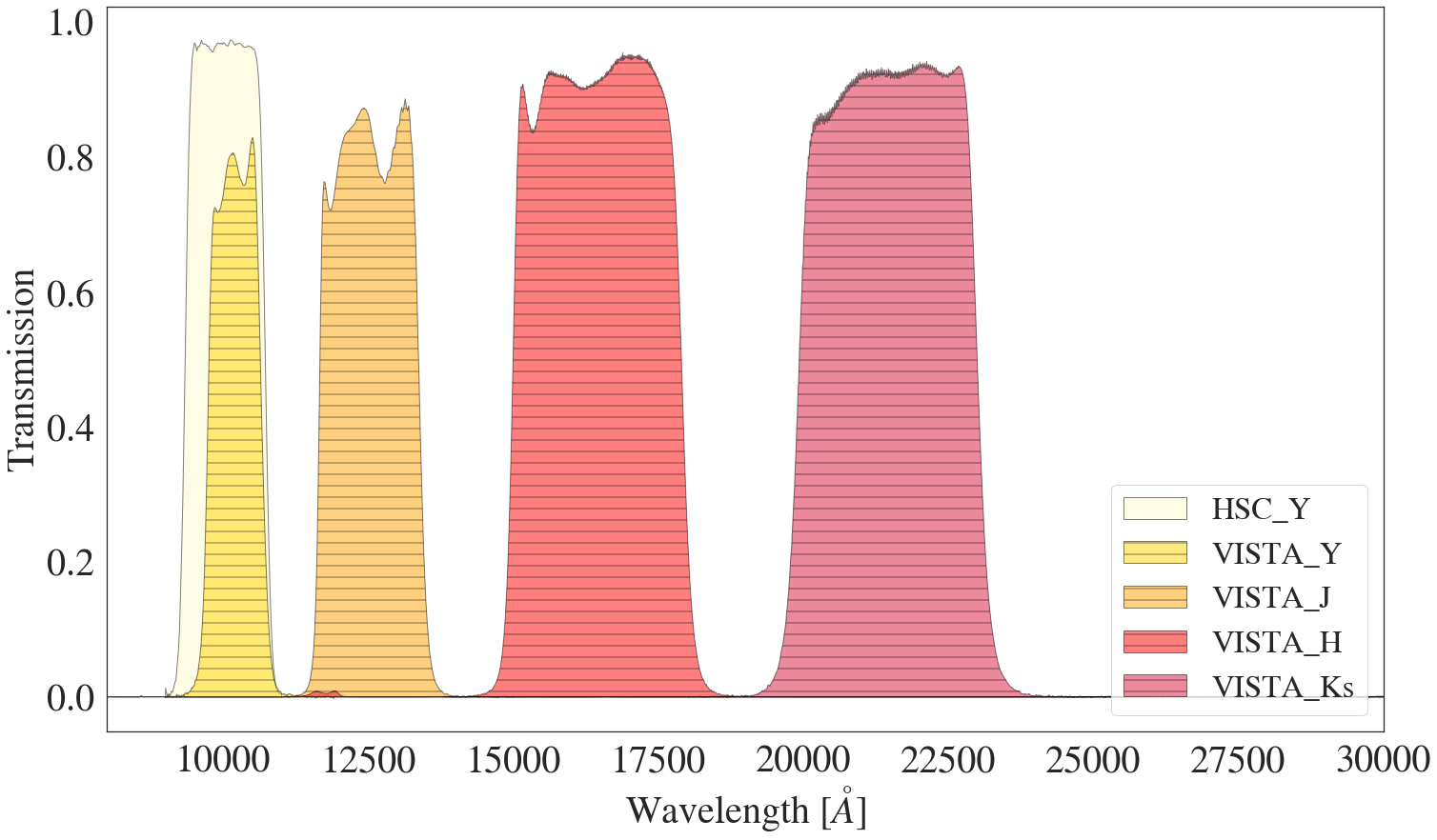}
\caption{Illustration of the band-pass and relative transmission curves of the UV/optical/NIR photometry used for computation of photo-z in this work. For clarity, we do not show the WISE band-passes.}
\label{fig:filters}
\centering
\end{figure*}
Because of its size, the field is well populated by stars, AGN, clusters and nearby galaxies. Each eROSITA working group has developed independent methods for the identification of sources of interest, and in Sections ~\ref{sect:other_methods} and ~\ref{section:in_overdensity} a comparison is made with two  main source classes: stellar coronal emitters and  clusters of galaxies, with the ultimate goal of consolidating the counterparts and classifying them at the same time.  
Section~\ref{section:properties} deals with the multiwavelength properties of the counterparts  and the characterisation of their Galactic or extra-galactic nature. Section~\ref{section:photoz} presents and discusses the photo-z computed with Le PHARE \citep[][]{Ilbert06,Arnouts99}, including a comparison with DNNz (Nishizawa et al., in prep), an independent method based on machine learning. Section~\ref{section:catalogs} describes the released data. The basic properties of the point-source eFEDS population based on redshift, photometry, and X-ray flux are presented in Section~\ref{section:discussion}. The conclusions in Section~\ref{section:conclusions} close the paper, with a forecast of the results and challenges that we will face when working on the eROSITA all-sky  survey.\\

The description of the Catalogs that we release is provided in the Appendix, together with the list of templates used for computing the photo-z.

Throughout the paper we assume AB magnitudes unless differently stated. In order to allow direct comparison with existing works from the literature of X-ray surveys, we adopt a flat $\Lambda$CDM cosmology with $h=H_0/[100\,\mathrm{km\, s}^{-1} \mathrm{Mpc}^{-1}]=0.7$; $\Omega_M$=0.3; $\Omega_\Lambda$=0.7.

\section{Supporting ancillary data \label{sec:supporting}}

For studies of X-ray sources (taken singularly or as a population) the entire spectral energy distribution (SED) needs to be constructed and the redshift determined. Only rarely can redshift be obtained directly from X-ray spectra, and it is instead  routinely obtained either via optical/Near-Infrared spectroscopy or via photometric techniques. However, for that to work the counterparts to the X-ray sources need to be determined first. Deep and homogeneous  multi-wavelength data are therefore a pre-requisite for any complete population study of an X-ray survey.

The main challenge is that, in survey mode, eROSITA has a Half Energy Width (HEW) of 26\,arcsec\,\footnote{i.e., comparable to the XMM Slew Survey:  \url{https://www.cosmos.esa.int/web/xmm- newton/xmmsl2- ug}} \citep[][]{Predehl2021}, which makes the identification of the correct counterparts not at all trivial \citep[e.g., see the review of][]{Naylor13, Salvato18a}, especially if considering in addition that wide field multi-wavelength homogeneous surveys are very difficult to obtain, with very few exceptions \citep[see the Legacy imaging survey supporting the Dark Energy Spectroscopic Instrument, DESI;[]{Dey19}. By construction, the eFEDS field is placed in an area fully encompassing the GAMA09 equatorial field \citep{Driver09}, and rich with ancillary photometric and spectroscopic data (Merloni et al., in prep.).
Below we list and describe in more detail the surveys that have been used in this work. Table~\ref{tab:photometry} summarises the depth in each filter, while Figure~\ref{fig:filters} shows the coverage of the ancillary data in wavelength.

\begin{table*}[t]
\begin{center}
\begin{tabular}{llll}
\toprule
Bands         & Survey   & Depth(AB mag) & Reference \\
              &          & various indicators &      \\
\midrule 
FUV, NUV    & GALEX & 19.9, 20.8   & \cite{Bianchi14} \\
{\it u,g,r,i} & KiDS & 24.2, 25.1, 25.0, 23.7 & \cite{Kuijken19}\\
{\it g,r,i,z,y}    & HSC & 26.8, 26.4, 26.4, 25.5, 24.7   &\cite{Aihara18b} \\
{\it g,r,i}    & LS8 &  24.0, 23.4, 22.5 & \cite{Dey19}\\
{\it z}, J,H,K & KiDS/VIKING &23.1, 22.3, 22.1, 21.5, 21.2 & \cite{Kuijken19}\\
J,Ks   & VISTA/VHS  & 21.1, 19.8  & \cite{McMahon2013}\\
W1,W2,W3,W4 &LS8/WISE &21.0, 20.1, 16.7, 14.5 & \cite{Meisner2019}\\
\bottomrule
\end{tabular}
\end{center}
\caption{Photometry available for the counterparts identification and photo-z computation. For LS8 the required depth for DESI is listed. For Ls8/WISE the listed depth is taken from \citep{Meisner2019} and it is computed using WISE detected sources. Given that the photometry used here is forced photometry at the position of optically detected sources, the depth is higher. }
\label{tab:photometry}
\end{table*}

\subsection{Supporting the associations}
The identification of the counterparts is carried out using the  DESI Legacy Imaging Survey DR8 \citep[LS8; ][]{Dey2019}, for various reasons.
First of all, it covers homogeneously the field and it has sufficient depth, based on the expectation of the X-ray population optical properties \citep{Merloni12}. In addition,
the survey provides
together with Gaia also the AllWISE {\it tractor} \citep[][]{Lang14} photometry extracted at the position of the optical sources.  Finally, the survey covers 14,000 square degrees of sky thus providing a sufficient number of sources external to eFEDS that can be used as training and validation samples for testing the association (see Section~\ref{section:counterparts}). The absolute astrometry of the LS8 catalogue is registered to the Gaia DR2 astrometric frame, with residuals typically smaller than 30 milliarcseconds\footnote{\url{https://www.legacysurvey.org/dr8/description/#astrometry}}. While the catalogued positional uncertainties of individual LS8 sources are often much larger than this systematic limit (especially toward fainter magnitudes), 
they are still extremely small relative to those of the eFEDS X-ray sources and are set to 0.1" for the entire survey.
Photometry and parallax measures from Gaia, which are optimised for point-like sources, are ideal for the identification of the stars in our sample. For this purpose, rather than the Gaia DR2 provided by LS8, the  EDR3\footnote{\url{https://www.cosmos.esa.int/web/gaia/earlydr3}} has been used \citep{GaiaEDR3}. 

\subsection{Supporting Photometric redshifts}
For the computation of the photo-z the following data sets have been used:
\begin{itemize}
\item {\bf GALEX}:
The NASA satellite GALEX has mapped the entire sky in Far and Near UV between 2003 and 2012, with a typical depth of 19.9 and 20.8 AB magnitude in FUV and NUV, respectively. We have used the catalog GR6/7 presented in \cite{Bianchi14} and available via Vizier. 
\item {\bf Kilo-degree Survey (KiDS)}\footnote{\url{http://kids.strw.leidenuniv.nl/}}: the survey mapped 1350 deg$^2$ in {\it u,g,r,i} bands using VST/OmegaCAM. The same area was also covered by the VISTA Kilo-Degree Infrared Galaxy Survey. \citep[VIKING,][]{Edge13} in Z,Y,J,H,K. 
We have used here the catalog presented in \cite{Kuijken19}; it has ZYJHK aperture-matched, forced photometry to the {\it ugri} source positions. About 65 deg$^2$ of sky  are shared between  KiDS/VIKING and eFEDS.

\item {\bf HSC S19A}:
The Hyper Suprime-Cam \citep[HSC;][]{Miyazaki} Subaru Strategic Program survey \citep[HSC--SSP;][]{Aihara18a} is an ongoing optical imaging survey with five broadband filters ($g$-, $r$-, $i$-, $z$-, and $y$-band) and four narrow-band filters \citep[see][]{Aihara18b}.
We utilised S19A wide data obtained from March 2014 to April 2019, which provides forced photometry for the five bands, with the $5\sigma$ limiting magnitudes as listed in Table ~\ref{tab:photometry} \citep[see][]{Aihara18b,Aihara19}.
The astrometric uncertainty is approximately 40 milliarcseconds in rms.

\item{\bf VISTA/VHS}:
The entire Southern hemisphere has been observed by VISTA in Near Infrared  and at least for J and Ks the depth is thirty times the depth of 2MASS \citep[][]{McMahon2013}.
We have used the DR4 data available via Vizier.

\item {\bf WISE}:
The Wide-field Infrared Survey Explorer \citep[WISE;][]{Wright10}, over the course of one year scanned the entire sky in the 3.4, 4.6, 12 and 22 $\mu$m bands (hereafter W1, W2, W3, W4). Afterwards, the survey continued  with  observations only in W1 and W2.  The  photometry in W1, W2, W3, W4 from LS8 includes all five years of publicly available WISE and NEOWISE reactivation \citep[][]{Meisner2019} and it is measured using the {\sc {Tractor}} algorithm \citep[][]{Lang14} at the position of {\it grz} detected sources.
\end{itemize}

\subsection{Optical spectroscopy \label{subsection:zspec}}
The eFEDS field has previously been observed by several spectroscopic surveys, 
most notably GAMA, SDSS, WiggleZ, 2SLAQ, LAMOST. Many of the existing spectra 
are of high enough quality that we can use them for science applications, 
in particular where we just need redshift and basic classification 
(i.e. deciding between star, QSO, or galaxy).  However, a careful collation 
and homogenisation of the existing spectroscopy catalogues is first 
needed to provide a reliable compendium of these data.

The largest body of spectroscopic redshift information comes from the SDSS survey \citep[][]{York2000, Gunn2006, Smee2013, Abdurrouf2022}, totalling more than 68k spectra of 61k science targets within the outer bounds of the eFEDS field.
We have collected archival public data from SDSS phases I-IV
\citep[][]{SDSSdr16}, as well as the results of the recent dedicated SPIDERS (Spectroscopic identifications of eROSITA sources) campaign \citep[][Merloni et al, in prep]{Comparat2020}, within SDSS-IV \citep[][]{Blanton2017} 
following-up eFEDS X-ray sources. A small team
formed from among the authors have visually inspected all of the
SDSS 1D spectra lying in the vicinity of eFEDS X-ray sources, correcting occasional pipeline
failures, and grading the spec-z onto a common normalised quality (\texttt{NORMQ}) scale between 3 and -1.
One can interpret \texttt{NORMQ} as follows: spec-z having \texttt{NORMQ}=3 are those with `secure' spectroscopic redshifts,
those with \texttt{NORMQ}=2 are `not secure' (although a large fraction are expected to be at the correct redshift),
spec-z with \texttt{NORMQ}=1 are `bad' (e.g. low SNR, problematic extraction, dropped fibres), and
those with \texttt{NORMQ}=-1 are `blazar candidates'. For completeness, we also retain SDSS-DR16 spectroscopic redshifts in the eFEDS field that do not lie near eFEDS X-ray detections, but only when they satisfy all the following criteria: \texttt{SN\_MEDIAN\_ALL}>2.0, \texttt{ZWARNING} = 0,  \texttt{SPECPRIMARY} = 1, 0 < \texttt{Z\_ERR} < 0.002.
An exhaustive description of the SDSS dataset within the eFEDS field will be
presented separately by Merloni et al. (in prep).

We also gathered published spectroscopic redshifts and classifications (hereafter `spec-z') from
the literature where they overlap with the eFEDS footprint, with the detailed breakdown presented in  Table~\ref{tab:available_specz}.
In order to gather spec-z from smaller surveys that might only contribute a few redshifts each, we have 
also queried the Simbad database \citep[as of 05/03/2021,][]{Wenger00}, in the vicinity of the eFEDS X-ray counterpart positions.

\begin{table*}[t]
\begin{center}
\begin{tabular}{@{}cccc}
\toprule
         Spectroscopic Survey & Quality threshold & $N_\mathrm{specz}$ & Data Release \& Reference \\
\midrule 
SDSS  & see text & 46837 & up to DR17; Merloni et al., in prep\\
GAMA  & NQ$\geq$4 & 26318 & DR3; \citet[][]{Baldry18}\\
WiggleZ & Q$\geq$4 & 13466 & Final DR; \citet[][]{Drinkwater18}\\
2SLAQ &  q\_{z2S}=1 &  953 & v1.2; \citet[][]{Croom09}\\
6dFGS & 4 $\leq$ q\_cz $\leq$ 6 &  365 & Final DR; \citet[][]{Jones09} \\
2MRS & non-null velocity  & 152 & v2.4; \citet[][]{Huchra12}\\
LAMOST & snrr $>$ 10, z $>$ -1, 0.0 $\leq$ z\_err $<$ 0.002 & 55866 & DR5v3; \citet[][]{Luo15}\\
Gaia RVS & non-null velocity & 15568 & DR2; \citet[][]{GaiaDR2}\\
Simbad & non-null redshift & 3915 & as of 05/03/2021; \citet[][]{Wenger00}\\
\midrule
Total unique objects &  & 143637 & \\
\bottomrule
\end{tabular}
\caption{Spectroscopic redshift measurements considered} within the eFEDS footprint (126<RA<146.2\,deg, -3.2<Dec<+6.2\,deg). $N_\mathrm{specz}$ is the number of spectroscopic redshift measurements that pass the quality threshold (applied to columns provided in the originating catalogue). For Simbad, the number of entries is limited to objects lying within 3\,arcsec\, of the optical coordinates of counterparts to eFEDS sources. Note that some astrophysical objects appear in two or more redshift catalogues.
\label{tab:available_specz}
\end{center}
\end{table*}
For the purposes of this work, we place greater weight on purity rather than completeness. Therefore, where the parent 
survey catalogues include some metric of quality/reliability, we have applied strict criteria to retain only the most secure spec-z information. The filtering criteria
applied to the original catalogues, and the number of spec-z
considered from each catalogue listed in Table~\ref{tab:available_specz}.
We assume that after these quality filtering steps, all the archival spec-z are `secure' (i.e. \texttt{NORMQ}=3),
except Simbad for which we adopt \texttt{NORMQ}=2, meant in this case to be interpreted as `not yet proven to be secure'.

All these spec-z were progressively collated into a single catalogue, with a single redshift and classification per sky position, using a match in coordinates between the coordinates listed for the nine input spectroscopic catalogs. An appropriate search radius (in the range 1-3\,arcsec) was chosen according to the expected positional fidelity and/or fibre sizes associated with each input spectroscopic catalogue. After this de-duplication step we are left with 143637 unique entries over the eFEDS field, of which 108834 are `secure' (i.e \texttt{NORMQ}=3).

\section{Counterparts identification: Methodology}\label{section:counterparts}
Because of the large PSF of the eROSITA telescopes and the small number of photons associated with typical X-ray detections, the $1\sigma$ rms positional uncertainties of individual X-ray sources can be several arcseconds. Specifically, in eFEDS the mean positional error is 4\farcs7\, and extends above 20\,arcsec\, only for a handful of sources \citep[][]{Brunner2021}. For the expected optical/infrared magnitude distribution of X-ray sources at the depth of the eFEDS \citep[see e.g., ][]{Merloni12, Menzel16}, the sky density of the relevant astrophysical source populations is often large. For this reason, the identification of the  true associations cannot be determined solely by closest neighbour searches, as there will be several potential counterparts within the error circle of any given X-ray source. Taking this into account, the identification of the counterparts of eFEDS point-like sources has been performed using two independent methods.

{\sc{nway}} \citep[][]{Salvato18a}, based on Bayesian statistics, and {\sc{astromatch}} \citep[][]{Ruiz18} based on the Maximum Likelihood Ratio \citep[MLR;][]{Sutherland_Saunders1992}, have been specifically developed to identify the correct counterparts to X-ray sources, independently of their Galactic or extragalactic  nature. In order to assess the probability (or likelihood) of an object to be the correct counterpart to an eFEDS sources, the two methods take first into account the separation between the sources, their positional accuracy and the number density of the sources in the ancillary data.
The difference between the methods resides then in the adoption of specific features able to distinguish an X-ray emitter (regardless its Galactic or extragalactic nature) from a random source in the field. Both methods determined the features (priors) using a representative training sample constructed using secure counterparts to X-ray sources detected in 3XMM. In the case of {\sc{nway}} the prior is determined also by comparing the features of the sources in the training sample with the features of the field sources present within 30\,arcsec\, from the 3XMM sources. The disentangling power of the priors was then tested on a blind validation sample of 3415 {\it Chandra} sources with secure counterparts, where the accuracy of the {\it Chandra} position was made "eFEDS-like" (see Subsection~\ref{subsection:blind}). The detailed description of the construction of training, validation and respective fields sample is presented in Appendix~\ref{Appendix:TrainingAndBlindSample}. Here we provide a short description of {\sc{nway}}  and {\sc{astromatch}} and how their respective priors have been determined.

\subsection {{\sc{nway}} enhanced  with  photometric priors defined via machine-learning}
In addition to astrometry, i.e. the separation between an X-ray source and a candidate counterpart, the associated positional uncertainties and the number densities of the sources in the two catalogs,  the photometry  of potential counterparts is valuable information to determine whether or not they are associated to a given X-ray detection. Traditionally, the likelihood ratio associated to angular distance was multiplied by a factor accounting for the magnitude distributions and the sky density of a population of X-ray sources and background objects \citep[e.g][]{Brusa2005,Brusa2007,Luo10}. In {\sc{nway}}, this idea was re-formulated in the Bayesian formalism, in the following way. 

Given some data $D$, the posterior association probability $P(H\mid D)$ is related to the prior probability
$P(H)$ via the likelihood $P(D\mid H)$,  $P(H\mid D) \propto P(H) \times P(D\mid H)$. $P(H)$ is computed from the source densities in each catalogue. If photometric information (or any other feature, in fact) is used, then the likelihood becomes:  $P(D\mid H)=P(D_\phi\mid H) \times P(D_m\mid H)$ where $D_\phi$ and $D_m$ refer to the astrometric and photometric information, respectively. For any possible association, the modifying factor $P(D_m\mid H)$ is computed from the feature (e.g., magnitude or colour) $m$ of the counterpart candidate and from the expected distribution of this observable for X-ray sources and field (non X-ray) sources. We call such factors "priors" to {\sc nway}, as they enter as {\it a priori} information in the ultimate matching process. These priors are posteriors previously learned from other data. For further details on the formalism we refer to \citet{Salvato18a} and the {\sc{nway}} documentation\footnote{\texttt{https://github.com/JohannesBuchner/NWAY}}.

In order to take full advantage of the LS8 ancillary catalogue, we have extended this approach for the eFEDS counterpart identification. Instead of using a subset of magnitudes, colours and their associated distributions, we have trained a Random Forest classifier \citep[\texttt{sklearn} implementation,][]{Pedregosa11} on a large number of features to reliably map the available Legacy DR8 information to real X-ray sources and real field objects the details of which are described in next subsection. The trained classifier is then used to predict the probability of all counterpart candidates to be X-ray emitting  taking into account also  the spatial information, as described in subsection~\ref{subsection:apply_NWAY}. This probability is directly used to compute $P(D_m\mid H)$.  
In the following section, we describe the definition of the features in the training sample.

\subsubsection{Random Forest prior: training and performance}

From the 3XMM training sample described in Appendix~\ref{Appendix:TrainingAndBlindSample}, we have extracted a set of photometric and astrometric features. The training features are listed and described in Table~\ref{tab:rf_features}. X-ray sources are flagged as target class "1", field objects as target class "0". About $15\%$ of the 3XMM training samples (61821 sources) are randomly extracted for testing purposes and not further considered in the training procedure. The baseline model is composed of 200 trees, allowing decision split points if at least 8 samples are left in each branch. All of the 22 features can be used for the decision-tree building, which makes use of bootstrap samples of the training set. 

By construction the training sample is highly imbalanced, since the field objects strongly outnumber the X-ray sources. We therefore opted for a weighting scheme, automatically adjusting weights of training examples for the class imbalance.

\begin{figure}[ht]
\centering
\includegraphics[width=9cm]{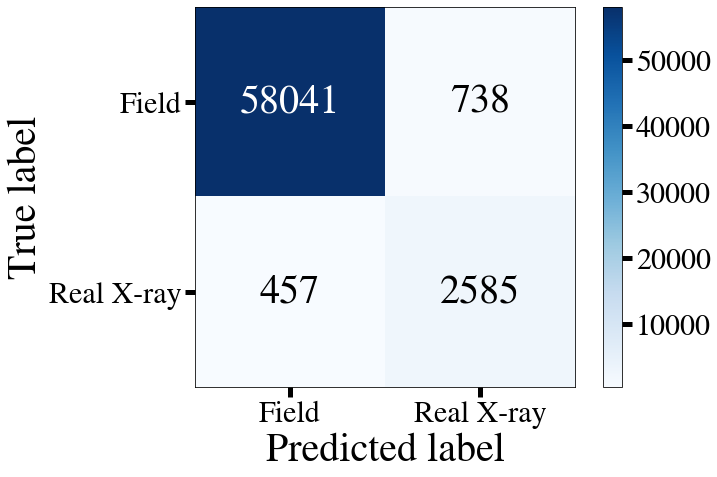}
\caption{Confusion matrix resulting from the random forest prediction on an independent test set. X-ray sources are labelled as "Real X-ray" while field objects as "Field". Numbers on the right downward diagonal correspond to correctly predicted classes. At this step the separation between the sources and the X-ray position are not considered.}
\label{fig:rf_test_cm}
\centering
\end{figure}

The trained model is evaluated on the test set, resulting in the confusion matrix presented in Figure~\ref{fig:rf_test_cm}. We note that the cut in the class prediction for the presented confusion matrix is made at $p_{\rm X-ray}=0.50$, where $p_{\rm X-ray}$ is the predicted probability that a counterpart candidate is X-ray emitting. Since {\sc{nway}} uses the continuous predicted probability as modifying factor for the likelihood $P(D\mid H)$, real counterparts with rare or untypical photometric features, i.e with $p_{\rm X-ray}\lesssim 0.50$, may still be selected by the algorithm if the astrometric configuration favours them. 
We obtain a high recall fraction of $2585/(2585+457)=85\%$,
while the fractional leakage of contaminating field objects remains low: $738/(738+58041)=1\%$.We note that $p_{\rm X-ray}$ is only computed from the photometric and proper-motion properties of the LS8 sources. In particular it does not depend on coordinates and positional uncertainties. As discussed in the previous section, this allows us to split the likelihood of a match into independent astrometric and photometric terms: $P(D\mid H)=P(D_\phi\mid H) \times P(D_m\mid H)$. $P(D_m\mid H)$, the photometric term, is directly related to $p_{\rm X-ray}$.
\begin{table}[ht]
\begin{tabular}{@{}ll@{}}
\toprule
Feature                     & Description                                            \\ \midrule
flux\_*/mw\_transmission\_* & deredenned flux in \it{g,r,z,W1,W2} \\
gaia\_phot\_*\_mean\_mag    & original GAIA phot. in \it{G, G$_{\rm bp}$, G$_{\rm rp}$}             \\
snr\_*                      & S/N for \it{g,r,z,W1,W2,G,G$_{\rm bp}$,G$_{\rm rp}$}        \\
$\sqrt{pmra^2+pmdec^2}$     & Gaia proper motion                                     \\
parallax                    & Gaia paralllax                                         \\
\it{g-r, r-z, z-W1, r-W2}       & dereddened colours                   \\ \bottomrule
\end{tabular}
\caption{Final list of LS8 training features used to model the photometric prior in {\sc NWAY}.}
\label{tab:rf_features}
\end{table}
\subsubsection{{\sc{nway}} association run \label{subsection:apply_NWAY}}
Using the trained model, we predict $p_{\rm X-ray}$ for all LS8 sources in the eFEDS field. We then run the {\sc{nway}} matching procedure using the ratio $p_{\rm  X-ray}/(1-p_{\rm  X-ray})$ for $P(D_m\mid H)$. This is done by adding $p_{\rm X-ray}$ as a column to the LS8 catalog and activating it as a prior column in {\sc nway} with the {\texttt{--mag} option}. We set a radius of 30\,arcsec\, from each eFEDS X-ray source, considering all LS8 sources within this radius. This may appear to be a relatively large maximal separation, given the mean eFEDS positional error of 4.5\,arcsec\,; however, we want to account also  for the largest positional uncertainties  of a few objects in the eFEDS source catalogue \citep[see][]{Brunner2021} and the use of a large search radius minimises the probability of missing counterparts that are widely separated from the X-ray centroid position. The sky coverage of eFEDS and LS8 are respectively 140\, deg$^2$ and ${N_{\rm eFEDS} \times \pi \times (30'')^2 - A_{\rm overlap}}$ where ${N_{\rm eFEDS}}$ is the number of eFEDS sources (point-like or extended) and $A_{\rm overlap}$ the overlap area of neighbouring search windows around the X-ray sources. As described in the Appendix of \citet{Salvato18a}, the area coverages are used for computing the number densities, which in turn are used in the computation of the probability for an eFEDS source to have a counterpart (p\_any) and the probability for each source in LS8 to be the right counterpart (p\_i). These are the two quantities that are then used for assigning a counterpart to an eFEDS source: while the LS8 source with the highest p\_i is considered to be the best available counterpart, we use the p\_any value to decide whether the identification of the counterpart is reliable (see ~\ref{subsection:thresh} for details).

\subsection{The MLR approach}
The maximum likelihood ratio (MLR) statistic for the correct pairing of sources from multiple catalogs  was introduced in the seminal work of \cite{Sutherland_Saunders1992} and is widely used, although mostly pairing only two catalogs.
For sources detected at two different wavebands and separated by angular distance $r$ on the plane of the sky, the likelihood ratio provides a measure of the probability that the two sources are true counterparts normalised by the probability that they are random alignments. Quantitatively this is estimated as:

\begin{equation}\label{eq:age:lr}
LR =\frac{q(\overrightarrow{m}) \cdot f(r)}{n(\overrightarrow{m})}, 
\end{equation}
\noindent where $q(\overrightarrow{m})$ is the prior knowledge on the properties of the true associations, such as the distribution of their apparent magnitudes at given spectral window, their colours and/or the spatial extent of the observed light in a given waveband. The collection of all possible source properties for which a prior probability can be estimated is represented by the vector $\overrightarrow{m}$. The quantity $n(\overrightarrow{m})$ is the sky density of all known source populations in the parameter space of $\overrightarrow{m}$.  It measures the expected contamination rate from background/foreground sources that are randomly projected on the sky within distance $r$ off a given position. The probability that the true associations are separated by distance $r$ is measured by the quantity $f(r)$. This depends on the positional uncertainties of matched catalogues. 

For the MLR applied to the eFEDS X-ray sources a multi-dimensional prior is used that combines knowledge of the optical and mid-infrared colours/magnitudes of X-ray sources as well as their optical extent, i.e. point-like vs extended. 

The version of MLR applied to the eFEDS work is based on the {\sc{astromatch}}\footnote{\texttt{https://github.com/ruizca/astromatch}} implementation. This tool has been specifically designed to deal with the complexity of wide-area surveys that contain a very large number of sources. The HEALPix Multi-Order Coverage map (MOC\footnote{\texttt{https://www.ivoa.net/documents/MOC/}}) technology is used to describe the footprint of a catalogue of astrophysical sources. The KD-tree library as implemented in the {\sc{astropy}} package \citep{astropy:2013, astropy:2018} is used to accelerate spatial searches of potential counterparts within a radius $r$ of a given sky position.  The core {\sc{astromatch}} functionality is expanded to enable the use of multidimensional priors. The version of  {\sc{astromatch}} adopted in this work is therefore a fork (\texttt{github.com/ageorgakakis/astromatch}) of the main development branch. 

Like for {\sc{nway}}, the optical counterparts are investigated out to a maximum radius of 30\,arcsec.  We assume that the positional uncertainties of the X-ray and optical catalogues follow a normal distribution. The quantity $f(r)$ is therefore represented by a Gaussian with $\sigma$ parameter estimated as the sum in quadrature of the X-ray and optical positional uncertainties. 

The priors are generated using the training sample defined in Appendix ~\ref{subsection:3xmmref}. 
The LS8 photometric properties of the sources in that sample were explored to identify parameter spaces, in which they separate from the general LS8 field population.
After some experimentation we opted for the following 3 independent priors:
\begin{itemize}
\item A space that includes the WISE colour $W1-W2$, the WISE magnitude $W2$ and the optical extent of a source. For the latter we use the LS8 parameter {\sc{type}}, which provides information on the optical morphology of sources. In our application we only differentiate between optically unresolved ({\sc{type="psf"}}) and optically extended ({\sc{type$\neq$"psf"}}) populations.

\item A space that includes the optical/WISE colour $r-W2$, the optical magnitude $g$ and the optical extent of a source. For the latter we use the Legacy-DR8 parameter {\sc{type}} as explained above. 

\item The distribution of the Gaia $G$ magnitudes listed in the LS8 catalogues. This is to identify X-ray sources associated with very bright counterparts. 

\end{itemize}

The distribution of the training sample  sources in the parameter spaces above is used to define two 3-dimensional and one 1-dimensional independent priors. These  are provided as input to the {\sc{astromatch}} code, together with the distribution of the sources in the field population  when computing the association for eFEDS.
For a given eFEDS source all the potential associations within the search radius of 30\,arcsec\, are identified. Each of them is assigned one LR value for each of the 3 priors using Equation~\ref{eq:age:lr}.The LS8 source with the highest value of LR from one of the three priors is considered to be the counterpart.

\section{Comparing {\sc{nway}} and {\sc{astromatch}} on a validation sample \label{section:nway_astromatch_validation}}
In order to compare completeness and purity of {\sc{nway}} and {\sc{astromatch}}, the same setting adopted for identifying the counterparts to eFEDS has been used for determining the best counterparts to a blind, validation sample  of 3415 counterparts to {\it Chandra} sources (see Appendix~\ref{Appendix:TrainingAndBlindSample}). This validation sample was used as a truth table to test the performance of {\sc{nway}} and {\sc{astromatch}} for finding counterparts and to define the \texttt{p\_any} and \texttt{LR\_BEST} thresholds  above which a counterpart is considered secure. 

\subsection{The eROSITA-like validation sample \label{subsection:blind}}

The {\it Chandra} sources were assigned eROSITA positional errors by randomly sampling from the astrometric uncertainties listed in the core eFEDS source catalogue. We account for the flux dependence of these uncertainties by matching any given Chandra source with a certain flux from 0.5-2\,keV to only those eFEDS sources with similar 0.6-2.3\,keV flux within a margin of 0.5\,dex. The flux transformation between the Chandra and eFEDS spectral bands is small, e.g. about 2\% for a power-law spectral energy distribution with $\Gamma=1.9$, and is ignored. The positional uncertainty, $\sigma$, assigned to each of the {\it Chandra} sources can be split into a right-ascension and a declination component. It is assumed that these two uncertainties are equal and therefore $\delta RA$ = $\delta Dec = \sigma/\sqrt{2}$. Under the assumption that both the $\delta RA$  and $\delta Dec$ are normally distributed, the total radial positional uncertainty follows the Rayleigh distribution with scale parameter $\sigma$.

Instead of directly using the assigned $\sigma$ as the astrometric error to be applied to the Chandra positions to make them resemble the eFEDS astrometric accuracy, we prefer to add further randomness to the experiment. For each {\it Chandra} source the assigned $\sigma$ is treated as the scale factor of the Rayleigh distribution and a deviate is drawn, which represents the positional error. This is applied to the sky coordinates of the optical counterpart of the {\it Chandra} source and the new offset position is taken as the centroid of the X-ray source in the case of an eFEDS-like observation. 

\subsection{Probability thresholds definition}\label{subsection:thresh} 

The identification of the LS8 counterparts to the {\it Chandra} eFEDS-like sources was performed by {\sc{nway}} and {\sc{astromatch}} using the same setup adopted for the real eFEDS observation.
 The resulting catalogue of best counterparts was matched  with the true associations, providing a direct comparison between the methods, and, at the same time providing  a measure of the false-positive identification rate of the eFEDS counterpart catalogue.
 
\begin{figure}[t]
\centering
\includegraphics[width=8cm]{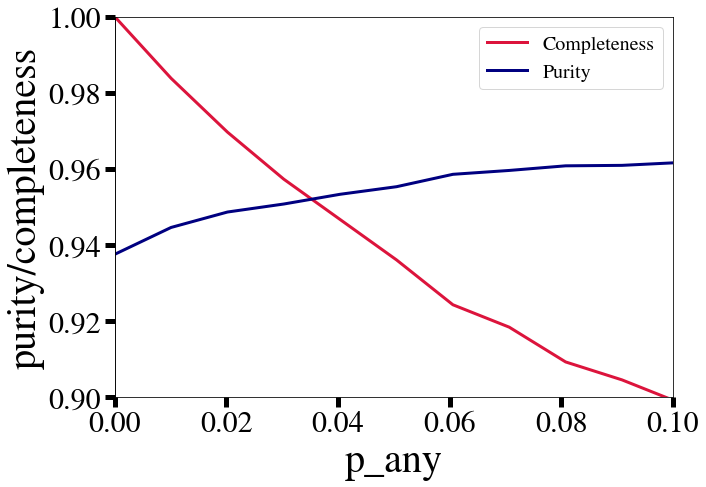}
\includegraphics[width=8cm]{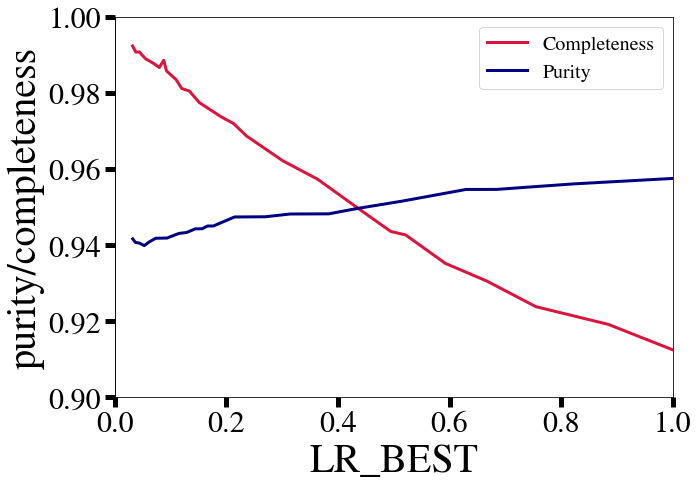}
\caption{Purity (blue solid line) and completeness (red solid line) as a function of {\texttt{p\_any}} for the association of the CSC2 eROSITA-like validation sample made with {\sc{nway}} (top panel) and as a function of {\texttt{LR\_BEST}}, for the {\sc{astromatch}} (bottom panel).}
\label{fig:purity}
\end{figure}

First, we compared the primary identifications returned by {\sc{nway}} and {\sc{astromatch}} to true identifications stored in the validation sample.  {\sc{nway}} and {\sc{astromatch}} identify correctly 3216/3394 (95\%)  and 3024/3394 (89\%) of the sources, respectively. {\sc{nway}} has a higher success rate. Additionally, NWAY has a smaller fraction of sources with a second possible counterpart (115 sources against 367).
Another way to look at the results is to compare purity and completeness for the two methods.
At any given value of \texttt{p\_any}/\texttt{LR\_BEST} we define as {\it purity} the fraction of sources with the correct identification. In addition we define as {\it completeness} the fraction of sources for which we can assign a counterpart (see Figure~\ref{fig:purity}).
Both methods have very high purity and completeness, with {\sc{nway}} providing a sample that is purer, consistent with the fact that very few sources have a second possible counterpart, in addition to the correct one.  This, combined with the success rate, makes {\sc{nway}} the more robust method for determining the counterparts. Its strength comes first of all  from the capability to account for complicated priors involving multiple features (essentially resembling an entire SED, together with other physical properties), from different catalogs at the same time. Furthermore, the Bayesian statistics upon which   {\sc{nway}} is based, also allows accounting for sources that are lacking one or more of the features.

Similarly to what is traditionally done in Maximum Likelihood \citep[see e.g.,][]{Brusa2007}, the intersection between the completeness and purity can be used for defining a threshold above/below which the counterparts is considered reliable. This corresponds to 0.035 for \texttt{p\_any} and 0.45 for \texttt{LR\_BEST}.  

-------------------------------------------------------------------------------------------------------------
\section{Determination of the counterparts to eFEDS sources \label{section:comparison}}
While for the large majority of the cases the two methods select the same counterparts, there are cases where they do not agree or where they identify multiple likely associations. In the following we describe the procedure that we have adopted for the final assignment of the counterparts.

Then, after the consolidation of the counterpart, we describe  a further test for consistency that was done by comparing the results of the association with an independent method, HamStar (Schneider et al., 2021), which is tuned for the identification of Galactic coronal X-ray emitters (Section~\ref{sect:other_methods}).  Important to note is that the same process is then repeated for the 246 sources in the eFEDS hard point-source catalog. From now on, all numbers and descriptions are given for the main sample, unless specified otherwise.

\subsection{Comparison of counterparts from {\sc{nway}} and {\sc{astromatch}}\label{section:nwayLR}}

For  24193/27369 (88.4\%) eFEDS point like sources in the main sample, {\sc{nway}} and {\sc{astromatch}} point at the same counterpart and disagree for 3176 (11.6\%)  of the cases. The numbers are quoted at this stage regardless on the {\texttt{p\_any}} or {\texttt{LR\_BEST}} thresholds, that will instead be used later in order to assign a flag for the quality of the proposed counterpart.

\begin{figure}[t]
\centering
 \includegraphics[width=9cm]{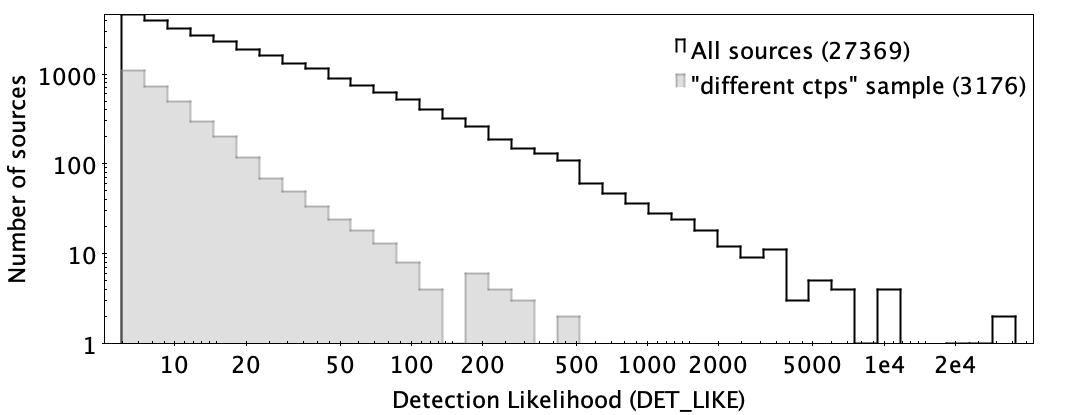}
\caption{Number of sources as a function of detection likelihood  for  the entire sample of eFEDS Main catalog sources (black histogram) and for the sources with {\sc{nway}} and {\sc{astromatch}} indicating different counterparts (grey shaded histogram).}
\label{fig:discDET}
\centering
\end{figure}

Table~\ref{tab:detml} summarises the number of eFEDS sources with the agreement/disagreement between the two methods as a function of detection likelihood of the X-ray source. 
Sources with low detection likelihood values have, on average, larger X--ray positional errors and a larger number of spurious sources is expected from simulations \citep[][]{Brunner2021, TLiu2021}. It is therefore not surprising that the largest discrepancies are observed at the lowest detection likelihoods (Figure~\ref{fig:discDET}).
In fact,  the disagreement drops from 11.6\% to 6.5\% when considering only eFEDS sources with \texttt{DET\_LIKE} grater than 10, suggesting that at low detection likelihood a fraction of eFEDS sources might be spurious detections where {\sc{nway}} and {\sc{astromatch}} assign a different "field" source. The notion that these are "field" sources is also supported by the fact  that for about 50\% of eFEDS sources with \texttt{DET\_LIKE} below 10 and  with different counterparts, both \texttt{p\_any} and \texttt{LR\_BEST} are below threshold. 

\begin{table}
\centering
\begin{tabular}{@{}lr|rr@{}}
\toprule
Sample        & Number & \multicolumn{2}{c}{Counterparts}       \\
\midrule
                     &  & "same" & "different"                         \\
\midrule
DET\_LIKE\_0$>6$ &  27369  & 24193 & 3176 (11.6\%) \\ 
DET\_LIKE\_0$>8$ &  21410  & 19162 & 1795 (8.4\%) \\ 
DET\_LIKE\_0$>10$ &  17574  & 16435 & 1136 (6.5\%) \\ 
 \\ 
\bottomrule
\end{tabular}
\caption{Comparison of matches between {\sc{nway}} and {\sc{astromatch}}. In the last column the fraction of the "different ctps" with respect to the whole sample is also reported.}
\label{tab:detml}
\end{table}

Table~\ref{tab:match_rel} summarises the comparison between the two methods also taking into account the reliability of the associations.
In this table we further split the sample with the  same counterparts ("same ctps" for brevity) in two subsamples: one for which the proposed counterparts are the only associations suggested by both methods ("single solutions"; 86.3\% of the entire sample) and one for which, although both methods point to the same associations, at least an additional counterpart at lower significance exists from at least  one methods ("multiple solutions; 2.1\% of the entire sample).  

The different priors  and the different methods used for assigning the counterparts explain the selection of different counterparts in the "different ctps" sample. {\sc{astromatch}} uses three priors, but they are used each independently and for any given eFEDS source the counterpart is assigned by the prior with the higher probability. Instead, {\sc{nway}}  uses all the features at the same time and the best counterpart is the one that mimics best the training sample in a multidimensional space. We consider this second method more reliable and for this reason we decided to list always as primary the counterpart suggested by {\sc{nway}}, unless \texttt{LR\_BEST} is above threshold and \texttt{p\_any} is not.  

Interestingly, we note that, in the "different ctps" sample, for about 25\% of the cases the primary counterpart assigned by one method is the secondary counterpart assigned by the other.

\subsection{Assigning a quality to the proposed counterparts \label{section:qualCTP}}

As a consequence of the discussion above, each counterpart in the catalog has been flagged as following ([number] refers to the number of sources in the category): \\

\begin{itemize}
\item \texttt{CTP\_quality}=4:  when {\sc{nway}} and {\sc{astromatch}} agree on the counterpart and both \texttt{p\_any} and \texttt{LR\_BEST} are above threshold [20873 sources; black in Table~\ref{tab:match_rel}];

\item \texttt{CTP\_quality}=3: when {\sc{nway}} and {\sc{astromatch}} agree on the counterpart but only one of the methods assigns the counterpart with a probability above threshold [1379 sources; blue in Table~\ref{tab:match_rel}];

\item \texttt{CTP\_quality}=2: when there is more than one possible reliable counterpart.
 This includes a) all the sources in the "different ctps" sample with at least one probability above threshold
 and b) the sources in the "same ctps" sample with possible secondary solutions [2522 sources in total;{ cyan in Table~\ref{tab:match_rel}]}. This latter case, due to the  low spatial resolution of eROSITA, implies that both sources are contributing to the X-ray flux. 
 A supplementary catalog with the properties of the secondary counterparts for these 2522 sources is also released (Section~\ref{section:catalogs}). 

\item \texttt{CTP\_quality}=1: when {\sc{nway}} and {\sc{astromatch}} agree on the counterpart but both \texttt{p\_any} and \texttt{LR\_BEST} are below threshold [1370 sources;  purple in Table ~\ref{tab:match_rel}]; Note that a probability below threshold does not necessarily imply a wrong counterpart. It could also indicate that the counterpart is correct but its features are not sufficiently mimicking those in the training sample.

\item \texttt{CTP\_quality}=0: when {\sc{nway}} and {\sc{astromatch}} indicate different counterparts and both \texttt{p\_any} and \texttt{LR\_BEST} are below threshold [1225 sources; red in Table~\ref{tab:match_rel}].
\end{itemize}
Counterparts with quality 4,3,2 are considered reliable (90.5\% of the main sample and 93.9\% of the hard sample), while sources with quality 1 or 0 are considered unreliable (9.5\% of the main sample and 6.1\% of the hard sample).

\begin{table}
\centering
\begin{tabular}{@{}l|cc@{}}
\toprule
\multicolumn{3}{c}{"Same ctps", single solutions -- 23622  (86.3\%)}          \\
\midrule
 & {\texttt{p\_any}} &  \texttt{p\_any}       \\   &    >0.035     &      <0.035             \\
\midrule
\texttt{LR\_best}$>0.45$ &  {\bf 20873 (0.763)} & {\bf \textcolor{blue}{561 (0.020)}} \\ 
\medskip
\texttt{LR\_best}$<0.45$ &  {\bf \textcolor{blue}{818 (0.030)}}  & {\bf \textcolor{purple}{1370 (0.050)}}  \\ 
  \toprule
\multicolumn{3}{c}{"Same ctps",  multiple solutions -- 571  (2.1\%)}          \\
\midrule
 & {\texttt{p\_any}} &  \texttt{p\_any}            \\   
    &    >0.035     &      <0.035                \\ 
    \midrule

\texttt{LR\_best}$>0.45$ &  {\bf \textcolor{cyan}{505 (0.018)}} & {\bf \textcolor{cyan}{7 (3x$10^{-4}$)}} \\ 
\medskip
\texttt{LR\_best}$<0.45$ &  {\bf \textcolor{cyan}{59 (0.002)}}  &  - \\ 

   \toprule
   \multicolumn{3}{c}{"Different ctps"  -- 3176  (11.6\%) }          \\
\midrule
           & \texttt{p\_any} &  \texttt{p\_any}   \\      
 & >0.035 &  <0.035 \\ 
\midrule
\texttt{LR\_best}$>0.45$ &  {\bf \textcolor{cyan}{1243 (0.045)}} & {\bf \textcolor{cyan}{226 (0.008)}} \\ 
\texttt{LR\_best}$<0.45$ &  {\bf \textcolor{cyan}{478 (0.017)}}  & {\bf \textcolor{red}{1225 (0.045)}}   \\ 
\medskip
HamStar$^1$ &  {\bf \textcolor{cyan}{4 (2x$10^{-4}$)}}  &  - \\ 
 \bottomrule
 \end{tabular}
\caption{Counterparts quality summary. Comparison of matches with {\sc{nway}} and {\sc{astromatch}} as a function of their respective thresholds, split between "same counterparts" (for both cases of single and multiple counterparts) and "different counterparts" classes; in parenthesis the fractions of the total sample. The numbers in each box are colour-coded by their \texttt{CTP\_quality} value (see text for more details): Thick black = 4; thick blue = 3; thick cyan = 2; thick purple = 1 and thick red = 0. $^1$ Objects for which HamStar would point to a different counterpart with \texttt{p\_stellar}>0.95 are by definition sources with \texttt{CTP\_quality}=2 (see text for details).}
\label{tab:match_rel}
\end{table}

\subsection{Comparison with an independent association method tuned to stars: HamStar \label{sect:other_methods}}

The content of the eFEDS point-source catalog has also been analysed in order to specifically identify stellar coronal X-ray emitters with sufficiently well-defined properties. This method, called 'HamStar' in the following, is based on the properties expected for this type of star; the details  are presented in \citep[][]{Schneider2021}. In short, HamStar performs a binary classification between stellar coronal emitters and other objects. This classification is based on the concept of eligible stellar counterparts, i.e., the match catalog contains only stellar objects that may reasonably be responsible for the X-ray sources. Specifically, the parent sample that HamStar uses includes only sources from Gaia EDR3 that:

\begin{itemize}
\item are brighter than 19th magnitude in G band (implied by the stellar saturation limit of $L_{X}/L_{\rm bol}\lesssim10^{-3}$ and the depth of eFEDS);
\item have accurate magnitudes in all three Gaia photometric bands (to apply colour-dependent corrections);
\item have a parallax value at least three times larger than the parallax error (to select only genuine stars).
\end{itemize}

Then, a positional match between sources in eFEDS and the eligible stellar candidates is made, considering all sources within 5$\sigma$ of the positional uncertainty of the eFEDS source as possible stellar counterparts. 
Finally, the matching probabilities of all possible counterparts are adjusted based on the value of the two dimensional Bayes map at the counterpart?s Bp-Rp colour and ratio between X-ray to G-band flux. With the HamStar algorithm, 2060 eFEDS sources are expected to be stellar \citep[][]{Schneider2021}. The vast majority of them have a unique Gaia counterpart, and only 83 eFEDS sources have two possible eligible counterparts. 

Of the 2060 eFEDS sources with a counterpart from HamStar,  1883 have the counterpart identified in this work that is less than 2\,arcsec from the counterpart proposed by Hamstar and we assume to be the same source. We visually inspected the cutouts of the 29 sources for which the separation between the counterpart proposed by Hamstar and this work is between 2 and 3\,arcsec, and concluded that for 9 sources the counterparts are the same but the sources are heavily saturated in LS8 so that the coordinates are not sufficiently precise.  
This corresponds to an  92\%  agreement; incidentally, this value corresponds almost exactly to the expected reliability and completeness of HamStar \citep[][]{Schneider2021}. All these sources will be then classified as "Secure Galactic" in Section~\ref{section:properties}.

HamStar applies well-understood X-ray-to-optical properties of stars to a well-defined subsample of Gaia sources. On the other hand, the training samples used by {\sc{nway}} and {\sc{astromatch}} include various classes of X-ray emitters: stars and compact objects, AGN and galaxies, including the bright ones at the  centre of clusters (BCG).
We considered the prior defined by {\sc{nway}} and {\sc{astromatch}} more representative of the population of X-ray emitters at large and decided  to keep the counterpart assigned in the previous section  rather than changing counterparts for the  177 sources for which the methods point to different counterparts. However, we degrade the \texttt{CTP\_quality}, because of the presence of an alternative solution. Interestingly only 4/211 sources were considered secure, with \texttt{CTP\_quality}==3, while all other counterparts had already a low \texttt{CTP\_quality}.


\subsection{Separation and magnitude distribution of the counterparts}

\begin{figure}
\centering
\includegraphics[width=9cm]{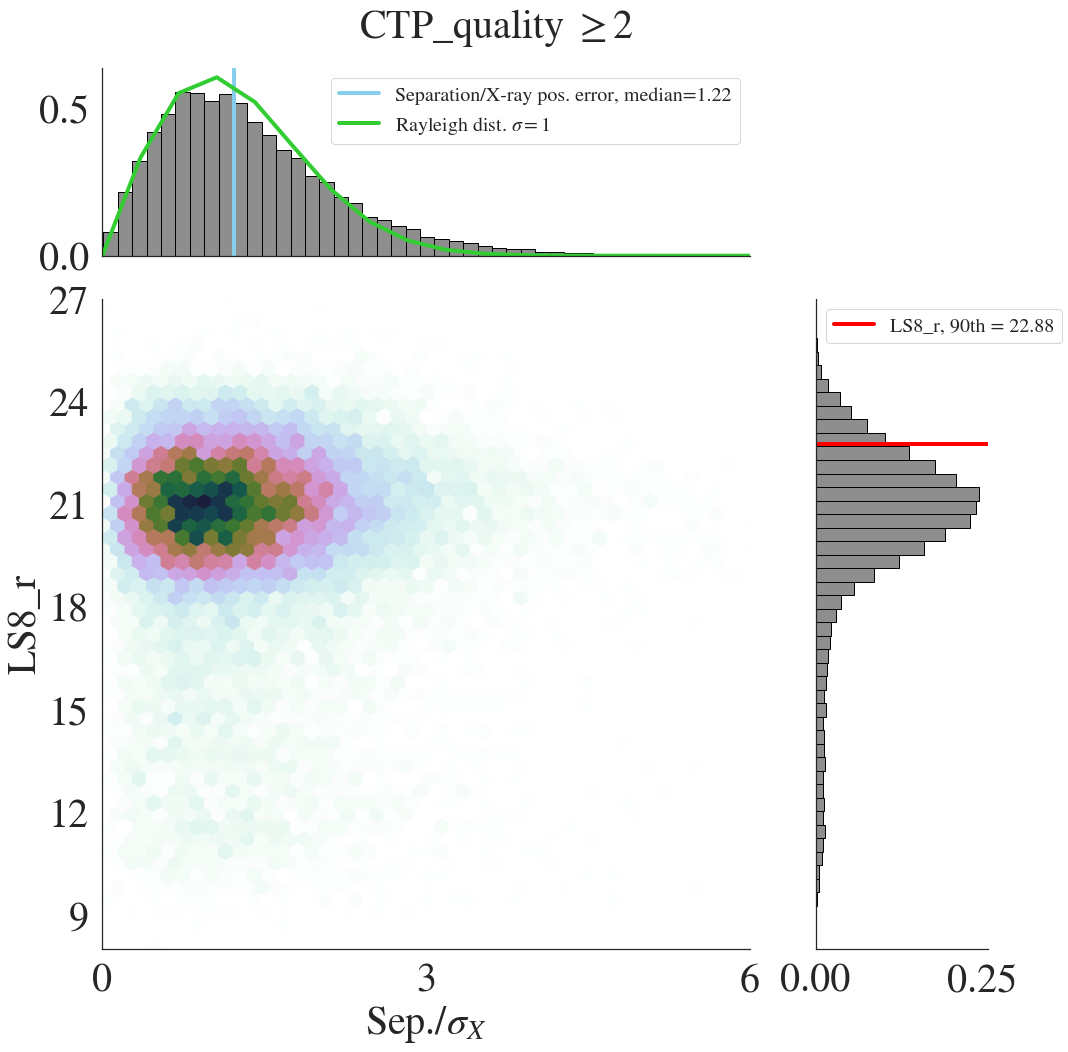}
\caption{Separation between X-ray position and the selected counterpart normalised by the 1-dimensional positional error of the X-ray source, as a function of its {\it r} band magnitude for the sources with secure counterparts ({\texttt{CTP\_quality}}$\ge$2). The hexagons are colour coded linearly according to the counts in the specific bins. The marginal histograms have linear y-axis. The 90-th percentile of the r-band magnitude distribution (22.88) and the median of the normalised X-OIR separation (1.22) are also reported in the marginal 1D histograms. The 1$\sigma$ Rayleigh distribution expected for the normalised separations is overplotted in green.}
\label{fig:Xsep_2d}
\centering
\end{figure}

For 24427/24774 (98.5\%) of the sources having {\texttt{CTP\_quality}}$\ge$2, the separation between the X-ray position and the assigned LS8 counterpart is smaller than 15\,arcsec, with a mean of 4.3\,arcsec. As might be expected, there is a trend for larger average X-ray-optical separations at smaller values of DET\_LIKE; lower detection likelihood sources typically have larger X-ray positional uncertainty \citep[see][]{Brunner2021}. The distribution of the observed X-OIR separations normalised by the X-ray positional uncertainty is shown in Figure~\ref{fig:Xsep_2d} as function of the {\it r} magnitude of the counterpart. The distribution is broadly comparable to the expectation of a Rayleigh distribution with scale factor = 1.

\begin{figure*}
\centering
\includegraphics[width=9cm]{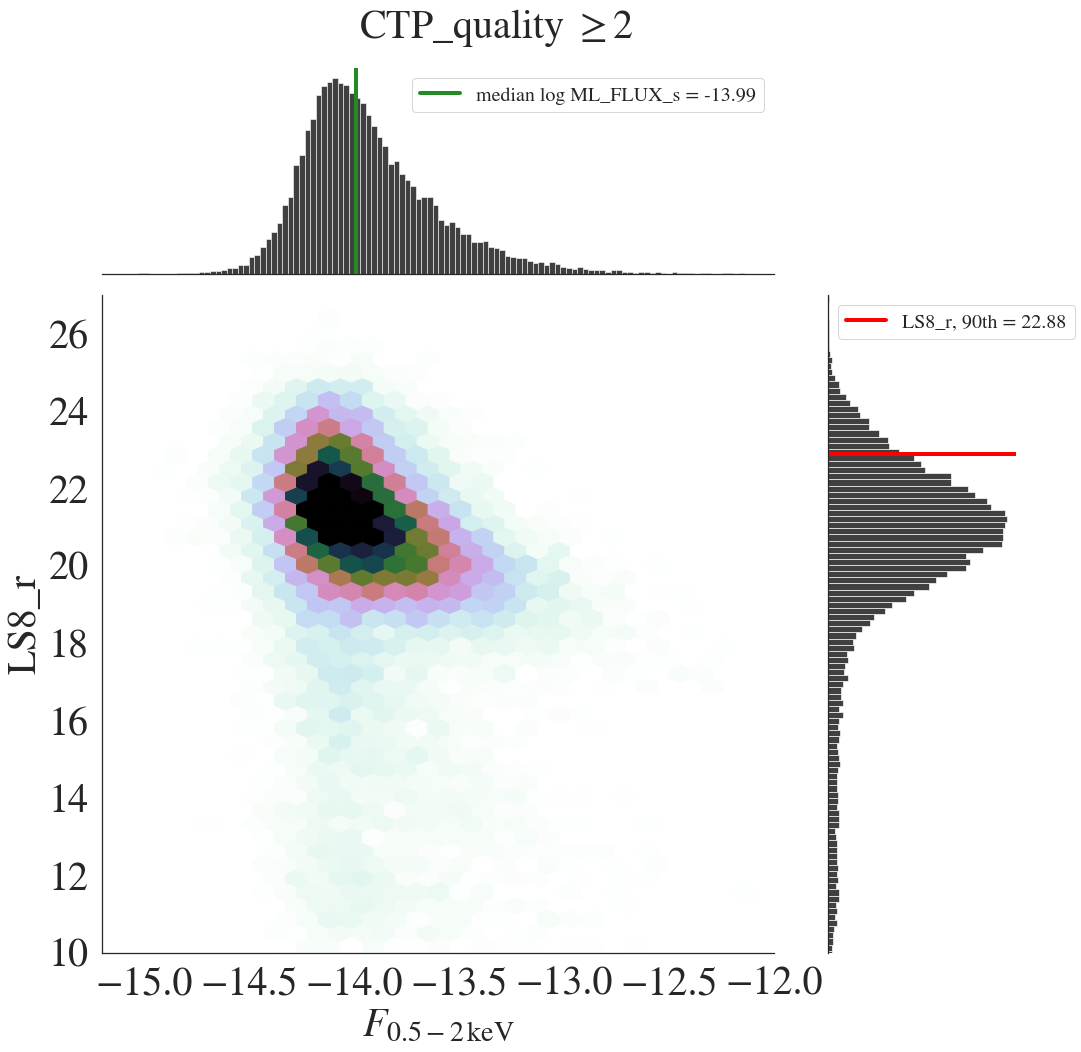}
\includegraphics[width=9cm]{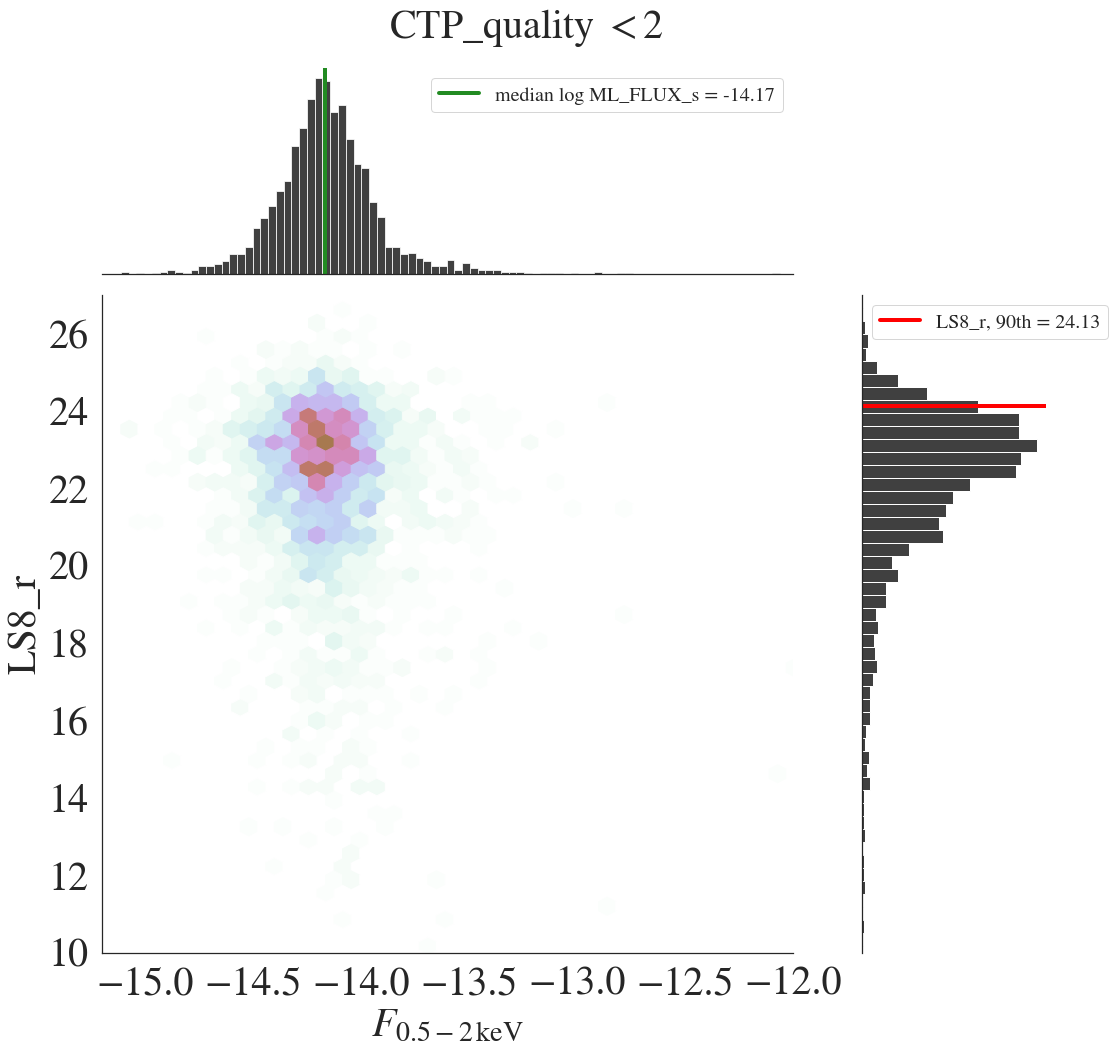}
\caption{Magnitude distribution of the counterpart vs the X-ray flux for sources with {\texttt{CTP\_quality}}$>=2$ (left) and {\texttt{CTP\_quality}}$<2$ (right). A cut at log ML\_FLUX\_s $> -20$ has been applied.} The magnitude distribution is clearly different. The green lines represent the median 0.5-2 keV flux 
(-13.99 and -14.17 respectively).
The red lines mark the 90-th percentile of the r-band magnitude distribution (22.88 and 24.13, respectively).
\label{fig:Xmag_2d}
\end{figure*}

In Figure~\ref{fig:Xmag_2d} we show the distribution of the sample in X-ray flux versus optical magnitude space, with the sample subdivided into those objects with more secure counterparts ({\texttt{CTP\_quality}}$\ge$2), and those with less reliable counterparts ({\texttt{CTP\_quality}}$\le$1). The less reliable counterparts tend to have fainter optical magnitudes for a given X-ray flux than the more secure counterparts.

\section{Source Characterisation and Classification \label{section:properties}}
After the identification of the counterparts, to understand physical processes and populations, the different classes of objects need to be separated. The most important separation is between extragalactic sources (galaxies, AGN, QSOs) and galactic sources (stars, compact objects, etc.). In the following, we describe how the classification of the sources was done and the validation tests performed.

\subsection{Galactic and extragalactic sources}
\begin{figure*}[t]
\centering
\includegraphics[width=0.8\textwidth]{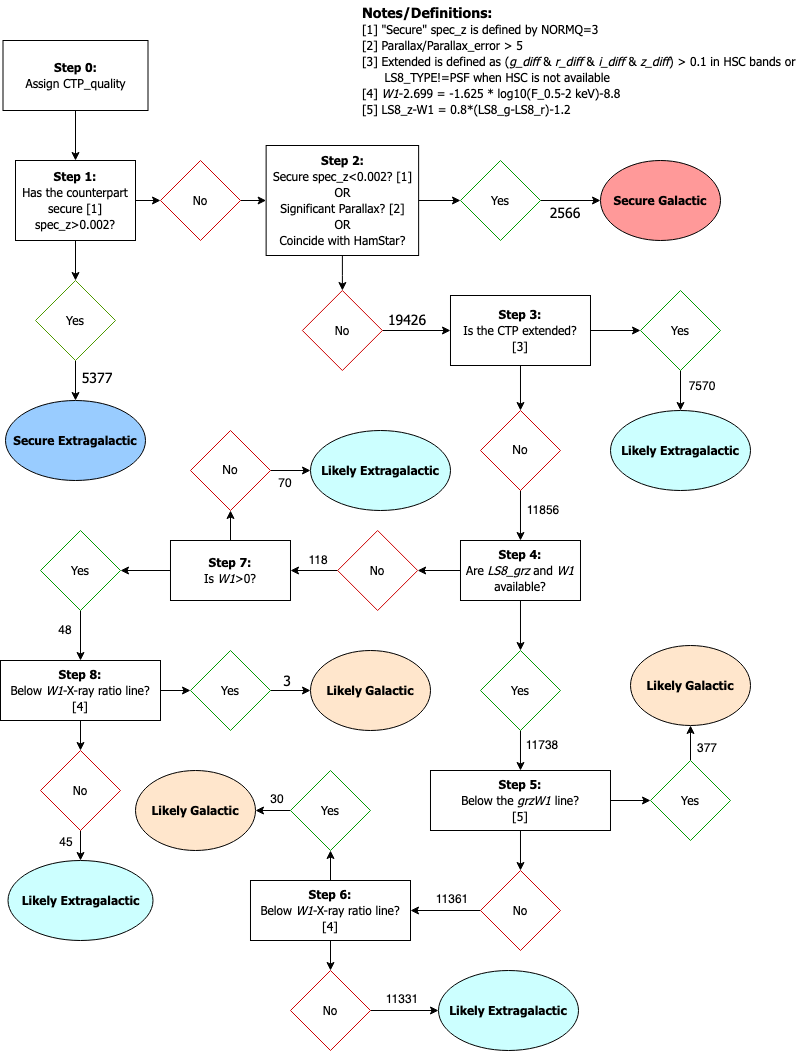}
\caption{Decision tree adopted for assigning each eFEDS point source to the Galactic or extragalactic classes.
First we classified the sources on the basis of the most secure methods (e.g., high confidence redshift) and then proceeded with less reliable methods (e.g., based on colours) on the sources remaining in the pool, creating less pure samples. Note that the numbers listed at each step include all sources, i.e., including  also those with an insecure counterpart ({\texttt{CTP\_quality}}<2)}.
\label{fig:CTP_class}
\end{figure*}

In order to classify sources in the most reliable way, we have used a combination of methods and various  information: spectroscopic, parallax measurements from Gaia, colours and morphology from imaging surveys.
None of the methods is infallible because they all depend on the quality of the data (e.g., SNR for spectra, depth and resolution of images) and because of the degeneracy in colour-redshift space for many of the sources. We have therefore adopted a multi-step approach: at each step we extract from the pool of sources those that can be  classified with high reliability either as {\it extragalactic} or {\it Galactic}.  Figure~\ref{fig:CTP_class} provides a graphical illustration of the decision tree  adopted for the classification, together with the number of sources in each the classes. The procedure is described below in detail.

We first apply the classification based on spectroscopy or large parallax. These can be considered primary methods as they are highly pure, but certainly not complete. The sources thus classified are defined  "Secure Galactic" or "Secure Extragalactic".  Briefly, we define "Secure Extragalactic" all sources with spectroscopic redshift $>$0.002 and \texttt{NORMQ=3} (STEP 1 in Figure~\ref{fig:CTP_class}) and "Secure Galactic" all sources satisfying at least one of the criteria: 1) spectroscopic redshift $<$0.002 and \texttt{NORMQ=3}; significant parallax  from Gaia EDR3 (above 3$\sigma$) or  agreement with Hamstar counterparts (STEP 2).

Next, from the sources still in the  pool, we extracted those that appear extended in the optical images. Depending on whether or not photometry from HSC is available, a source is defined as extended (EXT) if it satisfies:
\begin{equation}
\Delta mag = mag_{\rm Kron}-mag_{\rm psf}>0.1 
\label{eq:EXT1}
\end{equation}
simultaneously in {\it g,r,i,z} from HSC imaging data \citep[e.g.,][]{Palanque11}, or, when no photometry from HSC is available (either because the source is outside the field or because of saturated photometry):

\begin{equation}
{\texttt{LS8\_TYPE}} \neq {\rm PSF}. 
\label{eq:EXT2}
\end{equation}

The EXT sources were then flagged  as  "Likely Extragalactic" (STEP 3). This is considered a secondary classifier, given that, for example, in poor seeing conditions  point-like sources (or stellar binary systems) would  also be mis-classified as extended \citep[see discussion presented in][]{Hsu14}.

\begin{figure*}[h]
\centering
\includegraphics[width=0.41\textwidth]{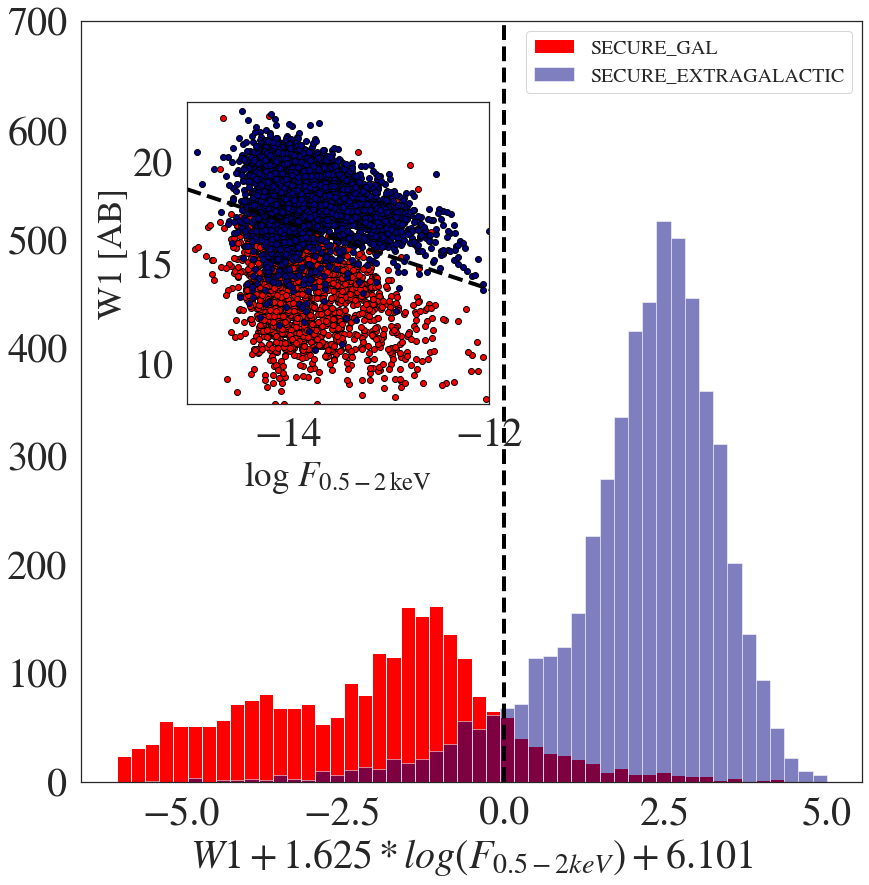}
\includegraphics[width=0.405\textwidth]{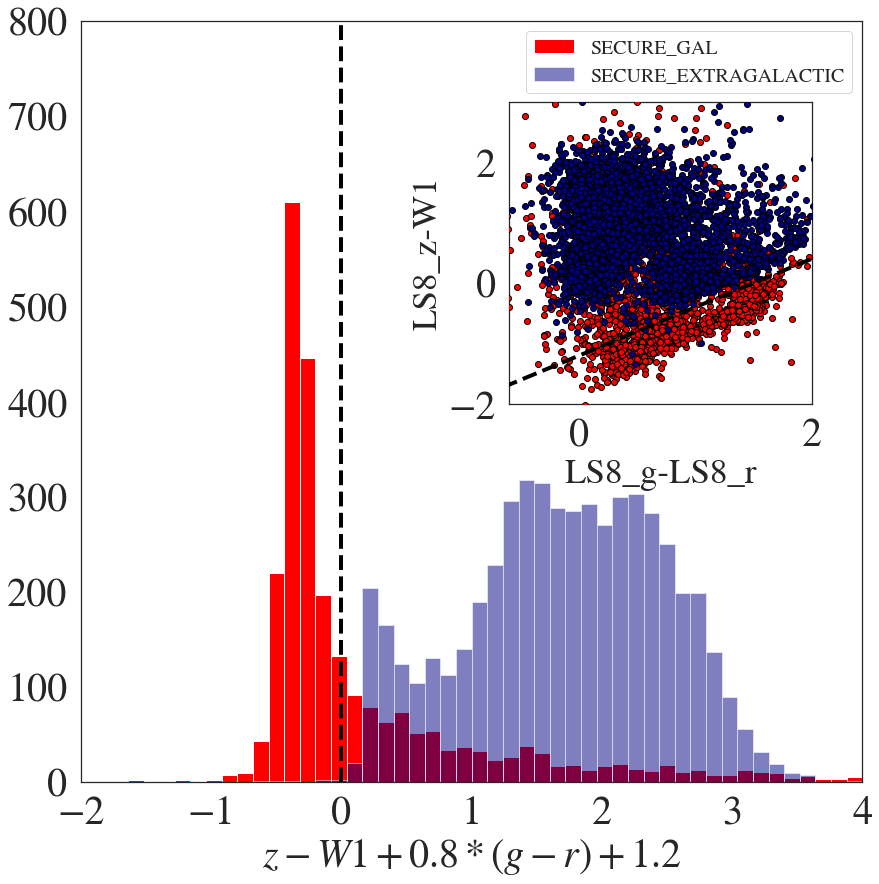}
\caption{The distribution of sources flagged as "Secure Galactic" (red) and "Secure Extragalactic" (blue) in the {\it g-r} vs {\it z-W1} (left) and W1 vs X-ray (right)  planes are used for determining a line separator (black line) to be used for classifying sources in steps 5, 6 and 8 of the flowchart presented in Figure~\ref{fig:CTP_class}. The line separator on the right has fewer Galactic sources that fall into the extragalactic locus. However, the line separator defined on the left  has  literally only a handful of extragalactic sources falling into the Galactic locus, making this classifier more efficient, provided the four photometric points are available.}
\label{fig:grzw1_wX_line}
\centering
\end{figure*}

The sources classified as "Secure" were then projected in the LS8 {\it z-W1} vs {\it g-r} plane (see inset in the left panel of Figure~\ref{fig:grzw1_wX_line}), following \cite{Ruiz18}. There, we have empirically defined a line separator described as:
\begin{equation}
 z-W1 - 0.8*(g-r)+1.2 = 0,
\end{equation}
\noindent which provides a sharp separation between secure Galactic and extragalactic sources, with a negligible fraction of  secure extragalactic sources lying below the separator (left panel of Figure~\ref{fig:grzw1_wX_line}). 
Then, for all the sources still in the pool and with available photometry from LS8 (STEP 4), we classify the sources below the line  as "Likely Galactic" (STEP 5). 
The remaining sources in the pool with available LS8 photometry  are classified as "Likely Galactic/Extragalactic" (STEP 6) depending on whether they fall below or above the line in the {\it W1} vs X-ray flux plane(see inset in the right panel of Figure~\ref{fig:grzw1_wX_line}), as defined in \cite{Salvato18a}:
\begin{equation}
W1+1.625*\log(F_{\rm 0.5-2 keV}) +6.101 = 0,
\label{eq:w1x}
\end{equation}
with W1 in Vega system and X-ray flux in cgs.
Originally, a similar line separator  was introduced by \cite{Maccacaro1988} using X-ray and optical bands and over time tested at different X-ray flux depth or at different wavelength \citep[e.g., NIR; see][]{Civano12}. This new line separates 'X-ray bright' AGN from 'X-ray faint' stars, and was constructed combining data from the deep COSMOS {\it Chandra} Legacy survey \citep[][]{Marchesi16} and ROSAT/2RXS \citep[][]{Boller16, Salvato18a}. It can be considered a good separator only after the extended, nearby extragalactic sources are taken into account (see right panel of  Figure~\ref{fig:grzw1_wX_line}). It has the advantage of generality, as the {\it W1} photometry and the X-ray fluxes are available virtually for all the eFEDS sources.
Finally, for the sources without complete information from LS8, we assume they are extragalactic (STEP 7), unless they are below the W1-X line defined in Equation~\ref{eq:w1x} (STEP 8).

\begin{figure}[ht]
\centering
\includegraphics[width=0.5\textwidth]{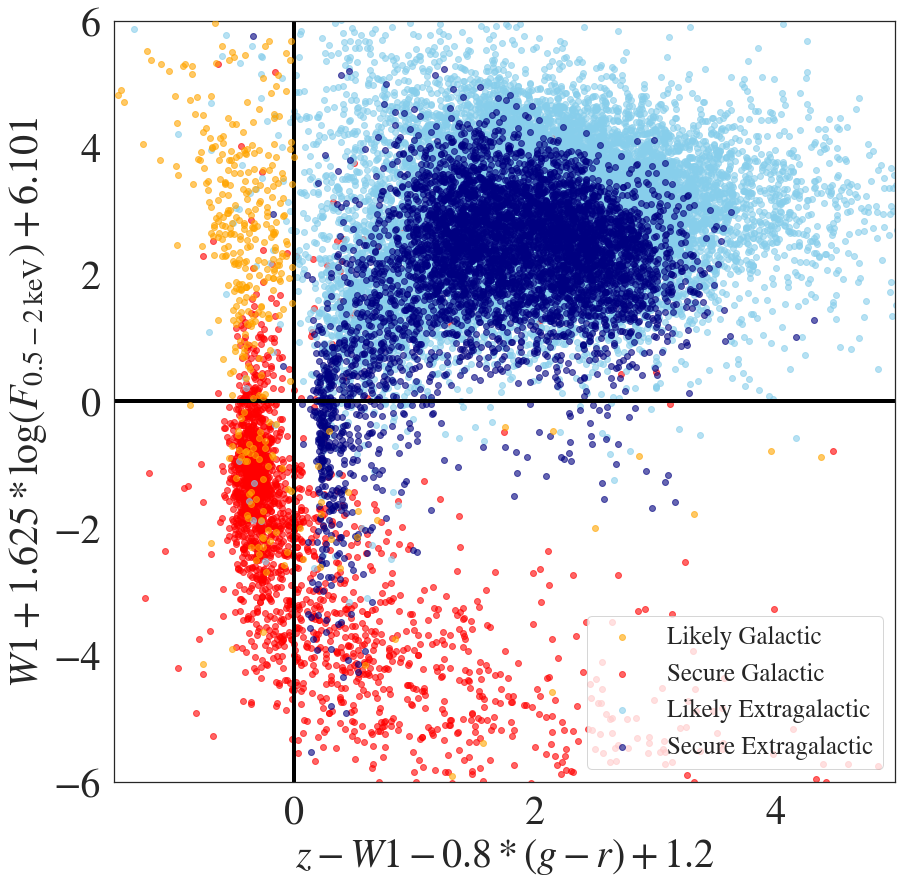}
\caption{The four eFEDS X-ray sources classes "Secure Galactic" (red), "Likely Galactic" (orange), "Secure Extragalactic" (blue) and "Likely Extragalactic" (cyan) defined in the flowchart presented in Figure~\ref{fig:CTP_class} distributed according to their distance from the two lines defined in Figure~\ref{fig:grzw1_wX_line}. Three of the four wedges thus defined contain extragalactic or Galactic samples that are up to 99\% pure (see text for details).}
\label{fig:color_class_lines}
\end{figure}

In this manner, a simple but reliable four-way classification scheme (Secure/Likely Galactic/Extragalactic) is achieved. The final distribution of the four classes of sources in the {\it g-r-z-W1} vs {\it W1-X} planes is shown in Figure~\ref{fig:color_class_lines}. The two line separators identify four wedges, two of which  can be used for defining  almost 100\% pure sub-samples of Galactic/extragalactic  Xray selected sources. The four wedges are: \
\begin{itemize}
    \item Top left: 724 sources, out of which  637 (87.9\%) are Galactic (463 and 428, respectively, only considering sources with reliable counterparts,  {\texttt{CTP\_quality}}>=2)\\
    \item Top right: 23874 sources, out of which 23809 (99.7\%) are extragalactic (21711 and  21647 for {\texttt{CTP\_quality}}>=2)\\
    \item Bottom left: 1391 sources, out of which  1373 (98.7\%) are Galactic (1337 and 1319 for {\texttt{CTP\_quality}}>=2)\\
    \item Bottom right: 1380 sources, out of which  479 (34.7\%) are extragalactic ( 1263 and  379 for {\texttt{CTP\_quality}}>=2).\\
    \end{itemize}

It is important to keep in mind that the order of the steps taken in the decision tree is crucial for limiting the mis-classification of the sources as much as possible. For example, the use of spectroscopic redshift in the first step allowed the identification of the bright and nearby extragalactic sources that would have been mis-classified as Galactic, in the {\it W1-X} plane. Similarly, the adoption of the high parallax from Gaia allowed the identification for secure galactic sources that would have been mis-classified as extragalactic in the {\it z-W1} vs {\it g-r} plane. 

In summary, the eFEDS Main sample comprises 24393 sources classified as Extragalactic (5377 "secure" and 19016 "likely") and 2976 classified as Galactic (2566 "secure" and 410 "likely". All these numbers are reported in Table 6.


\subsection{Validation of classification using external samples \label{section:validation_External}}

We have carried out sanity checks of the classification framework
against two external catalogues, whose members are expected to be 
almost completely extragalactic in nature: the Faint Images of the Radio Sky radio component
catalogue \citep[FIRST,][]{White97}, and the Gaia--unWISE AGN candidate
catalogue \citep[GUA,][]{Shu19}. Whilst the GUA sample is
derived from similar underlying datasets to those used in our own source classification logic,
the machine learning methods used by \citet{Shu19} make this at least a semi-independent test sample.  
Simple positional matches were
made against our best matching optical counterpart positions, with a
search radius of 3\,arcsec for the FIRST radio component
catalogue\footnote{we only considered the radio components and made no
attempt to handle complex sources appropriately}, and 1\,arscec for
GUA (GUA objects were considered when they had $\mathrm{PROB\_RF}>0.8$).
We examine the rate at which sources we classify as
Galactic or Extragalactic (both Secure and Likely) are matched to
objects in these external catalogues (see
Table~\ref{tab:classification_test}).  There is a very low rate of
apparent disparities between our classifications and those that may be
derived by matches to the external catalogues. For example, only
0.19\% of `Secure Galactic' sources have a radio counterpart in FIRST,
compared to 7.0\% of the `Secure Extraglactic' sample.  Likewise, only
0.19\% of `Secure Galactic' sources are matched to candidate AGN from \citet[][]{Shu19},
compared to 54\% of the `Secure Extragalactic'
sub-sample.


\begin{table}[ht]
\begin{tabular}{@{}lr|rr|rr}
\toprule
Ref.         & Total   & \multicolumn{2}{c}{Extragalactic} & \multicolumn{2}{c}{Galactic} \\
sample       & matches & Secure & Likely & Likely & Secure \\
\midrule 
All eFEDS    & 27369 &  5377 & 19016 &   410 &  2566 \\
& & & & &\\
FIRST        &   796 &   376 &   414 &     1 &     5 \\
GUA          &  6357 &  2924 &  3425 &     3 &     5 \\
\bottomrule
\end{tabular}
\caption{Comparison of our classification scheme against two (semi-)independent reference catalogues: the FIRST radio component
catalogue \citep{White97}, and the Gaia--unWISE AGN candidate
catalogue \citep[GUA,][]{Shu19}. The low rate at which our classification logic classifies
both radio sources and AGN candidates as being `Secure Galactic' or `Likely Galactic'
suggests that our classifications are robust.}
\label{tab:classification_test}
\end{table}


\subsection{Very nearby galaxies \label{subsection:nearbygal}}
Unlike what happens in pencil-beam surveys, within eFEDS there are numerous very nearby and thus resolved galaxies. \citet{Vulic2021} searched for eFEDS sources within the D25 ellipse of the sources in the Heraklion Extragalactic CATalogue (HECATE) of nearby galaxies \citep[][]{Kovlakas2021}. For the 100 HECATE galaxies with an eFEDS source nearby,  93/100  are consistent with the counterpart proposed here by the combination of  {\sc{nway}} {\sc{astromatch}}.  For the remaining 7 cases (ID\_SRC 7551, 12847, 2671, 22198, 17437, 29989 ,20952; see Figure~\ref{fig:HECATE})
the counterparts identified in this work fall within the HECATE galaxies but do not coincide with the  centre of the galaxy but rather with a source that could be either an ULX in the galaxy or an extragalactic source in the background. For these 7 sources, dedicated studies will be needed to identify the exact origin of the X-ray emission.

\section{Photometric Redshifts}\label{section:photoz}
Photo-z of AGN and X-ray selected sources in general have developed dramatically in the last 10 years, bringing the redshift accuracy and the fraction of outliers (usual quantities measured for assessing the quality of the photo-z) comparable to those measured for normal galaxies.
\begin{figure*}[t]
\centering
\includegraphics[width=8cm]{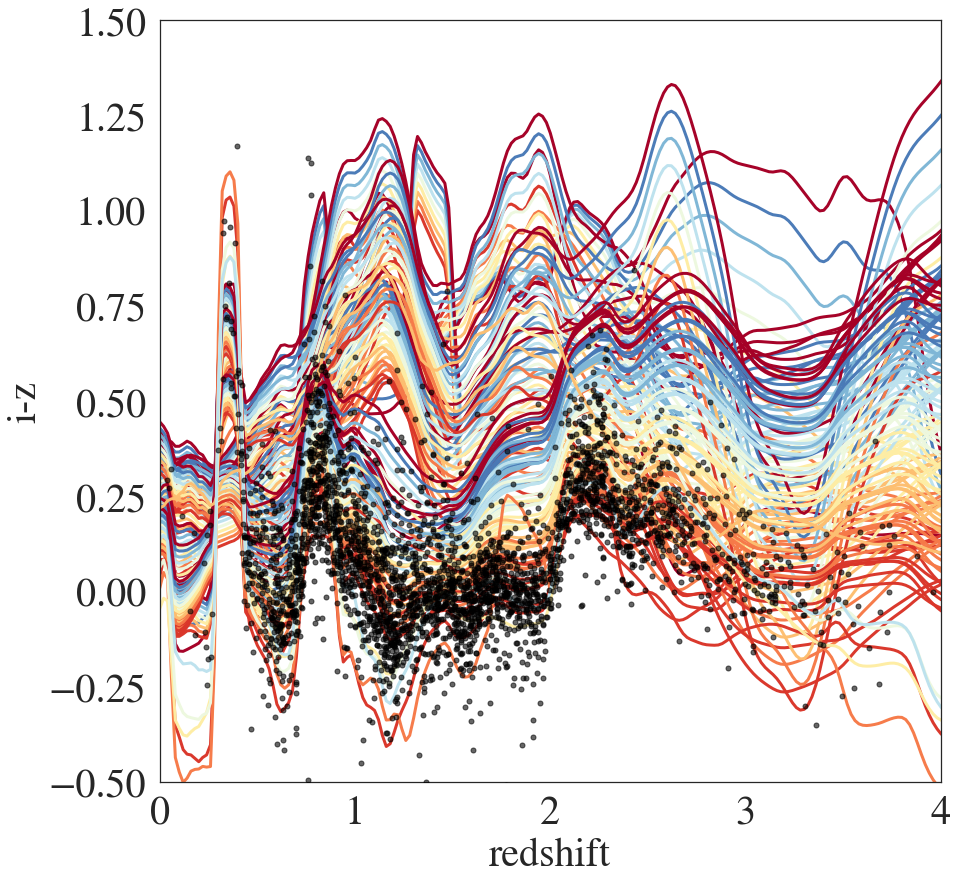}
\includegraphics[width=8cm]{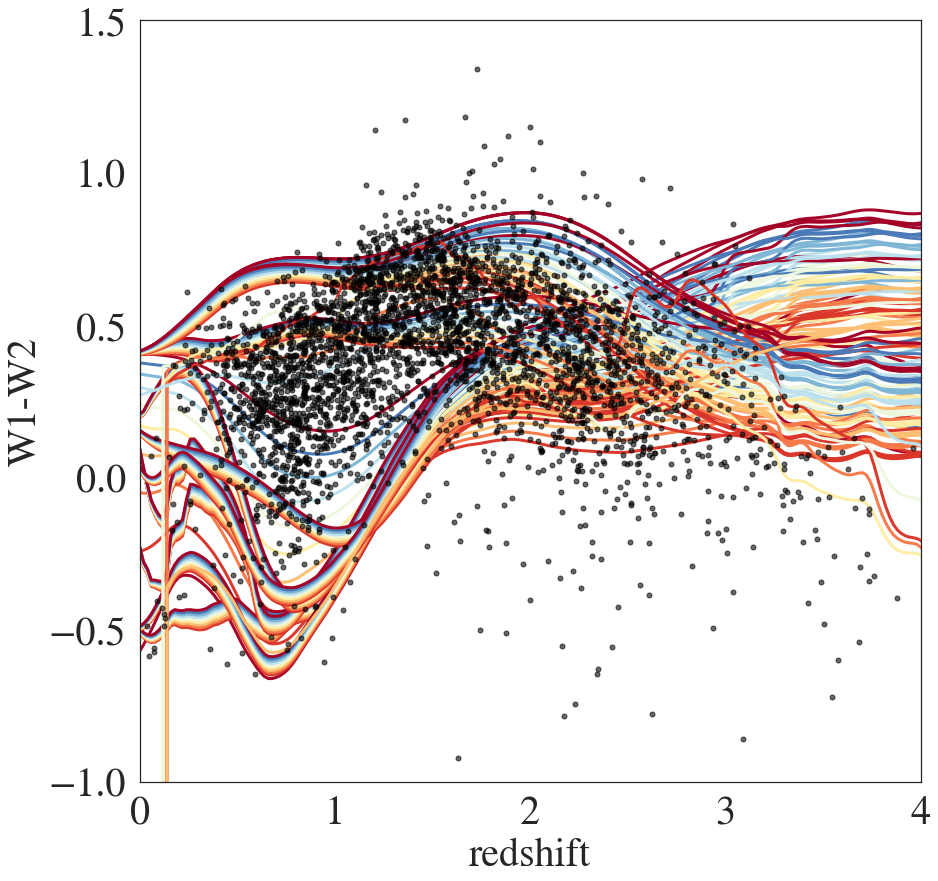}
\includegraphics[width=8cm]{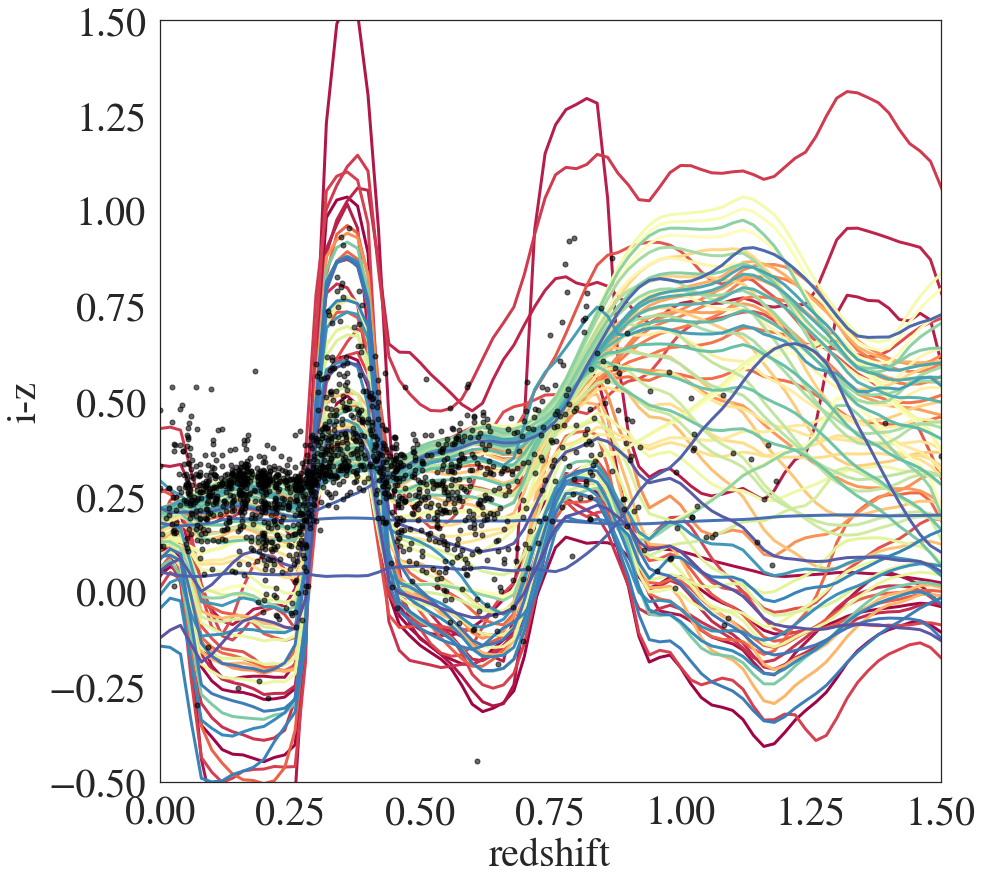}
\includegraphics[width=8cm]{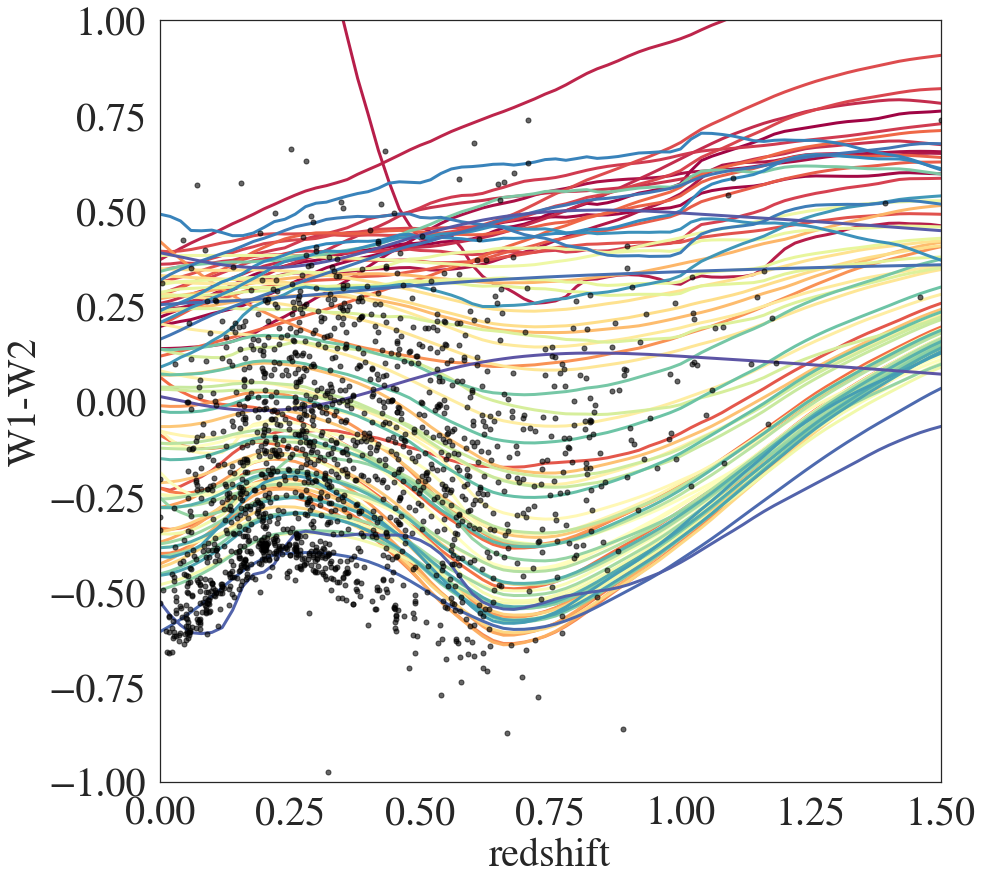}
\caption{i-z and W1-W2 colours of eFEDS extragalactic sources as a function of their (reliable) spectroscopic redshift (black points). Overplotted are the tracks of theoretical colours as a function of redshift derived from all the templates used in this work for the PLIKE (upper panels) and EXT (lower panels) samples, as listed in Appendix~\ref{appendix:templates}}. 
\label{fig:color_redshift}
\centering
\end{figure*}

Regardless of whether photo-z are computed via SED fitting or via Machine Learning, accurate photo-z for AGN are less straightforward to be obtained than for non-active galaxies \citep[see][for a review on the topic]{Salvato18b}. The main reason is that, for each multi-wavelength data point, the relative contribution of host and nuclear emission is unknown and redshift-dependent, with redshift being the parameter that we are trying to determine. To add to the difficulty, one should not forget the impact of dust extinction and variability, the latter an intrinsic property of AGN. This, especially for wide-area surveys where data  are taken over many years, can affect noticeably the accuracy of photo-z if not accounted for  \citep[e.g.,][]{Simm15}, as for example was possible to do in COSMOS \citep[][]{Salvato09,Salvato11,Marchesi16}. In eFEDS, we also have to face the issue that the photometry is not homogenised, and different surveys cover different parts of the field at different depth and with different ways of computing the photometry (Kron, Petrosian, apertures, model etc). In the following we describe the procedure adopted for computing photo-z using LePHARE \citep{Arnouts99, Ilbert06}. We then proceed with an estimate of the reliability of the photo-z and a comparison with DNNZ, an independent computation of photo-z using machine learning (Nischizawa et al., in prep).

\subsection{Photo-z Computation}
We computed the photo-z for the sources classified as extragalactic. In order to minimise systematic effects, we have used different types of photometry, depending on the survey; in particular, we have tried to avoid photometry derived from models for the extended and nearby sources. This is because usual models are good representation of point-like, disk-like and bulge-like sources, but are unable to represent, for example, a local Seyfert galaxy where nuclear and host components would be both contributing to the total flux.
For this reason, we have used total fluxes from GALEX;  Kron and cmodel photometry from HSC, depending on whether the source is extended or not (see below), and GAAP (Gaussian Aperture and Photometry) from KiDS+VIKING.  From VHS we have adopted Petrosian photometry as it appears to be more in agreement with the VISTA/VIKING photometry. 
All the photometry was corrected for Galactic extinction, using E(B-V) from LS8. Depending on whether the source is in the area covered by KiDS+VIKING, within HSC but outside KiDS and outside HSC, different bands were available\footnote{In particular, note that for HSC, in the S19A release available to us at the time of this work, photometry in {\it r2} and {\it i2}  filters is provided. However the filters have changed during the survey and depending on the coordinates of the sources the fraction of data obtained with the original or the new filters changes. In order to account for this at any location, we have adopted the filter that was used for obtaining at least 50\% of the data. This solution is not optimal and will  affect the quality of the photo-z  in some areas.}
\begin{figure}[h]
\centering
\includegraphics[width=0.45\textwidth]{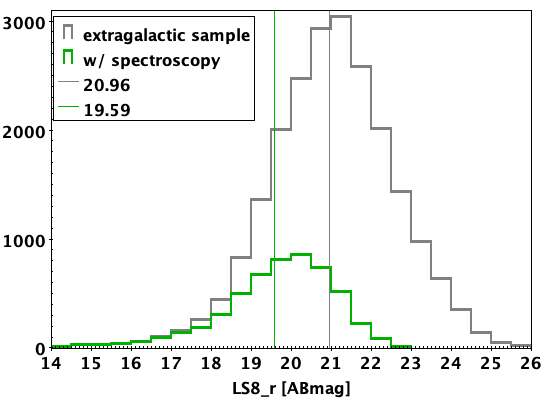}
\caption{Magnitude distribution for the entire extragalactic sample and the subsample for which reliable spectroscopy is available. The vertical lines indicate the mean values of the two samples.}
\label{fig:mag_zspec}
\centering
\end{figure}

\begin{figure*}[t]
\centering
\includegraphics[width=9cm]{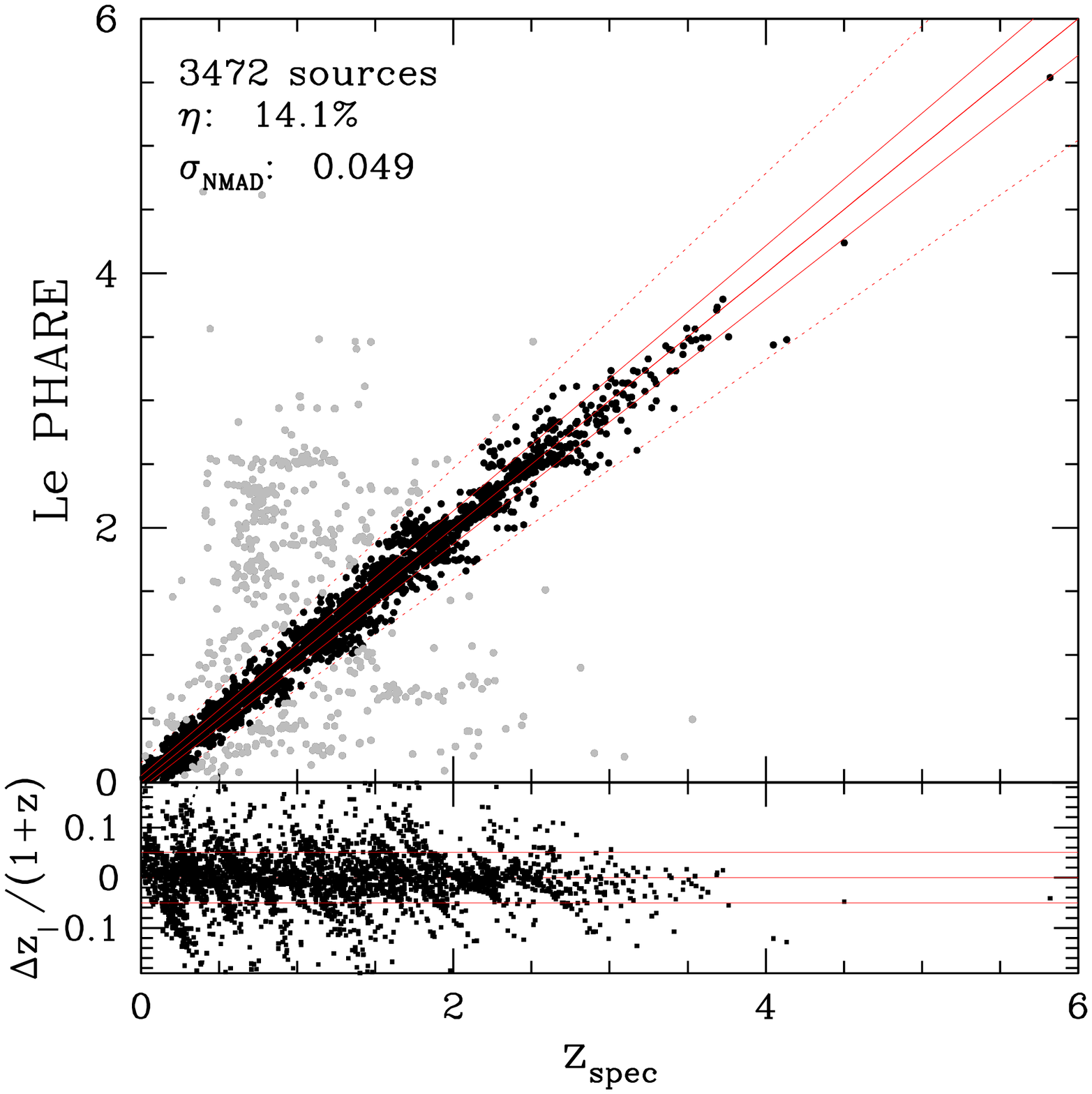}
\includegraphics[width=9cm]{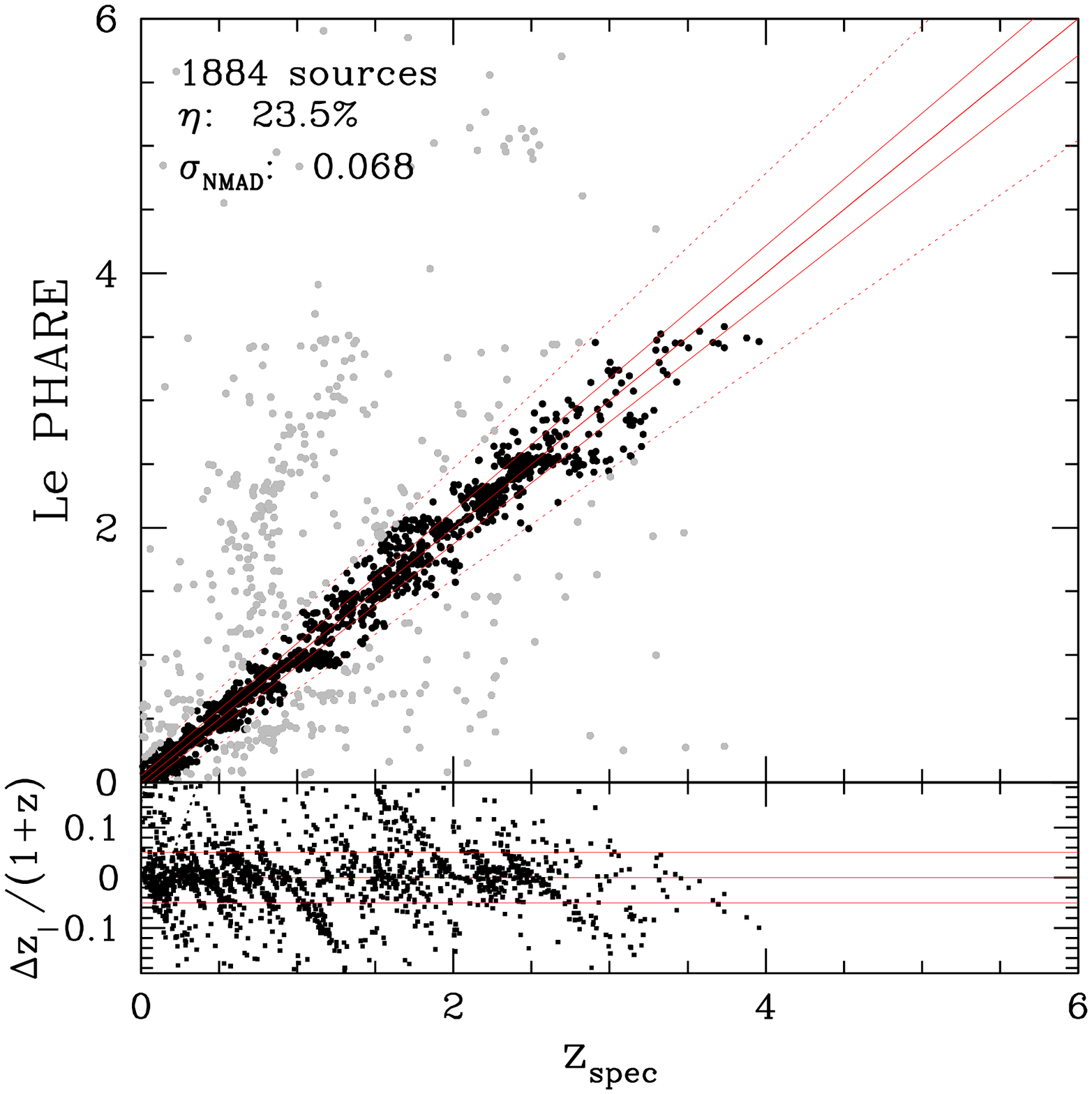}
\caption{spec-z vs photo-z  from {\sc{Le PHARE}} for the sources inside KiDS+VIKING (left panel) and inside HSC but outside KiDS+VIKING (right panel). The sources in gray are considered outliers while the red lines  correspond to 1) z\_phot = z\_spec (thick solid), 2) z\_phot = z\_spec $\pm$0.05 x(1+z\_spec ) (solid) and 3) z\_phot =z\_spec $\pm$ 0.15(1 + z\_spec ) (dotted).}
\label{fig:zszp_lephare}
\centering
\end{figure*}

The computation of the photo-z followed the procedure already outlined  in \cite{Salvato09,Salvato11,Fotopoulou12,Hsu14,Marchesi16,Ananna2017}, where sources are treated differently, depending on whether the optical images 
indicate them being extended (\texttt{EXT})
or a point-like/unresolved (\texttt{PLIKE}), following Section~\ref{section:properties}. This step is particularly important, as sources in the two samples are treated differently, using different priors and templates.

In addition, the fitting templates are selected on the basis of the X-ray depth and coverage of the surveys, keeping in mind that e.g., bright AGN will be mostly absent in a deep pencil-beam survey, characterised instead by  host-galaxy dominated sources. Given the similar X-ray depth, the libraries used in \cite{Ananna2017} for the Stripe-82X survey have been a good starting point for our work on eFEDS. However, recently a new library of templates for AGN and hybrids (AGN and host) was presented in \citet{Brown19}. The authors used photometry and archival spectroscopy of 41 AGN to create an additional set of 75 new hybrid templates. With respect to previous AGN templates they have the advantage that they are empirical for the entire wavelength coverage  and that the contribution  from the host and AGN components is fully taken into account when creating the final SED, including dust attenuation and emission lines.

eFEDS is particularly rich in sources with reliable spectroscopy (see Section~\ref{subsection:zspec}, allowing for a better tuning of the templates to be used for the  photo-z computation.
To optimize the template choice, the colours of all the sources with reliable spectroscopy were plotted as a function of redshift, together with the theoretical colours from all the templates available (Figure~\ref{fig:color_redshift} illustrates this for {\it i-z} and {\it W1-W2} for the \texttt{EXT} and \texttt{PLIKE} samples, respectively, for all the templates that ultimately were adopted in this work). In selecting the templates we tried to limit their number (to control degeneracy in the redshift solution), while at the same time compiling a list representative of the entire population. The sources that in Figure~\ref{fig:color_redshift} are outside the parameter space covered by the templates can be interpreted in various ways, from problems in the photometry of the specific objects due to blending with neighbours or variability or lack of certain features in the templates. While we will further investigate this latter possibility for the future eROSITA surveys, here we need to keep in mind that the figures  are representative of only two colours, while in selecting the templates we look at all the colours that our photometric set allows.

An important point  to keep in mind is the fact that despite being rich, the available spectroscopic sample is not representative of the entire eFEDS population as can be seen in Figure~\ref{fig:mag_zspec}. For this reason the final library should also include some templates for types of sources that are expected to be present in eFEDS, without being necessarily identified yet. In particular we have created a set of templates using the archetype of type 1 AGN from the counterparts of ROSAT/2RXS \citep[][]{Salvato18a} observed within SDSS-IV/SPIDERS  presented in \citep[][]{Comparat2020}, extended in UV and MIR with various slopes. For the non empirical templates  reddening was also considered, using the extinction law of Prevot \citep [][]{Prevot1984} with E(B-V) values from 0 to 0.4 in steps of 0.1. The selected templates  are presented in Appendix~\ref{appendix:templates}.

As output {\sc{Le PHARE}} provides the best value for the best photo-z together with the upper and lower 1,2,3 $\sigma$ error, the best  combination of template, extinction law and extinction value, the quality of the fit and the {\texttt{pdz}}, the latter being the  redshift probability distribution defined as  pdz$=\int F(z)\,dz $ between $z_{best} \pm 0.1(1+z_{best})$ with $z_{best}$ being the photo-z value corresponding to the best fit.  For photo-z computed with a sufficiently high number of photometric bands covering the entire SED \citep[e.g., XMM-COSMOS;][]{Salvato09} a high value of  {\texttt{pdz}} can be safely  translated in reliability of the photo-z. As it has been discussed in \citet{Brescia2019}, this is not the case when the photometric set is not rich, and {\texttt{pdz}} can be high also for a poor fit just because there are no sufficient constraints.  

\begin{table*}[h]
\centering
\begin{tabular}{@{}l|c|ccc||ccc}
\toprule
 AREA  & sample &\multicolumn{3}{c}{\sc{Le PHARE}} & \multicolumn{3}{c}{\sc{DNNz}} \\ 
    &   &N$_{spec}$/N$_{tot}$  & $\eta$ & $\sigma_{\rm NMAD}$ & N$_{spec}$/N$_{tot}$ & $\eta$ & $\sigma_{\rm NMAD}$\\
\midrule
\hline
\multirow{2}{*}{inside KiDS} & PLIKE & 2331/6808  & 17.8\% & 0.048 & 2280/6808&21.8\% & 0.045\\ 
 & EXT   &  1141/4598  & 6.7\% & 0.054 & 1139/4598 & 6.3\% & 0.032\\
 & TOTAL$^*$  &  3472/11406 & 14.1\% & 0.049 &3419/11406& 16.7\%& 0.039\\
  \hline
\multirow{2}{*}{outside KiDS} & PLIKE & 1399/8532  &28.6\%  &0.081  & 1325/8532& 34.9\%& 0.075 \\ 
 & EXT & 488/4221 & 8.6\% & 0.040 &483/4221 & 4.9\% &0.026\\ 
 & TOTAL$^*$ & 1887/12753  &23.8\%  &0.068 & 1808/12753& 27.2\% &0.052\\
 \hline
\multirow{2}{*}{outside HSC} & PLIKE & 15/145 & 42.9\% & 0.164 & N/A& N/A& N/A\\ 
 & EXT & 3/89 &0.0\% & 0.039  &N/A  &N/A  &N/A \\ 
 & TOTAL&  18/234   &33.3\%  & 0.122 & N/A & N/A & N/A\\ 
 \bottomrule
\end{tabular}
\caption{Fraction of outliers and accuracy for {\sc{Le PHARE}}  and {\sc{DNNz}} computed using the extragalactic sources with secure spectroscopic redshift, split by area. In each row the difference between the numerators in the N$_{spec}$/N$_{tot}$ columns provides the number of sources with spectroscopy for which {\sc DNNz} could not provide a photo-z, mostly because the HSC photometry is saturated for those sources.
}
\label{tab:photoz_accuracy_type}
\end{table*}

\subsection{Reliability of photo-z \label{section:z_reliability}}
The final comparison between photo-z and spec-z, considering \texttt{EXT} and \texttt{PLIKE} sources together, for the area within KiDS+VIKING and within HSC but outside KiDS+VIKING is shown in Figure~\ref{fig:zszp_lephare}.
We used the standard metrics for measuring the quality of photo-z \citep[see][for more details on definitions]{Salvato18b}: \
a) the fraction of outliers $\eta$: it highlights the fraction of sources with unexpectedly large errors  and it is defined as the fraction of sources for which $|z_{\rm phot}-z_{\rm spec}|/(1+z_{\rm spec}) > 0.15$ \citep[e.g.,][]{Hildebrandt2010}. \
b) accuracy $\sigma_{\rm NMAD}$ that describes the expected scatter between predictions and truths and it defined as $1.48 \; \times \; {\rm median}(| z_{\rm phot} -z_{\rm spec} |/(1+z_{\rm spec}))$ \citep[][]{Ilbert06}. 

\begin{figure*}[h]
\centering
\includegraphics[width=18cm]{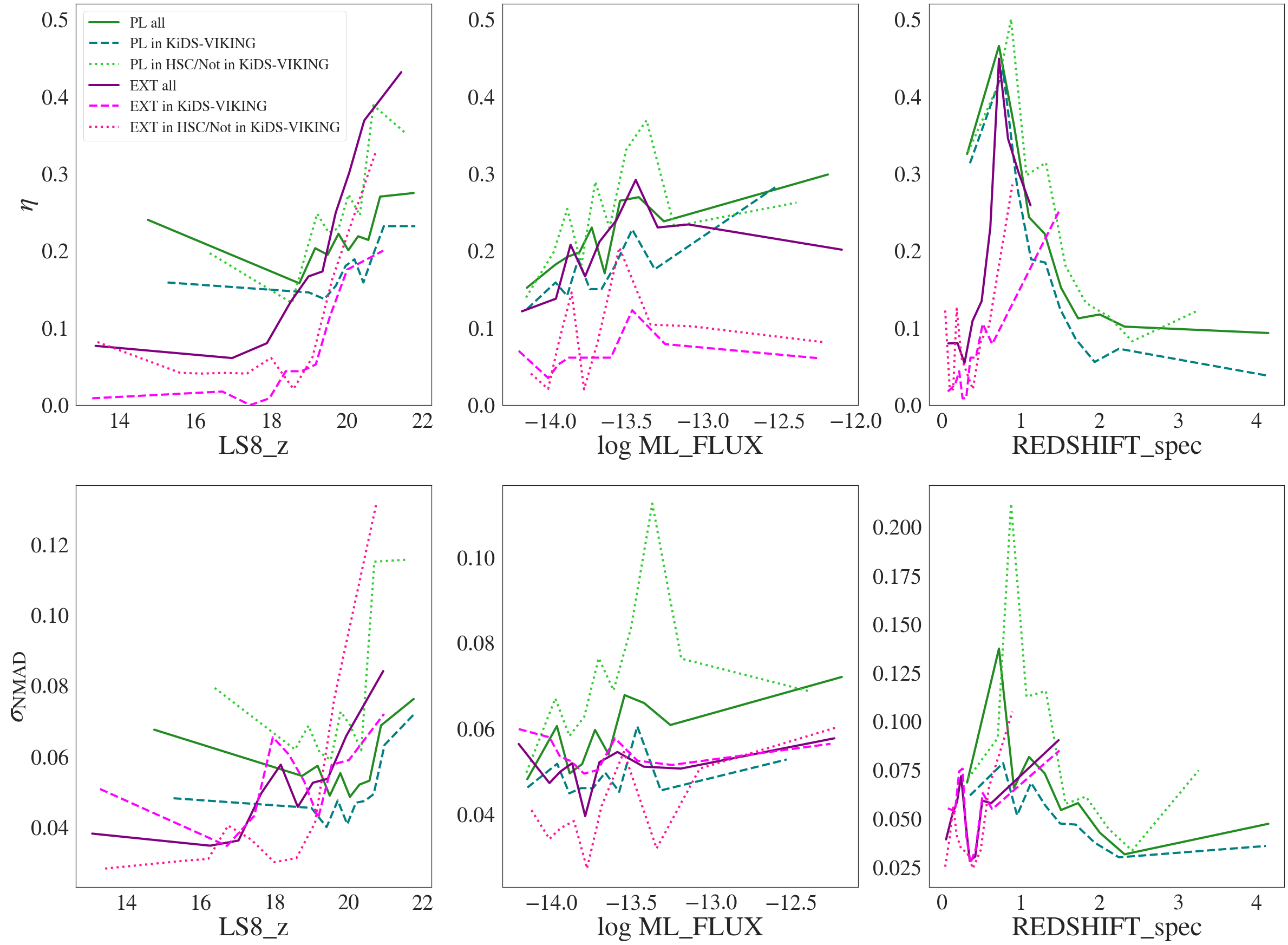}
\caption{Fraction of outliers (top) and accuracy (bottom) as a function of magnitude ({\it z} from LS8), 0.2-2.3 keV X-ray flux and spectroscopic redshift  split for type (EXT/PLIKE) and in area (with/without VIKING coverage). The x-axes are binned in equal numbers of elements taking the quantiles between [0,1] in steps of 0.1. In other words, all lines also account for the size of each spectroscopic sub-sample.}
\label{fig:photoz_etanmad}
\centering
\end{figure*}

The results are listed in Table~\ref{tab:photoz_accuracy_type}. Figure~\ref{fig:photoz_etanmad} shows the same results, but split as a function of {\it z}-band magnitude from LS8, X-ray flux and spectroscopic redshift.

Ideally, for the best computation of photo-z, in particular for sources dominated by emission lines as AGN,  photometry from broad band filters across the entire spectral range should be complemented by narrow band  and near infrared photometry and should be homogenised \citep[e.g][]{Salvato09,Salvato18b}.
While narrow-band photometry is not available, at least some of the surveys do provide homogenised photometry. For what concerns NIR photometry, the VISTA/VHS data are not sufficiently deep and how this impacts the photo-z is clearly visible in all the panels of Figure~\ref{fig:photoz_etanmad}, where the fraction of outliers is usually higher and the accuracy lower (high value of $\sigma_{\rm NMAD}$ in the area without VIKING coverage (dotted lines).
Not only are the NIR data shallow outside the KiDS+VIKING area; they are just a collection of photometric points computed in different ways, simply matched in coordinates. 
For this reason, based on the footprints shown in Figure~\ref{fig:AncillaryCoverage}, we can think of the photo-z in eFEDS as divided in three regions that reflect the quality of the available photometry: the inner area covered by deep forced photometry in KiDS+VIKING; the area that is within HSC but outside KiDS+VIKING for which  some  NIR  information is provided by the shallow VISTA/VHS; the area outside HSC for which the optical photometry is provided only by LS8. 

The lack of deep NIR data also creates an unusual number of sources at high-z (z>3), most of which are most likely incorrect. For example the number of sources with photo-z $>3$ is 188 within KiDS and 819 in the HSC area outside KiDS, despite the area being about the same size. Within  KiDS+VIKING, {\sc{{\sc{Le PHARE}}}} correctly estimates the redshift for  40 of the 55 (72.7\%) sources spectroscopically confirmed to be at redshift larger than 3. Most of these  high-z sources in excess  can be easily identified and flagged by noticing that they are characteri sed by  having high {\texttt{pdz}} despite being in the area outside the HSC, i.e., with a very limited  number of photometric points to be fitted (see Section~\ref{subsection:redshift_flag}).

Figure~\ref{fig:photoz_etanmad} also shows how the accuracy degrades and the fraction of outliers increases for the \texttt{PLIKE} that are  X-ray bright (top, second panels from the left). These are sources dominated by the AGN component with an SED close to a power-law and for which the lack of narrow-band photometry that would identify the emission lines, does not allow breaking of the degeneracy in the redshift solutions. However, in eFEDS there are only 47 extragalactic sources with an X-ray flux above $5\times10^{\rm -13}$ erg cm$^{-2}$ s$^{-1}$ and a reliable spectroscopic redshift is available for 39 of them, so that the low quality of the photo-z for these sources has only a limited effect. Finally, Figure~\ref{fig:photoz_etanmad} shows an undesired high fraction of outliers at low redshift, where photo-z values for normal galaxies usually are extremely accurate. We believe that the problem for AGN originate from the fact that both KiDS and HSC photometry are based on fitted models and not to total fluxes. Models are not able to account properly for the  contribution of the nuclear component that is comparable to the one from the host. Photometry from models can represent well sources at very low redshift, where the AGN contribution is negligible with respect to the host, and at high redshift, where the flux is dominated by the AGN component.

As already highlighted in the past it is always easier to obtain a reliable photo-z for galaxy dominated sources with the characteristic breaks in the SED. AGN dominated sources do suffer from degeneracy in the redshift solution, especially when little photometry is available, even within the KiDS+VIKING area (compare dashed lines for \texttt{EXT} and \texttt{PLIKE}).

\subsection{Comparison with {\sc{DNNz}} \label{subsection:DNNz}}
Within the HSC collaboration, the computation of photo-z is available in many flavors. The method that performs better on AGN is {\sc{DNNz}} (Nishizawa et al., in prep). It is based on machine learning and uses exclusively  HSC photometry, trained on the rich spectroscopic sample available for both AGN and normal galaxies within the entire HSC region (beyond the area in common with eFEDS). The {\sc{DNNz}} is based on the Multi-Layer Perceptron (MLP) that takes the cmodel flux, PSF matched aperture flux, and the second order moment size measured at five HSC filter bands as inputs, and takes posterior probability as an output. In total, $3\times 5$ inputs and output PDF is binned in 100 bins from $z=0$ to $z=7$. We have five hidden layers and each layer has 100 nodes where all nodes are fully connected to the nodes in the neighboring layers. With a 50k spectroscopic sample, it takes almost a single day to train this machine with NVIDIA GeForce RTX 2080Ti GPU.

One interesting feature of {\sc{DNNz}} is that it was trained for any type of extragalactic source, without any particular tuning for AGN. 
In Table~\ref{tab:photoz_accuracy_type}, the performances of {\sc{DNNz}} are directly compared with the output from {\sc{{\sc{Le PHARE}}}}. Remarkably, the accuracy of {\sc{DNNz}}  is in general higher than for {\sc{{\sc{Le PHARE}}}}, although with a higher fraction of outliers.

\begin{figure*}[t]
\centering
\includegraphics[width=6cm]{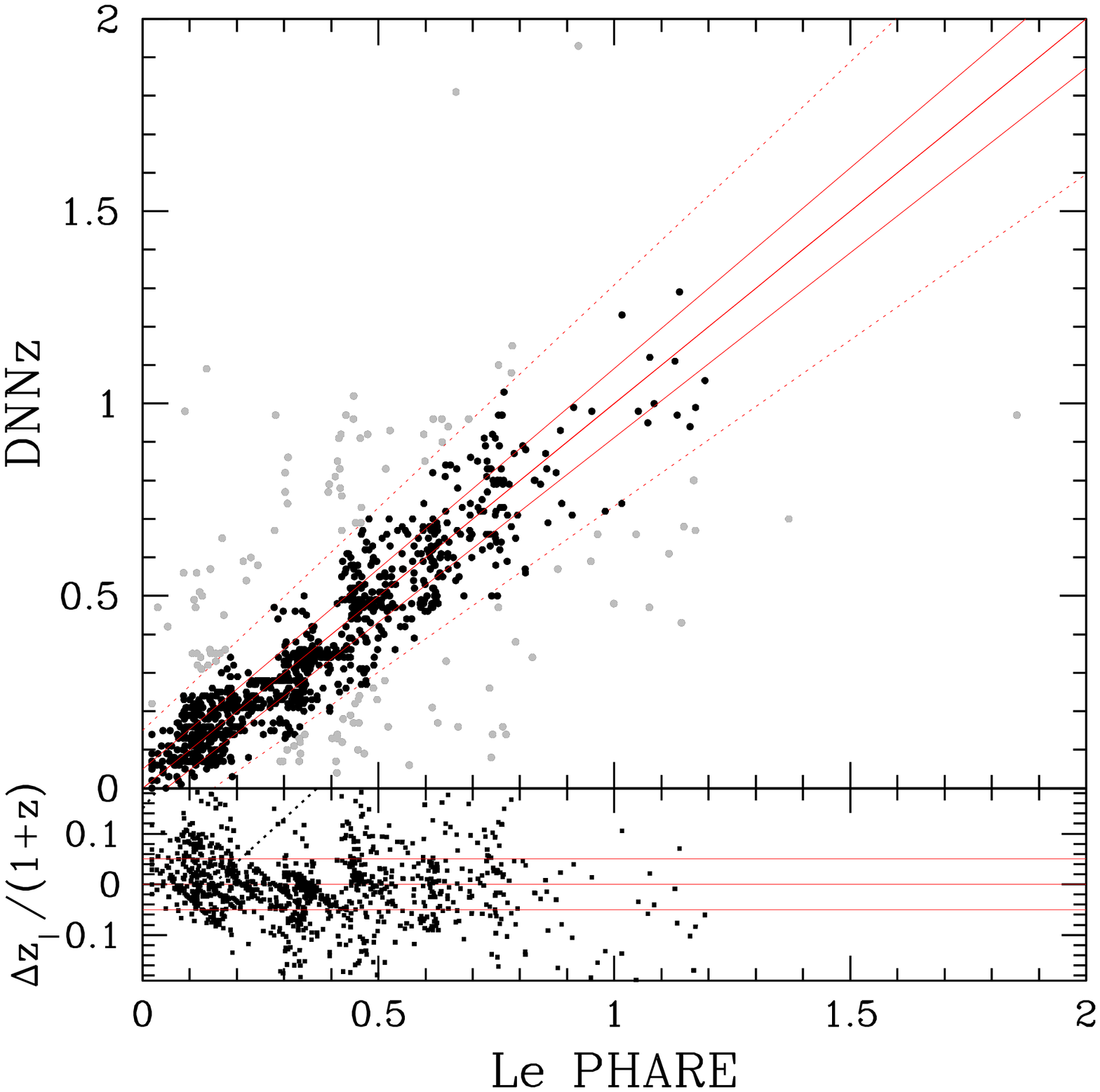}
\includegraphics[width=6cm]{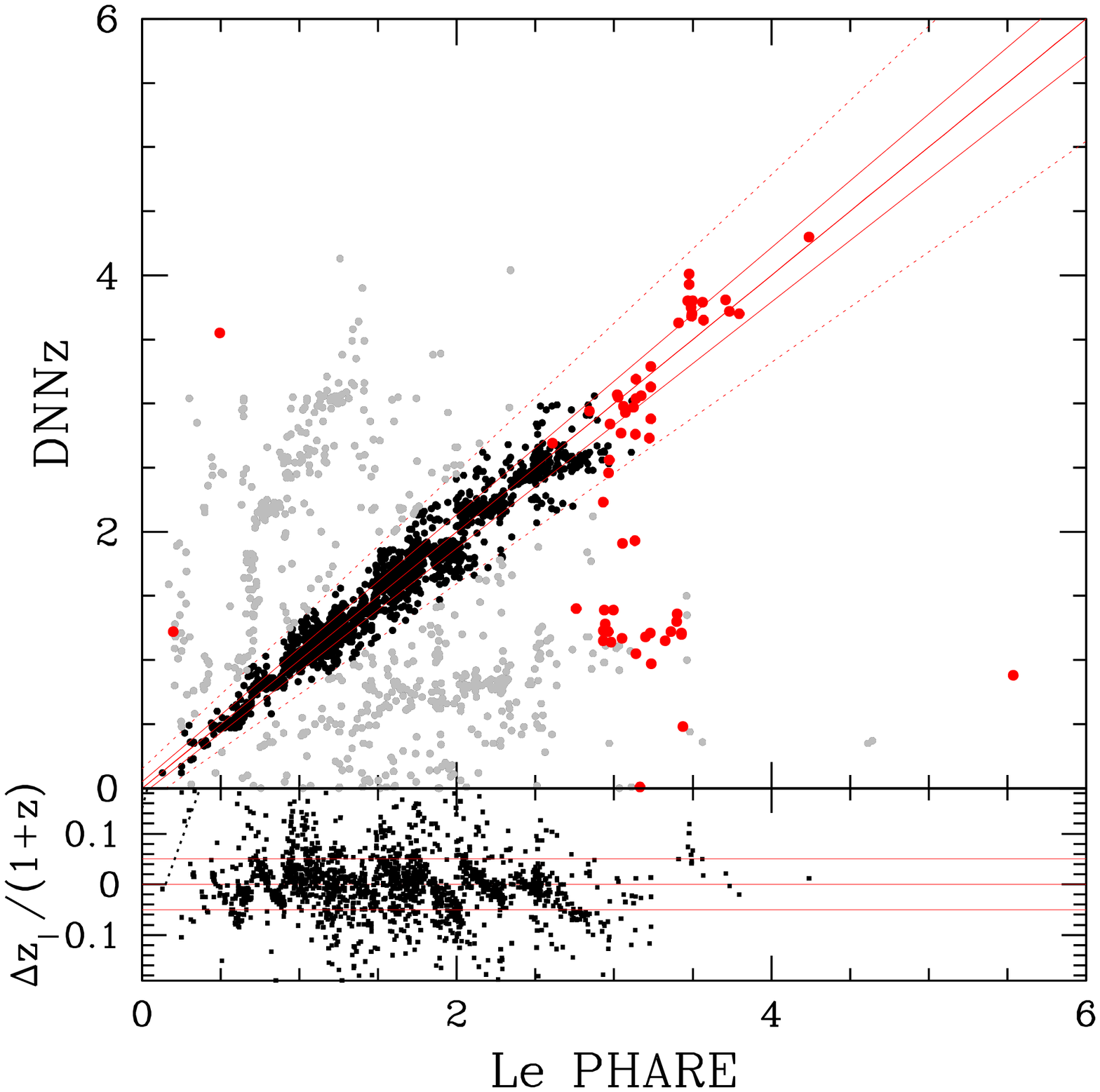}
\includegraphics[width=6cm]{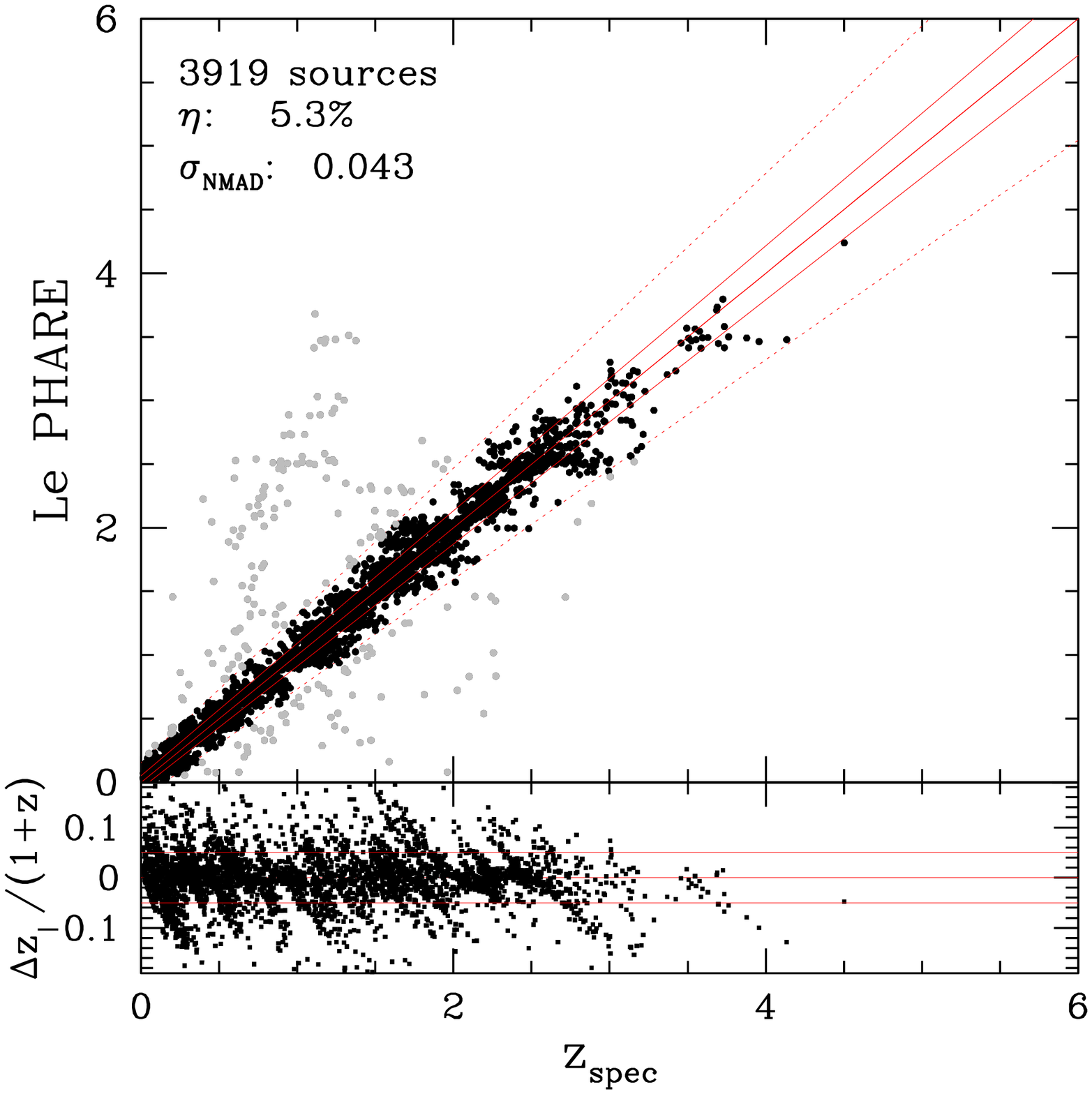}
\caption{Direct comparison between photo-z computed in this work with {\sc{Le PHARE}}  and {\sc{DNNz}}, within the HSC area for all the EXT (left panel) and PLIKE (central panel) sources. By construction true EXT sources should not have  spectroscopic redshift exceeding z $\approx$ 1. It is not possible to decide a {\it a priori} whether the photo-z are wrong, or whether the sources have been erroneously put in the EXT sample due to some issue of the photometry. In the middle panel in red we highlight the sources that have spectroscopic redshift higher than 3 (see text for details). 
Right panel: Comparison between photo-z from {\sc Le PHARE} and spec-z for the sources with {\sc Le PHARE} and {\sc DNNz} in agreement}.
\label{fig:zphot_DNNz}
\centering
\end{figure*}

Interestingly, despite using only HSC photometry also {\sc{DNNz}} shows a remarkable difference in the quality of the photo-z for the sources within or outside the area covered by KiDS+VIKING. This is probably due to the combined photometry from the filters {\it r} and {\it r2} and {\it i}  and {\it i2} that were changed during the survey. Most of the KiDS+VIKING area has been homogeneously observed  only in {\it i-} and {\it r-} band, while the rest of the area has a mixture of observations.
Taking this into account, we can compare {\sc{Le PHARE}} and {\sc{DNNz}} in the area within KiDS+VIKING and split by {\texttt{TYPE}}. Figure~\ref{fig:zphot_DNNz} shows how both  sets of photo-z suffer from some systematics (vertical and horizontal substructures) due on one side to the imbalance between galaxies and AGN in the training of {\sc{DNNz}} and on the other hand to the degeneracies in the solution for power-law dominated AGN and limited availability in photometry for {\sc{Le PHARE}}.

However, {\it when the photometry is sufficient and of good quality}, SED fitting can predict correctly the redshift of AGN also when higher than 3 (middle panel of Figure~\ref{fig:zphot_DNNz}; sources in red). This is a current limitation for photo-z computed via machine learning given the small sample of this kind of sources available for training (see Nishizawa et al., in prep.).

When the photo-z from {\sc{DNNz}} is available, for each source we can measure the mean photo-z between the values proposed by the two methods. Assuming this value as the right one, we  have that for 60.6\% of the extragalactic sources with {\texttt{CTP\_quality}}$\geq$2 {\sc{DNNz}} and {\sc{{\sc{Le PHARE}}}} agree ($|zp_{\rm {\sc{{\sc{Le PHARE}}}}}-zp_{\rm {\sc{DNNz}}}|<0.15\times (1 + mean(zp_{\rm {\sc{{\sc{Le PHARE}}}}},zp_{\rm {\sc{DNNz}}}))$. The comparison between {\sc{{\sc{Le PHARE}}}} and the spectroscopic redshift for 3919 sources with spec-z is shown in the third panel of Figure~\ref{fig:zphot_DNNz}; the fraction of outliers with respect to the spectroscopic sample is extremely small and the accuracy  very high, comparable to the accuracy routinely obtained for normal galaxies, using purely broad band photometry. 

Table~\ref{tab:comp_photoz} summarizes the result for {\sc{DNNz}} and {\sc{Le PHARE}} separately, within and outside  KiDS+VIKING. For the about 7500 sources for which the two methods provide results in disagreement, the spectroscopic sample does not help in discriminating the best photo-z, given that the spectroscopic sample is very small (756 and 553 sources in the two areas, respectively) and not representative  of the magnitude distribution in the sample (mean {\it r} value of the spectroscopic sample 20; mean {\it r} value of the sample for which {\sc{DNNz}} and {\sc{Le PHARE}} disagree 21.5. See also next section and Figure~\ref{fig:LS8_z_grade}). 
Photo-z derived via machine learning are well known to be very reliable only within the parameter space represented by the training sample, having little predictive power outside \citep[e.g.,][]{Brescia2019}. 
Keeping this in mind, we decided to rely on the prediction power of SED fitting and rely on the results from {\sc{Le PHARE}}. However, we report also the results from {\sc{DNNz}} and flag the sources for which {\sc{Le PHARE}} and {\sc{DNNz}} agree, or disagree, respectively (see next Section~\ref{subsection:redshift_flag}).


\begin{table*}[h]
\begin{center}
\begin{tabular}{@{}llll}
\toprule
\multicolumn{4}{c}{OUTLIERS FRACTION} \\ 
\midrule
         {\sc{{\sc{Le PHARE}}}} & &  in KiDS   & outside KiDS   \\
\midrule 
 &{\sc{Le PHARE}} and {\sc{DNNz}} agree    & 3.7\% [99/2663] & 8.5\% [107/1256]  \\

&{\sc{Le PHARE}} and {\sc{DNNz}} disagree    & 49.9\% [377/756] & 51.3\% [284/553]  \\
\toprule
 {\sc{DNNz}}  &       & in KiDS   & outside KiDS   \\
 \midrule       
&{\sc{Le PHARE}} and {\sc{DNNz}} agree    & 4.0\% [106/2663] & 6.8\% [85/1256] \\

&{\sc{Le PHARE}} and {\sc{DNNz}} disagree    & 56.1\% [424/756]& 64.0\% [354/553]  \\

\bottomrule
\end{tabular}
\caption{Fraction of outliers for the photo-z computed with {\sc Le PHARE} (top) and {\sc DNNz} (bottom), split by area, with respect to the number of sources with spectroscopic redshift for which the two methods agree (first row) or disagree (second row) following the definition in Section~\ref{subsection:DNNz}. The table clearly indicates that when {\sc Le PHARE} and {\sc DNNz} disagree, {\sc DNNZ} has a higher fraction of outliers among the spectroscopic sample, while when the two codes agree, the difference in fraction of outliers is marginal.Note that in the definition of agreement we used 1+mean({\sc LePHARE},{\sc DNNZ}). The small difference in the fraction of outliers for the two methods when they agree, depends on how close they are to the real spectroscopic value}.
\label{tab:comp_photoz}
\end{center}
\end{table*}

\subsection{{\texttt{CTP\_REDSHIFT}} and {\texttt{CTP\_REDSHIFT\_GRADE}} in the final catalog \label{subsection:redshift_flag}}

In the final catalog we report the spectroscopic redshifts (regardless of their reliability) and the photo-z from both {\sc{Le PHARE}} and {\sc{DNNz}}. In addition, for each source we summarise in the  two columns {\texttt{CTP\_REDSHIFT}} and  {\texttt{CTP\_REDSHIFT\_GRADE}} our best knowledge of redshift and its reliability.

The column {\texttt{CTP\_REDSHIFT}} lists original spectroscopic redshift when it is available and reliable ({\texttt{NORMQ}}=3). The redshift is set to 0 for all the sources that are classified as GALACTIC (either SECURE or LIKELY) or for which reliable redshift is not available. To the remaining sources we assign the photo-z from {\sc{Le PHARE}}.\\
 
Then in the column {\texttt{CTP\_REDSHIFT\_GRADE} } we provide a grade of confidence to the redshifts and the grades are:

\begin{itemize}
\item {\texttt{CTP\_REDSHIFT\_GRADE}=5 }: This is the higher grade, assigned to the sources with reliable spectroscopic redshift. Of the 6591 sources in this category 5377 are extragalactic sources and 1214 are Galactic (6465/6591 with {\texttt{CTP\_quality}} $\geq$ 2).

\item {\texttt{CTP\_REDSHIFT\_GRADE}=4 }: This is assigned to the sources for which the photo-z from {\sc{Le PHARE}} and {\sc{DNNz}}  agree (10949 in total, 9643 of which have {\texttt{CTP\_quality}} $\geq$ 2), because in the previous section we have demonstrated that for this subsample the fraction of outliers is very small and the accuracy very high. By construction, all the Galactic sources without spectroscopic redshift have {\texttt{CTP\_REDSHIFT\_GRADE}}=4 because {\sc{DNNz}} and {\sc{Le PHARE}} are set to zero and belong to this subsample (2995 sources).

\item {\texttt{CTP\_REDSHIFT\_GRADE}=3 }: This is assigned  to the sources with {\sc{Le PHARE}} and {\sc{DNNz}} in disagreement and  \texttt{pdz} > 40 (6741 in total, 6057 among the sources with {\texttt{CTP\_quality}} $\geq$ 2). The threshold at \texttt{pdz} > 40 has been set by looking at the fraction of outliers as a function of  \texttt{pdz} in the sample with spectroscopic redshift (see Table~\ref{tab:pdzThreshold}). At the same time we searched for the value of  \texttt{pdz}  that minimised the number of outliers and maximised  the number  of sources with z\_phot>4. The latter is suspiciously too high and this is due to the lack of deep photometry not only in the UV  but also in NIR: in fact the large majority if these high-z sources are concentrated in the area outside KiDS. 

\item {\texttt{CTP\_REDSHIFT\_GRADE}=2 }: This is assigned to the remaining sources with {\sc{Le PHARE}} and {\sc{DNNz}} in disagreement and \texttt{pdz} < 40, for which we are less confident about the photometric redshifts, In this group there are only 1326 sources, with 1092 having {\texttt{CTP\_quality}} $\geq$ 2.

\end{itemize}

\begin{figure}[!ht]
\centering
\includegraphics[width=9cm]{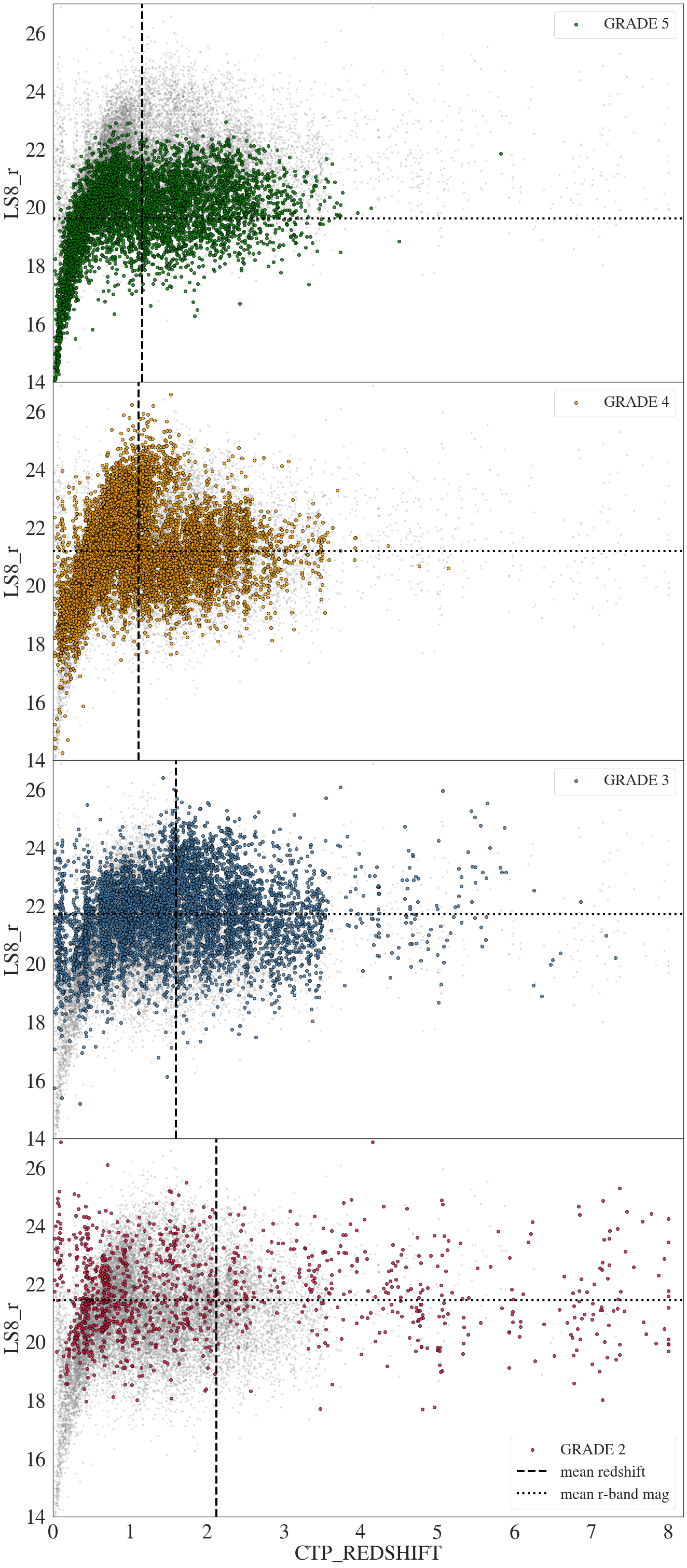}
\caption{eFEDS sources with reliable counterpart distributed in the magnitude vs. redshift plane, split among the various {\texttt{CTP\_REDSHIFT\_GRADE}} classes. For each panel also the mean value for redshift (vertical line) and magnitude (horizontal line) is indicated and the entire population is shown in light grey.}
\label{fig:LS8_z_grade}
\centering
\end{figure}

Figure~\ref{fig:LS8_z_grade} shows the distribution of the sources for each of the {\texttt{REDSHIFT\_GRADE}} in the magnitude redshift plane. Indicated are also the mean value of the redshift and magnitude for each of the subsamples.


\begin{table}[h]
\begin{center}
\begin{tabular}{ccccc}
\toprule
         pdz &  N. &
         N.  & N.   & N.  \\
         threshold  & sources &sources & sources & sources \\
         & & w/spec-z & w/spec-z & w/ z\_phot >4 \\
          & & &  \& outliers &  \\
\midrule
< 20  & 683 & 60/5287  & 46/60 & 103/386  \\
< 30  & 1058 & 93/5287  & 67/93 & 199/386  \\
< 40  & 1429& 134/5287  & 94/134& 230/386  \\
< 50  &1915 & 197/5287  & 121/197& 252/386  \\

\bottomrule
\end{tabular}
\caption{Properties distribution for secure counterparts classified as extragalactic sources with pdz lower than a certain threshold. The lower the pdz, the lower the quality of the fitting.}
\label{tab:pdzThreshold}
\end{center}
\end{table}

\begin{figure*}[!th]
\centering
\includegraphics[width=9cm]{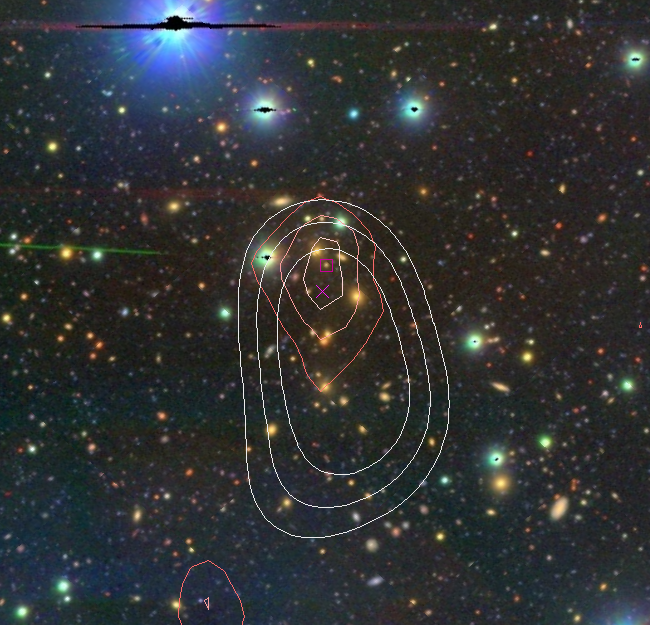}
\includegraphics[width=9cm]{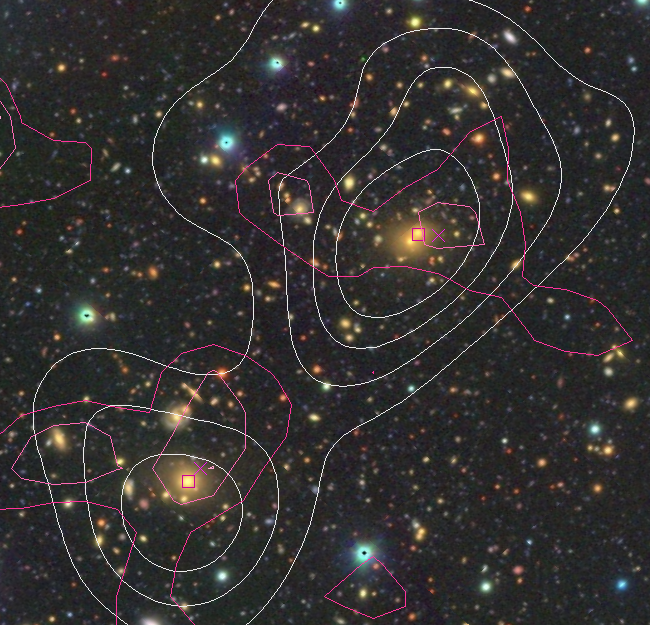}
\includegraphics[width=9cm]{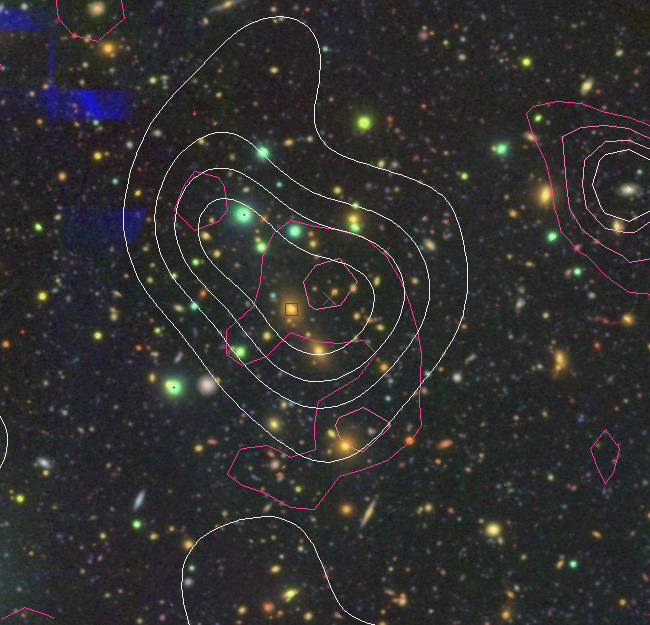}
\includegraphics[width=9cm]{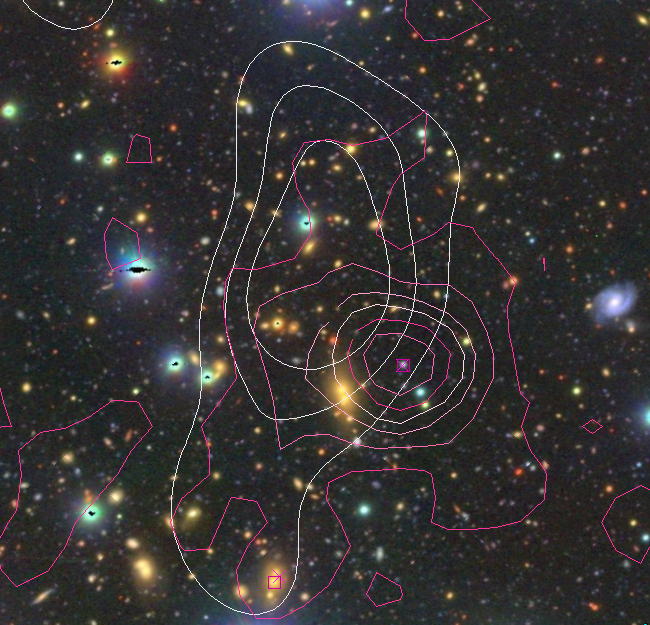}
\caption{from top-left,clock-wise:four examples of {\texttt{Cluster\_Class}}=5,4,3,2, as described in Section~\ref{section:in_overdensity}. X-ray contours are plotted in white, with the magenta cross indicating the X-ray position and magenta square indicating the counterpart selected in this paper, magenta contours indicate the red sequence galaxy density.  The HSC {\it g,r,i} cutouts are 5.5\arcmin$\times$5.5\arcmin in size.}
\label{fig:overdensity}
\end{figure*}
\begin{figure*}[!ht]
\centering
\includegraphics[width=\textwidth]{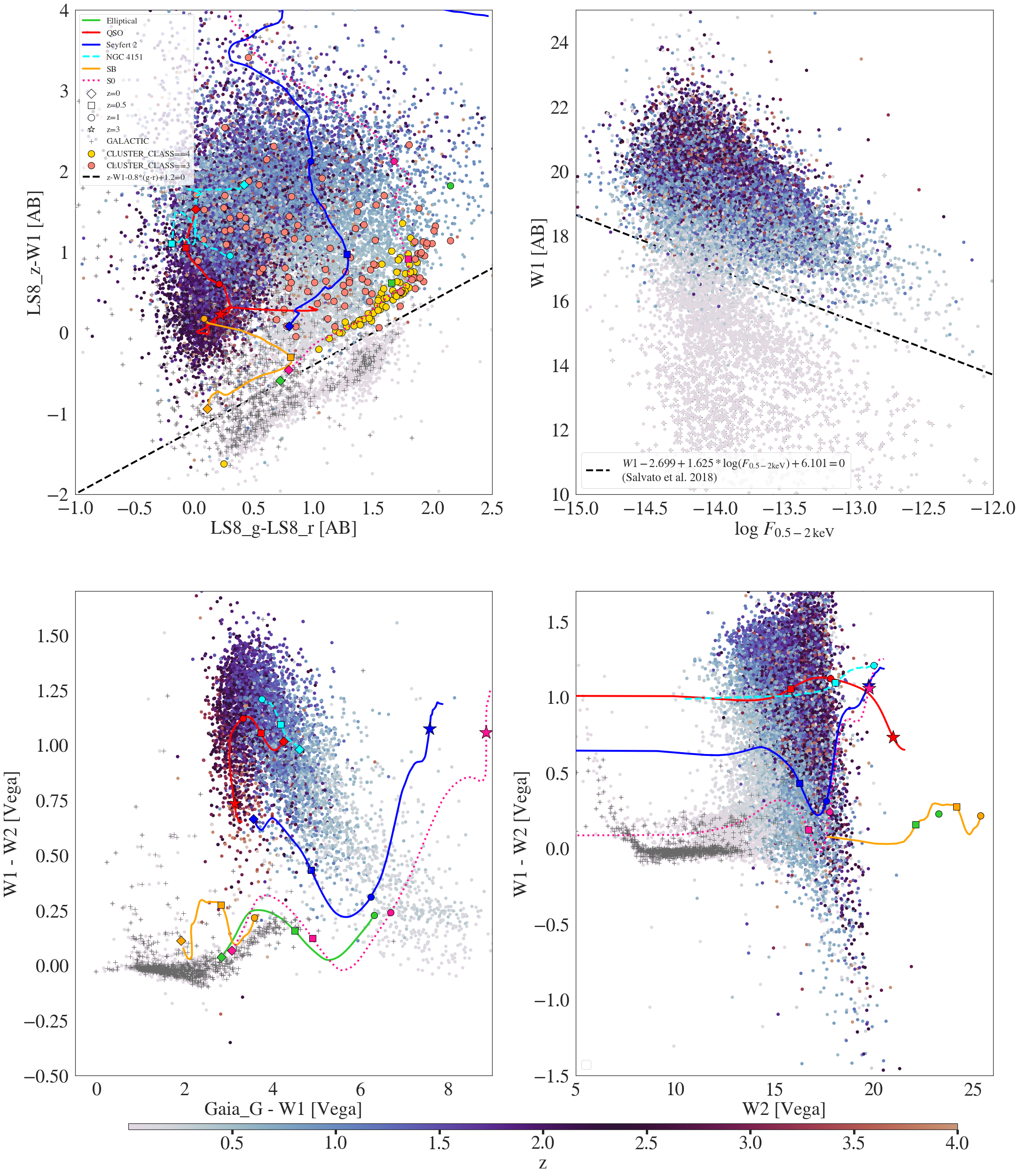}
\caption{The distribution of Galactic and extragalactic sources in eFEDS, colour coded by redshift in the parameter spaces defined by typical colours and fluxes. For the plots with optical and MIR colours, the track of templates characteristic of the population, also used in the computation of the photo-z are overplotted. The legenda on the top-left panel provides all the details for the four panels.}
\label{fig:z_population}
\centering
\end{figure*}
\begin{figure*}[!th]
\centering
\includegraphics[width=0.45\textwidth]{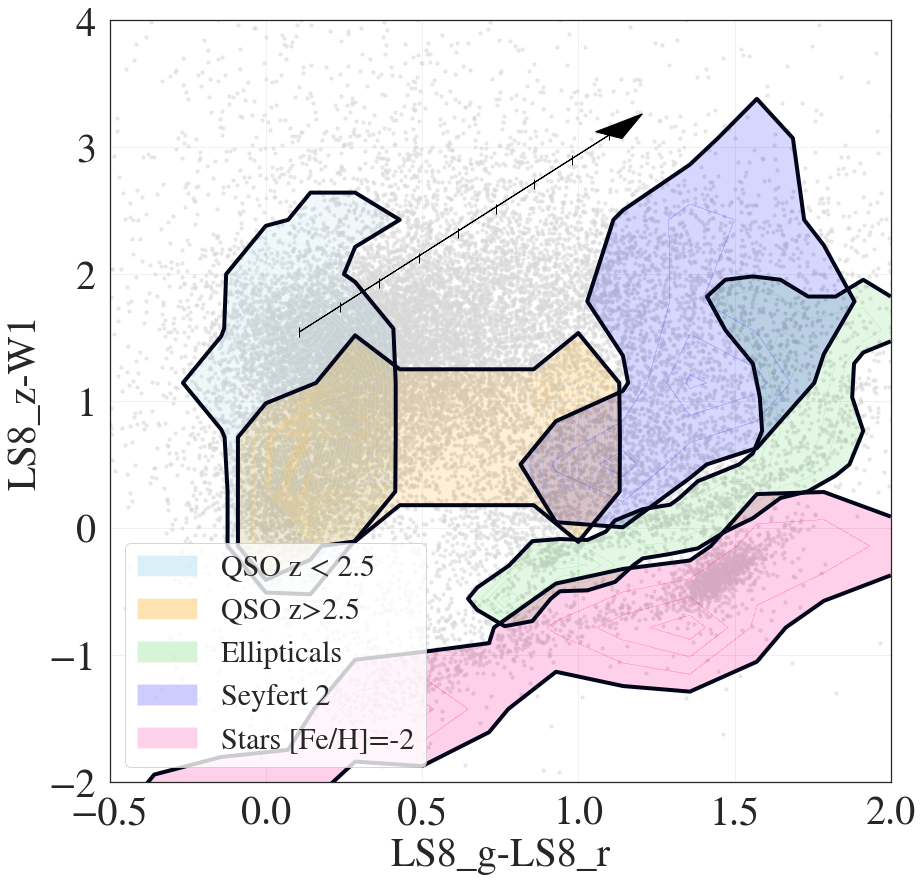}
\includegraphics[width=0.45\textwidth]{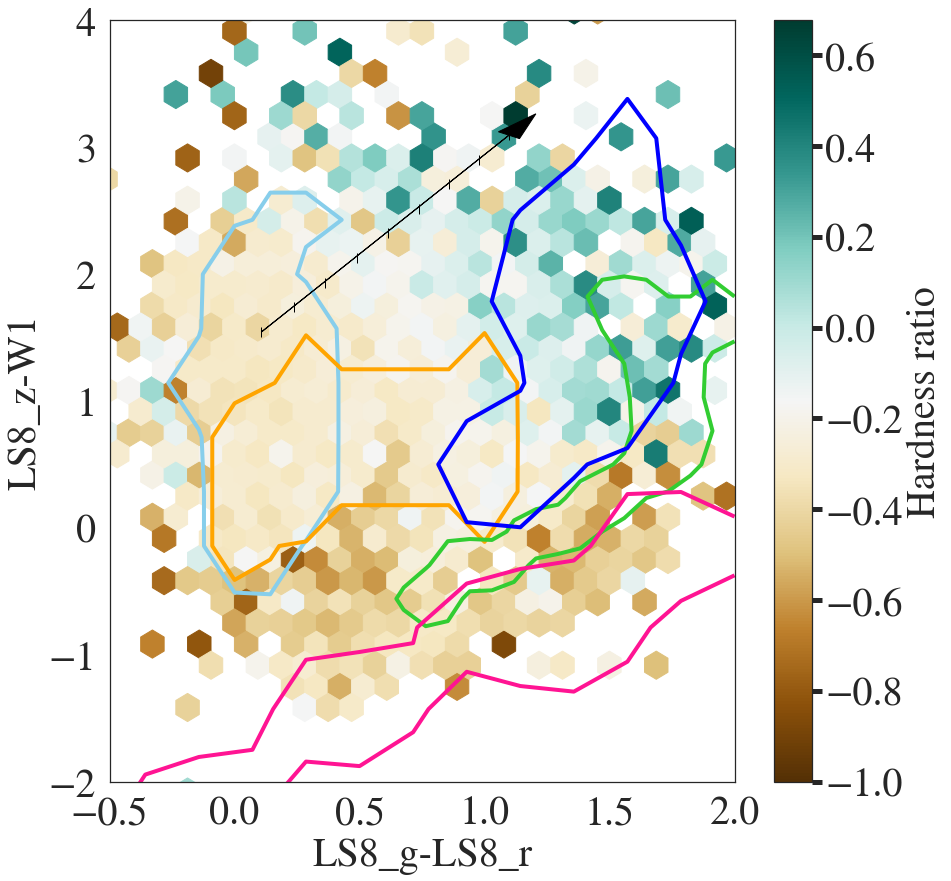}
\caption{\textit{Left}: Classes of X-ray emitting sources are shown in colour-colour space. The clouds for QSOs, elliptical galaxies and Seyfert 2 galaxies are derived from theoretical colour-redshift tracks computed with \texttt{LePhare} and a subset of SED templates used for the determination of photo-z in this work. Seyfert 2 and elliptical tracks were limited to z<1, while the QSO cloud is divided in z<2.5 and 2.5<z<4. The black arrow indicates the evolution of a theoretical QSO at z=0.2 in colour-colour space with increasing extinction. The ticks along this arrow are steps of $\Delta_{E(B-V)}=0.1$. The stellar cloud was derived from a 2D-histogram of MIST/MESA isochrones assuming [Fe/H]=-2 \citep{Choi2016,Dotter16,Paxton18}. \textit{Right:} The same classes are shown on top of the binned eFEDS data ($\mathrm{CTP\_quality\geq 2}$), colour-coded according to the average hardness-ratio (see text for details).} 
\label{fig:z_tracks}
\centering
\end{figure*}

\subsection{Flagging sources likely associated with clusters of galaxies in the point-like sample \label{section:in_overdensity}}

By construction the eFEDS X-ray point-source catalog is expected to have very little contamination from clusters of galaxies; still,  a small probability that a source is actually a cluster remains, as shown from the simulations we performed for eFEDS \citep[][]{TLiu2021}.
The reasons for clusters ending up in the point-like sample are manifold \citep[][]{Willis2021}. Most obviously, clusters with small apparent size or at low detection likelihood can fall below the thresholds used to define the extension of the X-ray source. In addition, clusters could leak into the point-like sample because of source splitting and superimposition of a bright point source and a cluster.

With this in mind we have run the multi-component matched filter cluster confirmation tool \citep[MCMF; ][]{MCMF1,MARDY3} on the eFEDS point source catalog. We run {\sc{MCMF}} as in the eFEDS extended sources catalog \citep[][]{Klein2021}, after adjusting some of the parameters (for example, limiting the area search from the X-ray position).
As in \citet{,Klein2021} we have defined 
a "contamination fraction", $f_\mathrm{cont}$, which expresses the probability for an optical concentration of red galaxies to be a chance alignment along the line of sight to the X-ray source.
This is the key selection criterion for selecting cluster candidates and immediately provides an estimate of the catalog contamination. A catalog created by selecting $f_\mathrm{cont}<a$ is expected to have contamination fraction of $a$, assuming the input catalog is highly contaminated.

Because of the high number density of sources in the point-like sample and/or the possibility that the emission from an actual cluster is split in many point sources,  it can happen that  many  close X-ray sources point to the same optical cluster. A simple cut in $f_\mathrm{cont}$ will therefore yield a much larger sample of sources than real clusters in that catalog, causing the contamination fraction to be much higher than expected. To compensate for that, for each eFEDS point source that is close to an optical overdensity,  an "environmental flag" is set to {\it true} for the source that is closest to the overdensity and that is at least 0.75 Mpc away from a cluster detected in the extent selected sample \citep[][]{TLiu2021,Klein2021} at similar redshift. Only when the flag is set to {\it true} is the point-source further considered as a candidate for being a cluster.

For these latter subgroup of sources, following \citet{Klein2021}, MCMF  assigns a redshift  to the cluster (via red-sequence). In addition, the photo-z of the counterpart to the point-like sources is recomputed assuming they are passive galaxies.
The two redshifts are then compared with the one computed by {\sc {LePhare}} as described in the previous section (Section~\ref{section:photoz}).

Combining all the information described above we define a new flag, \texttt{Cluster\_Class}, which indicates the possibility that an eFEDS X-ray (point-like) source is actually a cluster or  belongs to a cluster.

\begin{itemize}
    \item {\texttt{Cluster\_class}}=5: {\texttt{CTP\_QUALITY}}$\leq1$ \&  $f_\mathrm{cont}<0.2$  and the environmental flag set to {\it true}:  the counterpart {\sc{nway}} or {\sc {astromatch}} is considered unreliable and the X-ray emission is more likely associated to a cluster (top-left panel of Figure~\ref{fig:overdensity}; 120 cases).
    
    \item {\texttt{Cluster\_class}}=4: \texttt{CTP\_QUALITY}$\geq2$ \&  $f_\mathrm{cont}<0.2$ with the environmental flag set to {\it true},  the optical colours of the sources are typical of passive galaxies and the redshift computed with {\sc {LePhare}} coincides with the redshift of the optical cluster: the counterpart is reliable and the point source is a galaxy member (possibly the BCG) of the optically detected cluster (to-right panel in Figure~\ref{fig:overdensity}; 63 cases).
    
    \item {\texttt{Cluster\_class}}=3:  \texttt{CTP\_QUALITY}$\geq2$ \&  $f_\mathrm{cont}<0.2$ and the   environmental flag set to {\it true} and the redshift computed assuming an AGN template is consistent with the redshift of the optical cluster but the optical colours of the counterpart are not typical of a passive galaxy: the counterpart is correct and the source is a cluster member (bottom-left panel from the left of Figure~\ref{fig:overdensity}; 96 cases).
    
    \item {\texttt{Cluster\_class}}=2: \texttt{CTP\_QUALITY}$\geq2$ \&  $f_\mathrm{cont}<0.01$  and the  environmental flag set to {\it true} while the photo-z computed by the three methods are in disagreement: the counterpart is reliable and the source  (AGN) is just projected on a likely cluster (bottom-right panel of Figure~\ref{fig:overdensity}; 67 cases).
    \end{itemize}

 In all the other cases the X-ray emission is from a genuine point-source and the and it likely not from the extended, hot intercluster medium.
A dedicated effort is currently ongoing to confirm the secure clusters in the point source catalog as well as  characterising and measuring their X-ray and radio properties of the confirmed clusters \citep[e.g.,][]{Bulbul2021}.
\section{Data Release}\label{section:catalogs}
 The catalogs listing the properties of the counterparts to eFEDS point like sources in the Main and Hard selected samples \citep[][]{Brunner2021} associated with this paper are available via CDS/Vizier and via the web page at MPE dedicated to the eROSITA data release\footnote{\url{https://erosita.mpe.mpg.de/edr/eROSITAObservations/Catalogues/}}. The list of the columns and their description for the two samples is available in Appendix~\ref{appendix:catalogs_column}. Only the basic X-ray properties are listed here (columns 1-9). For the complete list please refer to the catalogs released in \citet{Brunner2021}. After the columns reporting the key X-ray  properties of the sources, columns 10-36 report the results of the counterpart (CTP) association  followed by the key parameters from {\sc{nway}}, {\sc{astromatch}} and Hamstar respectively.  Next (columns 36-49) we present the photometry from the recent Gaia EDR3 release, in the original photometric system, followed by all the collected photometry, corrected for extinction (columns 51-108). To be remembered is the fact that the HSC photometry  from S19A in {\it i-} and {\it r-} bands has been split in {\it i-, i2-} and {\it r-, r2-} and that Kron is listed for EXT sources while cmodel is listed for PLIKE (see Section~\ref{section:photoz}). Columns 109-117 list basic properties of the sources, like whether they are within KiDS, HSC etc., while columns 118-126 deal with all the information on the spectroscopy available  information related to  spectroscopy when available. The output parameters from {\sc{Le PHARE}} and {\sc{DNNZ}} are listed in columns 127-148.  The columns {\texttt{CTP\_REDSHIFT}} and {\texttt{CTP\_REDSHIFT\_GRADE}} summarise the redshift properties of the sources as discussed in Section~\ref{subsection:thresh}, while the column {\texttt{CLUSTER\_CLASS}} refers to the results presented in Section~\ref{section:in_overdensity}.
 
In addition to the catalogs, we can provide under direct request to the first author, the redshift distribution function  and SED fitting of each source in the catalog. An example is shown in Figure~\ref{fig:SED}.

\section{Discussion\label{section:discussion}}

The size and depth of the eFEDS X-ray survey, combined with ancillary data both in photometry and spectroscopy, allows us to paint a comprehensive picture of the average population of X-ray sources that contribute the bulk of the cosmic X-ray background (CXB) flux at energies $<10$ keV \citep[see e.g.,][]{Gilli07}, both in its Galactic and extragalactic content. The identification of the optical/IR counterparts, to a high degree of completeness and reliability, as discussed here, will facilitate detailed population studies of X-ray active stars, Galactic compact objects, and AGN. Here we briefly outline the main properties of our sample by examining in detail the distributions of the X-ray sources in various colour/redshift spaces.


\subsection{Population studies\label{section:population}}
Figure~\ref{fig:z_population} shows the distribution of all the eFEDS sources with a secure counterpart (\texttt{CTP\_quality}$\geq$2; see section~\ref{section:comparison}) 
in four different multi-band photometric spaces, chosen for their wide applicability to large areas of the sky.

The top panel shows sources in the {\it z-W1} vs. {\it g-r} space, colour-coded by their redshift. Overlaid are a few representative tracks of various classes of extra-galactic objects. Beside the clear separation between Galactic and extra-galactic objects already discussed in section~\ref{section:properties}, the X-ray points identify clear sequences of un-obscured QSO, obscured Seyferts and inactive galaxies. The inactive galaxies are best represented by the S0 and Elliptical tracks, suggesting that some of these are the sources that are associated (or confused) with a cluster. Indeed the sources indicated by a yellow circle in the top left figure, have {\texttt{CLUSTER\_CLASS}}=3,4 indicating that they belong to a cluster and in most cases they are the BCG (see Section ~\ref{section:in_overdensity}). These sources are best  fit by the template of a passive galaxy as the spectra for those available also suggest ( e.g., lack of emission lines from star formation, strong HK lines). However, some of the spectra together with the clear features from a non star-forming galaxy also reveal the presence of broad emission lines typical of AGN (see Bulbul et al., in preparation).

The top-right panel of Figure~\ref{fig:z_population} shows the distribution of points in the MIR (WISE {\it W1}) vs soft X-ray (0.5-2 keV) plane (same as Figure~\ref{fig:grzw1_wX_line}), originally introduced in \cite{Salvato18b}. 
X-ray bright objects above the dashed line are typically AGN, while most of the IR bright objects below the line are Galactic X-ray emitting stars, with some contamination from nearby extragalactic objects. These sources are rare, but given the size of eFEDS, they are a non-negligible number. Thus, when using this plot for other surveys one should not forget to account for the size of the survey. The larger the surveys the less efficient the line separator is.

The bottom-right panel shows the distribution of the sources in the Wise-only {\it W1-W2} vs {\it W2} colour-magnitude plane\footnote{For this plot we are using Vega System, so that user can compare the figure with similar ones done using AllWISE all sky}. This is widely used to classify point sources, as it easily separates stars, with {\it W1-W2} $\approx 0$, from QSOs, with {\it W1-W2} $> 0.5$, \citep[see e.g.][]{Wright2010,Assef2013}. Once more, the eFEDS X-ray selection reveals the full extent of the extragalactic (AGN) population with intermediate IR colours between AGN- and host-galaxy dominated, typical of either obscured (Seyfert 2) or low-luminosity AGN \citep[e.g.,][]{Merloni2016,Hickox2018}. 

Finally, the bottom-left panel shows the distribution of the eFEDS sources in the optical/MIR diagram defined by the "all-sky available" {\it G-W1} vs {\it W1-W2}, frequently used to separate QSO from stars in the Gaia catalog. As already pointed out in Section~\ref{section:properties}, 10\% of the Galaxtic sources are too faint to be detected by Gaia. This is even more true for the extragalactic sources: the plot shows only 57\% of the entire eFEDS sample. However, the plot shows insights on the population that the first eROSITA All-Sky Survey (eRASS1) will uncover. As expected, the X-ray selected eFEDS sources contain, beyond stars and (unobscured) QSO, a tail at high {\it G-W1} (i.e. bright MIR, faint optical magnitudes) typical of inactive galaxies and/or mildly obscured AGN.

This is indeed confirmed by comparing the location of the extra-galactic eFEDS sources in the {\it grzW1} plane with the X-ray hardness ratio measured from the X-ray counts in the bands where eROSITA is most sensitive. 
The right-hand panel of Figure~\ref{fig:z_tracks} shows the distribution of the sources in that plane, colour-coded by their average Hardness Ratio (defined as (H-S)/(H+S) where H and S are respectively the counts in the ranges 1.0-2.0 keV and 0.2-1.0 keV\footnote{The hardness ratio is calculated using the columns ML\_CTS\_b1, ML\_CTS\_b2 and ML\_CTS\_b3 in Brunner et al. 2021 catalog: \texttt{(ML\_CTS\_b3 - (ML\_CTS\_b1 + ML\_CTS\_b2))/(ML\_CTS\_b3 + (ML\_CTS\_b1 + ML\_CTS\_b2))}}), while the left-hand panel highlights the loci of the most common classes of sources based on the distribution of templates tracks. Indeed, the hardest sources in the eROSITA band populate the optical/MIR colour-space of Seyfert 2 galaxies and/or reddened QSO. A detailed discussion of the X-ray spectral properties of the AGN in the Main eFEDS sample will be presented in \citet{TLiu2021S}.


\begin{figure}[!htb]
\centering
\includegraphics[width=8cm]{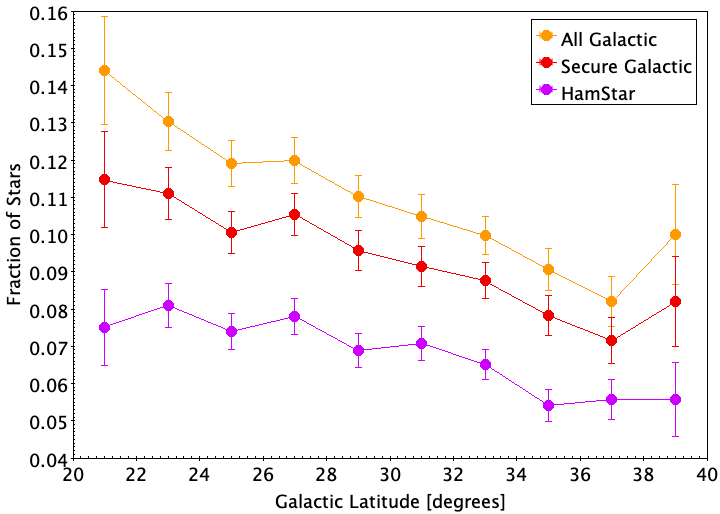}
\caption{Fraction of all X-ray sources with \texttt{CTP\_quality}$\ge$2 which are classified as Galactic as a function of Galactic Latitude. Purple symbols are for objects classified by the HamStar method; red symbols are for all "Secure Galactic" objects and orange for both "Secure" and "Likely" Galactic objects (see Section~\ref{section:counterparts} for the definition).}
\label{fig:gal_frac_latitude}
\end{figure}
\subsection{eFEDS stellar content \label{subsection:stars}}
The eFEDS field spans a wide range of Galactic latitudes (from about +20 to about +40). Reassuringly, the fraction of the X-ray sources which are classified as Galactic (see Section~\ref{section:counterparts} for the definition) increases towards low Galactic latitude, as shown in Figure~\ref{fig:gal_frac_latitude}. The figure also confirms that the priors adopted by Hamstar are not representative of all Galactic sources in eFEDS. In fact about 22.3\% of the Galactic sources identified by {\sc {nway}}/{\sc{astromatch}} are fainter than the 19th magnitude (with 10\% Gaia non detected).


\section{Conclusions}\label{section:conclusions}
We have presented the identification of the counterparts to the point sources in eFEDS listed in the Main and Hard catalogues \citet[][]{Brunner2021}, together with the study of their  multiwavelength  properties. eFEDS, having a limiting flux of $ F_{\rm 0.5-2 \, keV} \sim 6.5 \times 10^{-15} \, \mathrm{erg\, s^{-1}\, cm^{-2}}$ is a factor of $\sim50$\%  deeper than the final eROSITA all-sky survey, and because of that can be used also as a forecast for eRASS:8, not only for the population that eRASS:8 will reveal, but also for the challenges that are ahead of us with respect to counterpart identification and redshift determination.\\
$\bullet$ Counterpart identification: We have used {\sc{nway}} \citet[][]{Salvato18a} and {\sc {astromatch}}, each of them using, in addition to spatial information from eFEDS and LS8, a prior based on the properties of a training sample of 3XMM sources, and tested on a  validation sample of {\it Chandra} sources, made eFEDS-like in terms of positional accuracy. Each method has identified its own priors in a different way. For the validation sample {\sc {nway}}, identified correctly 95\% of the sources with only 2\% of sources having a second possible counterpart, against 89\%  and 10\% for {\sc{astromatch}}, but at the threshold adopted for p\_any and LR\_BEST, both methods  have very high completeness and purity (above 95\%). These remarkable  results, well above the predicted completeness and purity mentioned in \citet[][]{Merloni12}, are due to three important factors: the development of new methods for identifying the correct counterparts, large samples of X-ray detected sources with known counterparts, and the availability of sufficiently deep, homogenised, multi-wavelength photometry from optical to MIR over very wide areas from which to construct the SED of these sources to be used as training.  
In the next two years, by the time  eROSITA will have completed the final all-sky survey, the methods will continue to improve and the training/validation samples will increase in size. Most importantly, the coverage of the multiwavelength catalogs usable for the identification of the counterparts  will be larger. In fact, while the DESI Legacy Imaging Survey DR9 \citep[LS9; ][]{Dey2019}, just became publicly available, the work on DR10 has started. The survey will cover virtually all of the eROSITA-DE area of the sky at sufficient depth, thanks to the inclusion the DECam data taken via the DeROSITAS survey (PI A. Zenteno). We predict that the identification of the counterparts for the entire eRASS will be at least of the same quality of eFEDS, also in the Galactic plane, thanks to the inclusion of Gaia EDR3 recently released. \\ 
$\bullet$ Redshift determination: Given the lack of sufficiently deep NIR data outside the DES area \citep{SevillaNoarbe2021}, the possibility of obtaining reliable photometric redshifts via SED fitting will be low, at least until data from SpherEx \citep[][]{Dore18a} will be made available (launch planned for Summer 2024). However, as demonstrated in Nishizawa et al (in prep) and \citet{Borisov2021}, thanks to the increasing size and completeness of the spectroscopic sample that can be used for the training, reliable photometric redshifts for any type of X-ray extragalactic sources are now possible. For example, thanks to the spectroscopic follow-up of the eROSITA-DE sources planned via Vista/4MOST and SDSS-V/BHM, we will obtain redshifts for 80\% of the sources detected by eRASS:3, thus limiting the need of photo-z and, at the same time, assuring a high quality of photo-z that will use these spectroscopically confirmed sources as training.


\begin{acknowledgements}
    We are thankful to the anonymous referee that with a careful reading of the paper helped us  improving the manuscript. 
    M.S. thanks Olivier Ilbert for the feedback provided on the computation of the photo-z. J.W. acknowledges support by the Deutsche Forschungsgemeinschaft (DFG, German Research Foundation) under Germany’s Excellence Strategy - EXC-2094 -390783311.  M.K. acknowledges support by DFG grant KR 3338/4-1. B.M. acknowledges funding from  European Union’s Horizon 2020 research and innovation program under the Marie Skłodowska-Curie grant agreement No 860744 (BiD4BEST). PCS acknowledges support by DLR grants 50 OR 1901, 50 OR 2102. WNB acknowledges financial support from NASA grant 80NSSC19K0961 and the V.M. Willaman Endowment at Penn State.
    \newline
    This work is based on data from eROSITA, the soft X-ray instrument aboard SRG, a joint Russian-German science mission supported by the Russian Space Agency (Roskosmos), in the interests of the Russian Academy of Sciences represented by its Space Research Institute (IKI), and the Deutsches Zentrum für Luft- und Raumfahrt (DLR). The SRG spacecraft was built by Lavochkin Association (NPOL) and its subcontractors, and is operated by NPOL with support from the Max Planck Institute for Extraterrestrial Physics (MPE).
     \newline
    The development and construction of the eROSITA X-ray instrument was led by MPE, with contributions from the Dr. Karl Remeis Observatory Bamberg \& ECAP (FAU Erlangen-Nuernberg), the University of Hamburg Observatory, the Leibniz Institute for Astrophysics Potsdam (AIP), and the Institute for Astronomy and Astrophysics of the University of Tübingen, with the support of DLR and the Max Planck Society. The Argelander Institute for Astronomy of the University of Bonn and the Ludwig Maximilians Universität Munich also participated in the science preparation for eROSITA.
     \newline
    We have made use of TOPCAT and STILTS \citep[][]{Taylor05, Taylor06}   
    \newline
    Based on observations made with ESO Telescopes at the La Silla Paranal Observatory under programme IDs 177.A-3016, 177.A-3017, 177.A-3018 and 179.A-2004, and on data products produced by the KiDS consortium. The KiDS production team acknowledges support from: Deutsche Forschungsgemeinschaft, ERC, NOVA and NWO-M grants; Target; the University of Padova, and the University Federico II (Naples).
     \newline
    The Hyper Suprime-Cam (HSC) collaboration includes the astronomical communities of Japan and Taiwan, and Princeton University. The HSC instrumentation and software were developed by the National Astronomical Observatory of Japan (NAOJ), the Kavli Institute for the Physics and Mathematics of the Universe (Kavli IPMU), the University of Tokyo, the High Energy Accelerator Research Organization (KEK), the Academia Sinica Institute for Astronomy and Astrophysics in Taiwan (ASIAA), and Princeton University. Funding was contributed by the FIRST program from the Japanese Cabinet Office, the Ministry of Education, Culture, Sports, Science and Technology (MEXT), the Japan Society for the Promotion of Science (JSPS), Japan Science and Technology Agency (JST), the Toray Science Foundation, NAOJ, Kavli IPMU, KEK, ASIAA, and Princeton University. 
 \newline
This paper is based [in part] on data collected at the Subaru Telescope and retrieved from the HSC data archive system, which is operated by Subaru Telescope and Astronomy Data Center (ADC) at National Astronomical Observatory of Japan. Data analysis was in part carried out with the cooperation of Center for Computational Astrophysics (CfCA), National Astronomical Observatory of Japan.
\newline
The Legacy Surveys consist of three individual and complementary projects: the Dark Energy Camera Legacy Survey (DECaLS; Proposal ID 2014B-0404; PIs: David Schlegel and Arjun Dey), the Beijing-Arizona Sky Survey (BASS; NOAO Prop. ID 2015A-0801; PIs: Zhou Xu and Xiaohui Fan), and the Mayall z-band Legacy Survey (MzLS; Prop. ID 2016A-0453; PI: Arjun Dey). DECaLS, BASS and MzLS together include data obtained, respectively, at the Blanco telescope, Cerro Tololo Inter-American Observatory, NSF’s NOIRLab; the Bok telescope, Steward Observatory, University of Arizona; and the Mayall telescope, Kitt Peak National Observatory, NOIRLab. The Legacy Surveys project is honored to be permitted to conduct astronomical research on Iolkam D\'uag (Kitt Peak), a mountain with particular significance to the Tohono O’odham Nation.
\newline
Funding for the Sloan Digital Sky Survey IV has been provided by the Alfred P. Sloan Foundation, the U.S. Department of Energy Office of Science, and the Participating Institutions. SDSS acknowledges support and resources from the Center for High-Performance Computing at the University of Utah. The SDSS web site is www.sdss.org.
\newline
SDSS is managed by the Astrophysical Research Consortium for the Participating Institutions of the SDSS Collaboration including the Brazilian Participation Group, the Carnegie Institution for Science, Carnegie Mellon University, Center for Astrophysics | Harvard \& Smithsonian (CfA), the Chilean Participation Group, the French Participation Group, Instituto de Astrofísica de Canarias, The Johns Hopkins University, Kavli Institute for the Physics and Mathematics of the Universe (IPMU) / University of Tokyo, the Korean Participation Group, Lawrence Berkeley National Laboratory, Leibniz Institut für Astrophysik Potsdam (AIP), Max-Planck-Institut für Astronomie (MPIA Heidelberg), Max-Planck-Institut für Astrophysik (MPA Garching), Max-Planck-Institut für Extraterrestrische Physik (MPE), National Astronomical Observatories of China, New Mexico State University, New York University, University of Notre Dame, Observatório Nacional / MCTI, The Ohio State University, Pennsylvania State University, Shanghai Astronomical Observatory, United Kingdom Participation Group, Universidad Nacional Aut\'onoma de M\'exico, University of Arizona, University of Colorado Boulder, University of Oxford, University of Portsmouth, University of Utah, University of Virginia, University of Washington, University of Wisconsin, Vanderbilt University, and Yale University.

\end{acknowledgements}


\newpage
\begin{appendix}
\section{Construction of the training, validation and field samples \label{Appendix:TrainingAndBlindSample}}

In the following, we describe the construction of the training, validation and associated field samples used for determining the different priors adopted by 
{\sc{nway}} and {\sc{astromatch}} 
 for the determination of the counterparts and for assessing the reliability of the association, presented in Section~\ref{section:comparison}. 

\subsection{Reference sample selected from the 3XMM-DR8 serendipitous source catalogue (a.k.a. training sample)}
\label{subsection:3xmmref}
We start with the 3XMM-DR8\footnote{\url{http://xmmssc.irap.omp.eu/Catalogue/3XMM-DR8/3XMM_DR8.html}} catalogue of 
X-ray detections, and estimate the 0.5-2\, keV X-ray flux (and uncertainty) of each detection from the 
0.5-1, and 1-2\, keV band fluxes (and their uncertainties). We then select only those detections that meet all of the following X-ray quality criteria: 
\begin{enumerate}[i.]
\item have X-ray flux in the range probed by eFEDS ($F_{0.5-2\mathrm{keV}} > 2\times10^{-15}$\,erg\,s$^{-1}$\,cm$^{-2}$, see figure 9 of \citealt[][]{Brunner2021}), 
\item have X-ray positions that have been aligned with the optical frame and that have uncertainty smaller than 1.5\,arcsec, 
\item have a signal to noise ratio for $F_{0.5-2\mathrm{keV}}$ that is greater than 10, 
\item are consistent with being point-like at the resolution of {\it XMM-Newton}, 
\item have no close X-ray neighbours within 10\,arcsec, 
\item were not detected at the extreme off-axis angles, 
\item were detected in {\it XMM-Newton} exposures of at least 5\,ks,
\item were not labelled by the 3XMM pipeline as being confused, affected by high X-ray background or flagged as being problematic for any reason. 
\end{enumerate}
We then exclude any X-ray detections that lie in parts of the sky that are not representative of a well-chosen extragalactic survey field such as eFEDS, or where the optical imaging catalogue (LS8) is likely to be saturated/unreliable (due to very bright stars). Specifically, we {\em exclude} any X-ray detections that: 
\begin{enumerate}[i.]
\item lie near the Galactic plane ($|b| < 15$\,deg), 
\item lie near the Large or Small Magellanic clouds or M31 (within 5, 3 and 1\,degree radii respectively), 
\item lie within the disks of bright ($B_T < 12$) well-resolved galaxies from \citet[][]{deVacoulers1991}, 
\item lie closer than 3\,arcmin from any very bright star from the Yale Bright Star Catalog \citep[][]{Hoffleit1991paper}, or 
\item lie closer than 3\,arcmin from any Tycho-2 \citep[][]{Hog2000} star having $B_T < 9$ or $V_T < 9$. 
\end{enumerate}
After applying these criteria, we are left with a sample of 36276 unique point-like X-ray sources with median positional uncertainty 0.57\,arcsec\,. The X-ray flux distribution of the 3XMM-DR8-based training sample is broadly similar to that of the eFEDS main sample. We find that 92\% of the training sample have 0.5--2\,keV fluxes in the range 5x10$^{-15}$ - 1x10$^{-12}$\,erg\,s$^{-1}$\,cm$^{-2}$, and median flux is 1.7x10$^{-14}$\,erg\,s$^{-1}$\,cm$^{-2}$. The equivalent metrics for eFEDS main sample are 91\% and 1.0x10$^{-14}$\,erg\,s$^{-1}$\,cm$^{-2}$ respectively.

We then carried out a positional match of this X-ray  training sample to the LS8, initially considering optical/IR objects that 
lie within 5\,arcsec\, of the X-ray positions. We used {\sc{nway}} \citep[][]{Salvato18a} to carry out this cross-match, using only astrometric information and number densities ({\sc{nway}} basic mode, i.e. without any magnitude or colour priors). We retain only the X-ray sources with very secure unique optical counterparts. Specifically, we require that we consider only X-ray sources with >90\% probability of having an optical/IR counterpart, and only cases where the best optical/IR counterpart is at least 9 times more probable than the next best possibility (\texttt{p\_any} > 0.9, \texttt{p\_i} > 0.9)\footnote{in {\sc{nway}} \texttt{p\_any} is the probability, for each source in the primary catalog (eFEDS in this case) to have a counterpart in the secondary catalogs; then, for each source in the secondary catalogues, \texttt{p\_i} gives the probability to be the correct counterpart to the source in the primary catalogue (see more in the {\sc nway} manual and \cite{Salvato18a}}. As before, we can afford to be very strict with these criteria, since we primarily care about purity and not completeness. These cuts result in a 3XMM/LS8 reference sample of 20705 high quality X-ray/OIR matches.

We select a corresponding sample of non-X-ray emitting field objects from the LS8, using annular regions (15, 30\,arcsec\, radii) around each of the 20\,705 reference sample positions. The field sample was further filtered to remove any object that lies within 15\,arcsec\, of {\em any} 3XMM-DR8 source. This field sample contains just under 396\,000 entries.

\subsection{Reference sample selected from the Chandra Source Catalogue v2.0 (a.k.a. validation sample) \label{subsection:csc2ref}}
A supplementary X-ray/OIR reference sample was derived from the  Chandra Source Catalogue v2.0\footnote{\url{https://cxc.cfa.harvard.edu/csc2/index.html}}.  We used the Web API to retrieve all 
CSC2 sources that satisfied the following X-ray quality criteria: 
i) Have $F_{0.5-2\mathrm{keV}} > 2\times10^{-15}$\,erg\,s$^{-1}$\,cm$^{-2}$ (estimated from the standard CSC2 `s' and `m' bands), 
ii) have high significance $> 6$, 
iii) have a signal to noise ratio on $F_{0.5-2\mathrm{keV}}$ that is greater than 5, 
iv) have X-ray positions with 95\% uncertainty ellipse radius smaller than 1.0\,arcsec\,, 
v) are consistent with being point-like at the resolution of Chandra, 
vi) were detected in Chandra exposures of at least 1\,ks,
vii) were not labelled by the CSC pipeline as being confused, affected by readout streaks, or piled up. 
Exactly the same sky region filtering criteria were applied to the CSC-based sample as were used to filter the 3XMM-based reference sample (see Section~\ref{subsection:3xmmref}). These criteria result in a sample of 6066 X-ray sources. 

We followed a similar process as before (Section~\ref{subsection:3xmmref}) to match the CSC2 sources 
to the LS8 catalogue, retaining only very secure matches (having \texttt{p\_any} > 0.9, \texttt{p\_i} > 0.9).
This results in a CSC-based X-ray/OIR reference sample that contains 3415 objects.


\begin{figure*}[t]
\centering

\includegraphics[width=5cm]{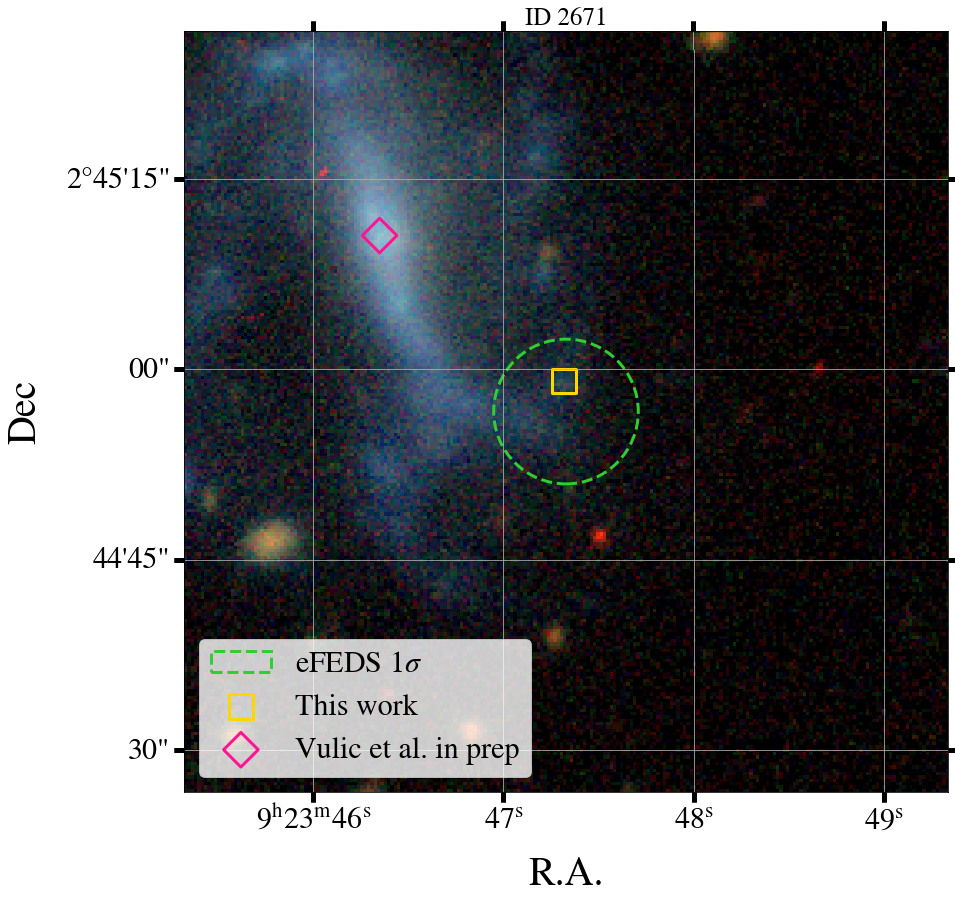}
\includegraphics[width=5cm]{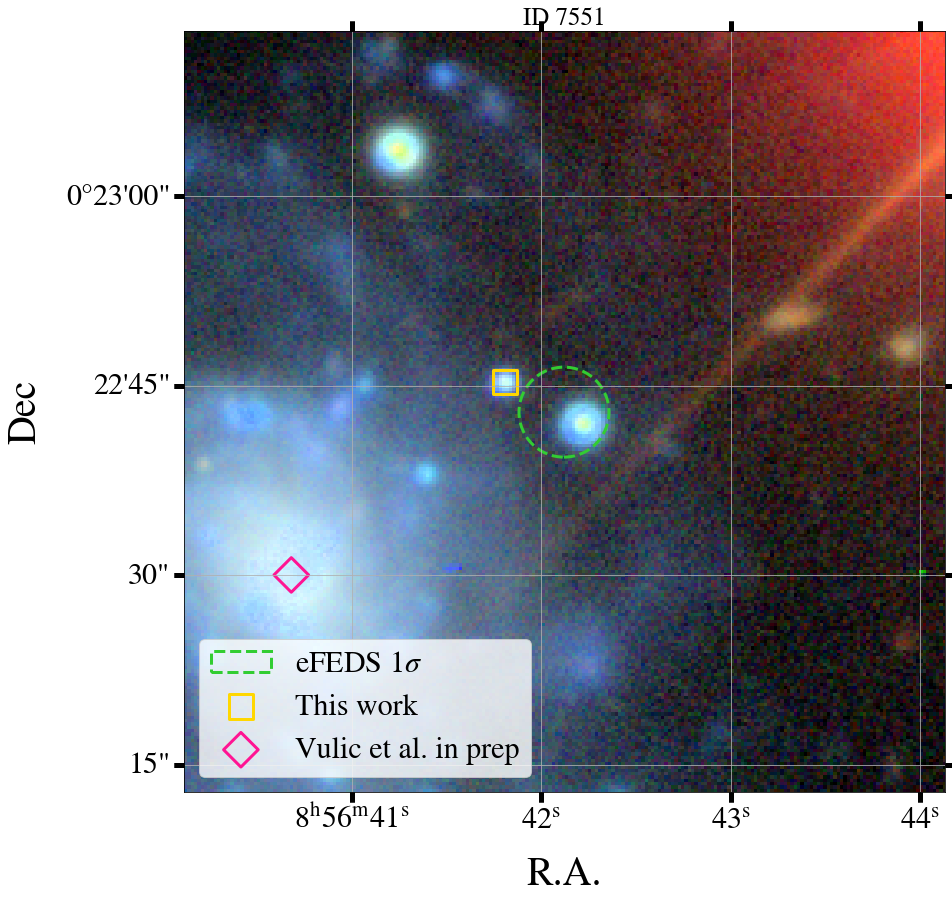}
\includegraphics[width=5cm]{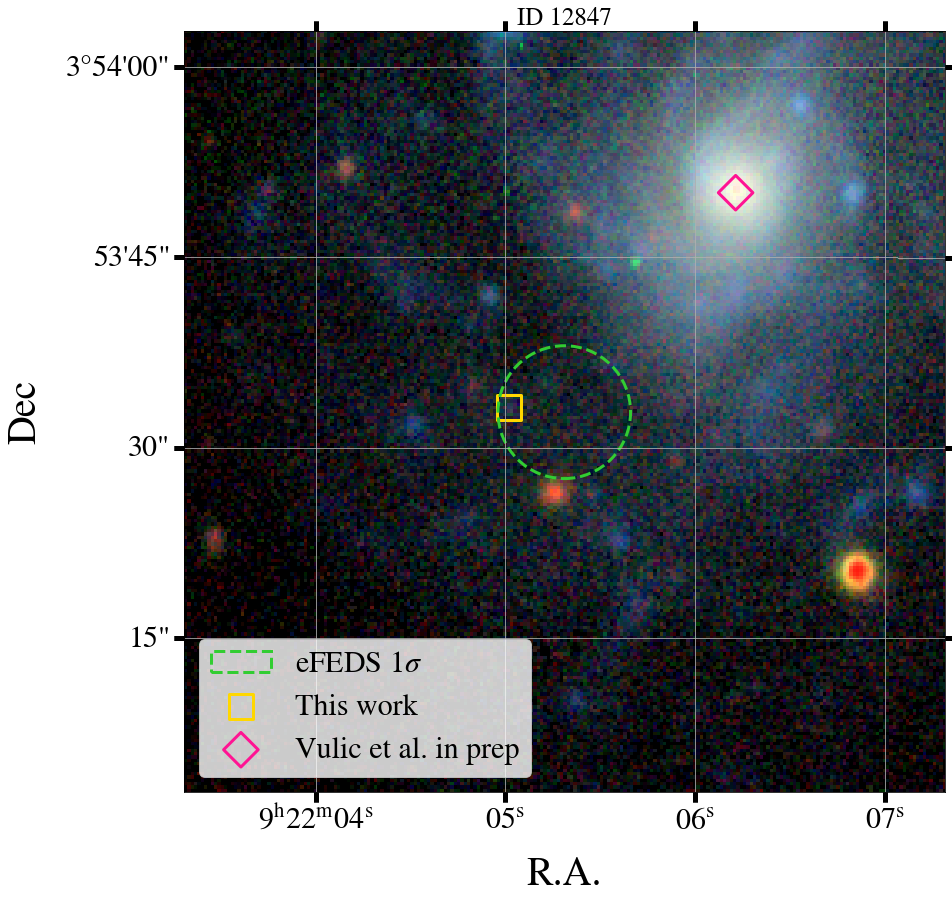}
\includegraphics[width=5cm]{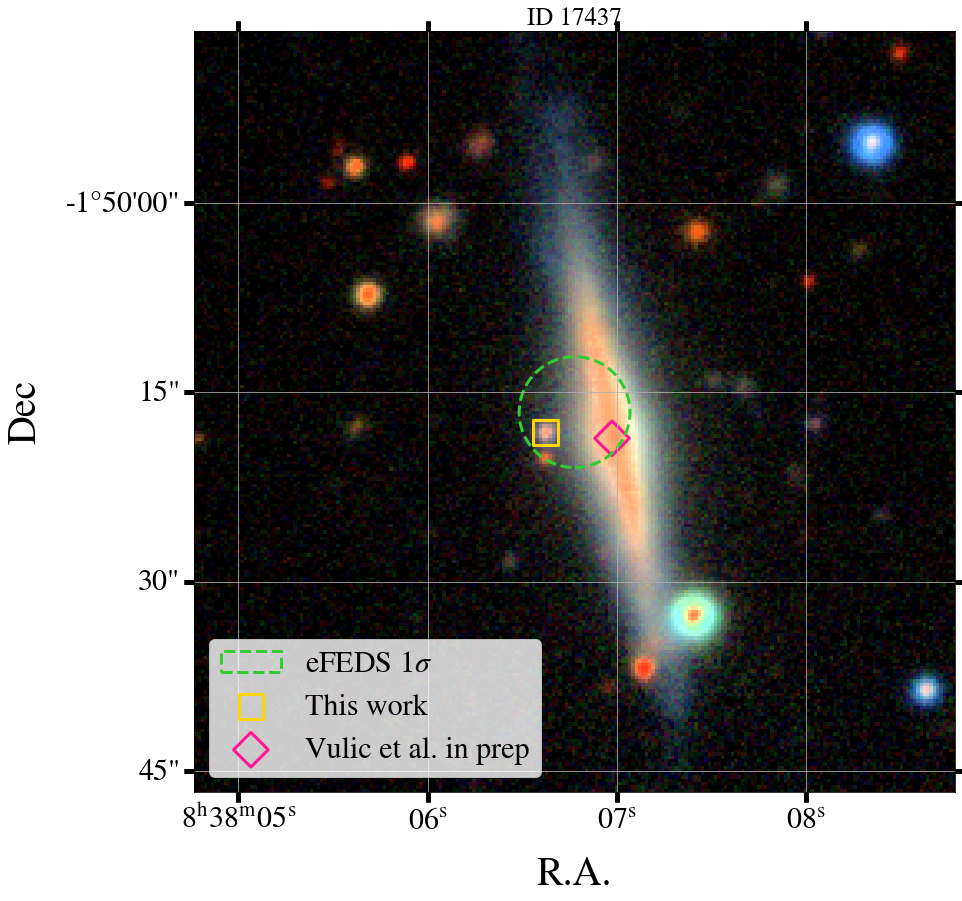}
\includegraphics[width=5cm]{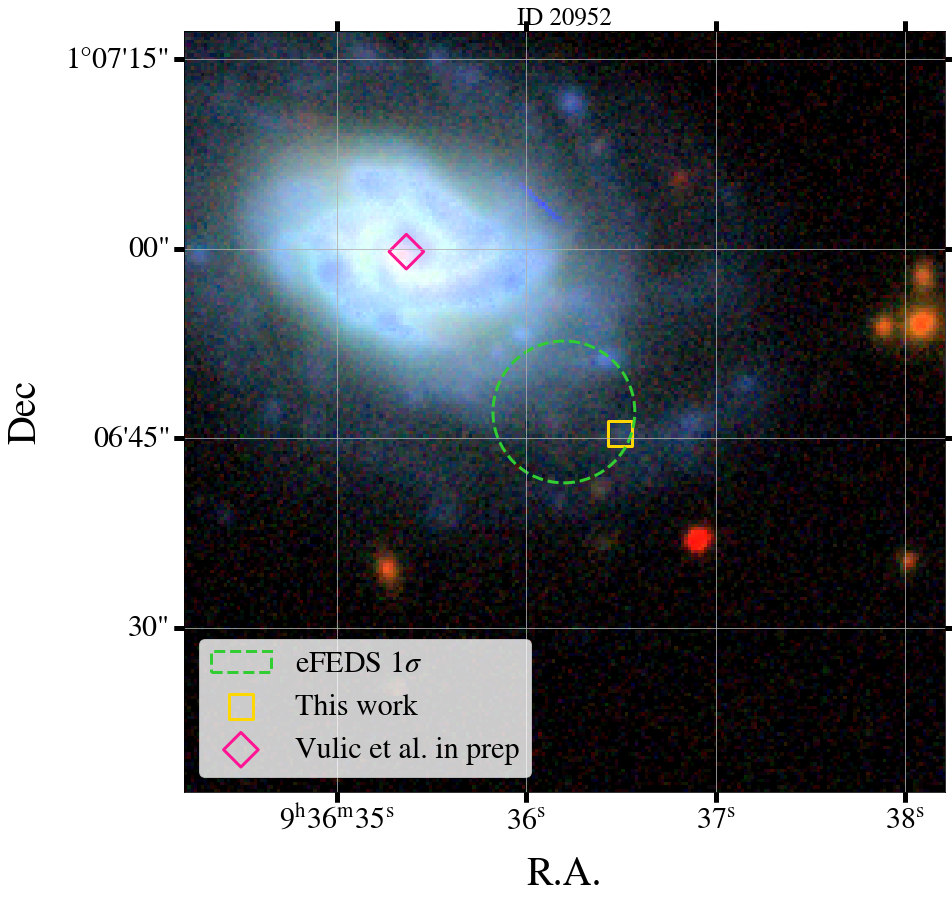}
\includegraphics[width=5cm]{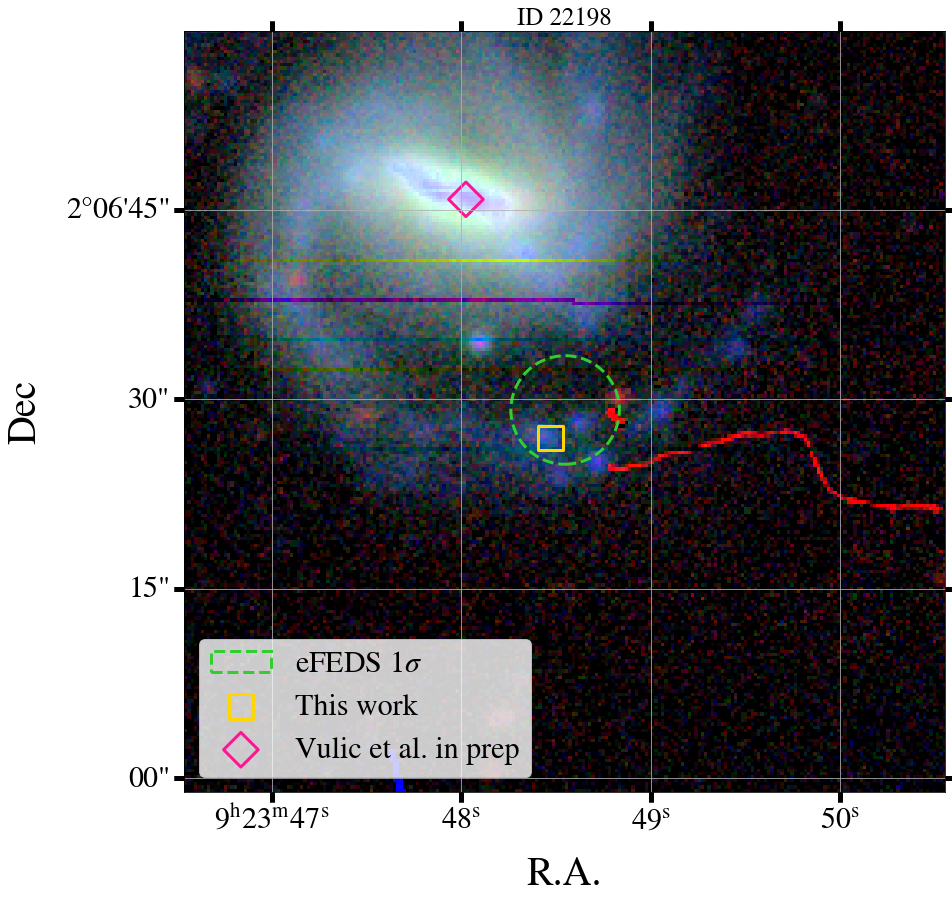}
\includegraphics[width=5cm]{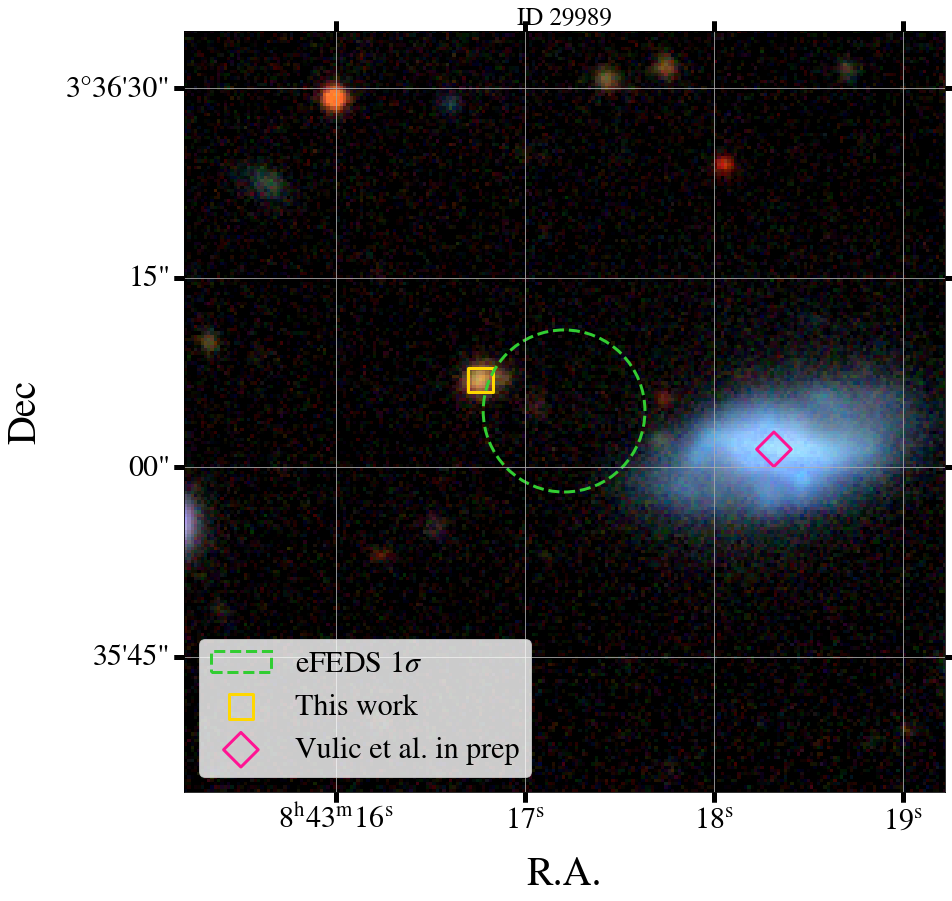}

\caption{The seven sources for which our counterpart fall within an HECATE galaxy \citep[see][]{Vulic2021} but is probably pointing to a background extragalactic source. The RGB images are 1\arcmin $\times$1\arcmin. In the cutouts the position of our proposed counterpart, the  centre of the galaxy, the X-ray position and the positional error are indicated.} 
\label{fig:HECATE}
\centering
\end{figure*}

\section{Templates used for photo-z  \label{appendix:templates}}
As was discussed in the main text a key ingredient for computing the photometric redshifts is the use of the most appropriate templates, able to represent the population under analysis.
Here we list the templates that have been used for the EXT and PLIKE samples, respectively. For each of the templates we provide the name of the model, the corresponding number in the catalog and the reference to the paper that describes them.

\subsection{templates used for PLIKE \label{subsection:templatesPLIKE}}
The library of templates used for the PLIKE sample includes a mixture of SEDs from single objects and hybrids (different relative contribution of host and nuclear component, as introduced in \citet{Salvato09}. Templates 1 and 12-17 and 20 are originally from \citep[][]{Polletta2007}. However, as templates N. 18 and N. 19 (originally from SDSS-V\footnote{\url{http://classic.sdss.org/dr5/algorithms/spectemplates}}, template N.20 has been extended in the UV and presented in \citet{Salvato09}. The same template was then used for creating hybrids templates by mixing it with normal galaxies templates \citep[][]{Noll2004} with different degree of star formation, as presented in the \citet{Ananna2017}\footnote{Note that the templates are slightly different than in Ananna et al. in the UV part}. With a similar procedure the hybrid templates 2-10 where constructed combining an S0 and a QSO2 template, both from \citet{Polletta2007}. The hybrids were originally presented in \citet{Salvato09} and were successfully used in \citet{Salvato09, Salvato11, Marchesi16} among others. Finally, templates 24-29 are from the recent work of \citet{Brown19}.
\begin{enumerate}
\item S0
\item S0\_10\_QSO2\_90
\item S0\_20\_QSO2\_80
\item S0\_30\_QSO2\_70
\item S0\_40\_QSO2\_60
\item S0\_50\_QSO2\_50
\item S0\_60\_QSO2\_40
\item S0\_70\_QSO2\_30
\item S0\_80\_QSO2\_20
\item S0\_90\_QSO2\_10
\item CB1\_0\_LOIII4
\item Sb
\item Spi4
\item M82
\item I22491
\item Sey18
\item Sey2
\item pl\_QSOH
\item pl\_QSO\_DR2\_029\_t0
\item pl\_TQSO1
\item s250\_10\_pl\_TQSO1\_90
\item s180\_30\_pl\_TQSO1\_70
\item s800\_20\_pl\_TQSO1\_80
\item 3C120
\item MRK110
\item NGC5548\_64.00\_NGC4138
\item PG0052p251
\item NGC4151
\item NGC3783\_00.50\_NGC4725
\end{enumerate}

\subsection{templates used for EXT \label{subsection:templatesEXT}}
The templates used for computing the photo-z for the EXT sample are almost entirely taken from \citet{Brown19}. The list  of AGN templates includes SED of single objects (the name of the template is self explanatory) and hybrids constructed combining in different ratio the SED of two different galaxies and AGN.  Additionally we have added two templates of elliptical galaxies from \citet{Polletta2007} and two newly created templates of type 1 AGN. We have used the Type 1 archetype presented in \citet{Comparat2020} and obtained by the stacking of all type 1 sources that are counterparts to ROSAT/2RXS \citep[][]{Boller16, Salvato18a} that had an SDSS spectra. The SED was then extended in the MIR using the BQSO template presented in \citet{Polletta2007}. BQSO is similar to TQSO but with less enhanced MIR flux. This new hybrid was then extended in the UV with various slopes.
\begin{enumerate}
\item 2MASXJ13000533+1632151
\item Ark564
\item F16156+0146
\item F2M1113+1244
\item Fairall9
\item H1821+643
\item IRAS\_11119+3257
\item Mrk110
\item Mrk1502
\item Mrk231
\item Mrk290
\item Mrk493
\item Mrk590
\item Mrk817
\item NGC3227\_Central\_00.50\_NGC4569
\item NGC3227\_Central\_01.00\_NGC4569
\item NGC3227\_Central\_02.00\_NGC4569
\item NGC3227\_Central\_04.00\_NGC4569
\item NGC3227\_Central\_08.00\_NGC4569
\item NGC3227\_Central\_16.00\_NGC4569
\item NGC3227\_Central\_32.00\_NGC4569
\item NGC3227\_Central\_64.00\_NGC4569
\item NGC3516\_Central\_00.50\_NGC4826
\item NGC3516\_Central\_01.00\_NGC4826
\item NGC3516\_Central\_02.00\_NGC4826 
\item NGC3516\_Central\_04.00\_NGC4826
\item NGC3516\_Central\_08.00\_NGC4826
\item NGC3516\_Central\_16.00\_NGC4826
\item NGC3516\_Central\_32.00\_NGC4826
\item NGC3516\_Central\_64.00\_NGC4826
\item NGC3516\_Central
\item NGC3783\_Central\_01.00\_NGC4725
\item NGC3783\_Central\_02.00\_NGC4725
\item NGC3783\_Central\_04.00\_NGC4725
\item NGC3783\_Central\_08.00\_NGC4725
\item NGC3783\_Central\_16.00\_NGC4725
\item NGC3783\_Central\_32.00\_NGC4725
\item NGC3783\_Central\_64.00\_NGC4725
\item NGC4051\_Central\_00.50\_NGC3310
\item NGC4051\_Central\_00.50\_NGC4125
\item NGC4051\_Central\_00.50\_NGC4559
\item NGC4051\_Central\_01.00\_NGC3310\
\item NGC4051\_Central\_01.00\_NGC4125
\item NGC4051\_Central\_01.00\_NGC4559
\item NGC4051\_Central\_02.00\_NGC3310
\item NGC4051\_Central\_02.00\_NGC4125
\item NGC4051\_Central\_02.00\_NGC4559
\item NGC4051\_Central\_04.00\_NGC3310
\item NGC4051\_Central\_04.00\_NGC4125
\item NGC4051\_Central\_04.00\_NGC4559
\item NGC4051\_Central\_08.00\_NGC3310
\item NGC4051\_Central\_08.00\_NGC4125
\item NGC4051\_Central\_08.00\_NGC4559
\item NGC4051\_Central\_16.00\_NGC3310
\item NGC4051\_Central\_16.00\_NGC4125
\item NGC4051\_Central\_16.00\_NGC4559
\item NGC4051\_Central\_32.00\_NGC3310
\item NGC4051\_Central\_32.00\_NGC4125
\item NGC4051\_Central\_32.00\_NGC4559
\item NGC4051\_Central\_64.00\_NGC3310
\item NGC4051\_Central\_64.00\_NGC4125
\item NGC4051\_Central\_64.00\_NGC4559
\item NGC4051\_Central
\item NGC4151\_Central\_00.50\_NGC4125
\item NGC4151\_Central\_00.50\_NGC4579
\item NGC4151\_Central\_01.00\_NGC3310
\item NGC4151\_Central\_01.00\_NGC4125
\item NGC4151\_Central\_01.00\_NGC4579
\item NGC4151\_Central\_02.00\_NGC3310
\item NGC4151\_Central\_02.00\_NGC4125
\item NGC4151\_Central\_02.00\_NGC4579
\item NGC4151\_Central\_04.00\_NGC3310
\item NGC4151\_Central\_04.00\_NGC4125
\item NGC4151\_Central\_04.00\_NGC4579
\item NGC4151\_Central\_08.00\_NGC3310
\item NGC4151\_Central\_08.00\_NGC4125
\item NGC4151\_Central\_08.00\_NGC4579
\item NGC4151\_Central\_16.00\_NGC3310
\item NGC4151\_Central\_16.00\_NGC4125
\item NGC4151\_Central\_16.00\_NGC4579
\item NGC4151\_Central\_32.00\_NGC3310
\item NGC4151\_Central\_32.00\_NGC4125
\item NGC4151\_Central\_32.00\_NGC4579
\item NGC4151\_Central\_64.00\_NGC3310
\item NGC4151\_Central\_64.00\_NGC4125
\item NGC4151\_Central\_64.00\_NGC4579
\item NGC5548\_Central\_00.50\_NGC4138
\item NGC5548\_Central\_01.00\_NGC4138
\item NGC5548\_Central\_02.00\_NGC4138
\item NGC5548\_Central\_04.00\_NGC4138
\item NGC5548\_Central\_08.00\_NGC4138
\item NGC5548\_Central\_16.00\_NGC4138
\item NGC5548\_Central\_32.00\_NGC4138
\item NGC5548\_Central\_64.00\_NGC4138
\item NGC5548\_Central
\item NGC5728
\item NGC7469
\item OQ\_530
\item PG0026+129
\item PG1415+451
\item PKS1345+12
\item Ton951
\item Ell4\_A\_0 
\item Ell5\_A\_0
\item pl\_BQSO\_Co19\_sl-8
\item pl\_BQSO\_Co19\_sl-20
\end{enumerate}

\section{Release of PDZ and SED fitting}
\begin{figure}[h]
\centering

\includegraphics[width=9cm]{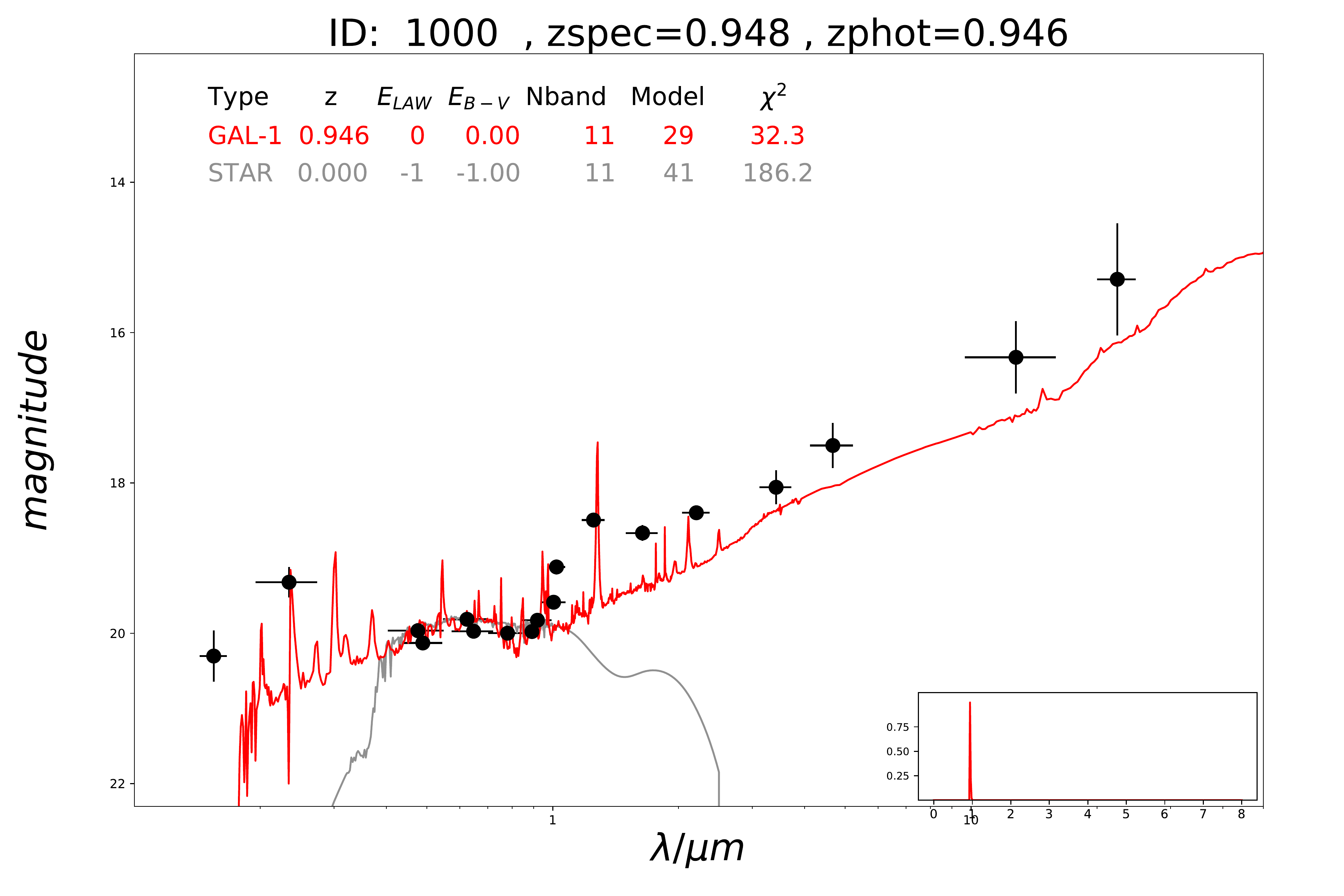}
\caption{Example of SED fitting and redshift probability distribution (in the inset). The photometric points and their errors are indicated with black points. The best extragalactic and Galactic templates are shown with a solid line. The plot and the data for creating the plots are available under request.}
\label{fig:SED}
\centering
\end{figure}

For each primary (and secondary, in case it exists) counterpart to the eFEDS point sources we make available under request the redshift probability distribution and the SED fitting as in Figure~\ref{fig:SED}.

\section{Columns description of released catalogs \label{appendix:catalogs_column}}
Most columns are common to both the Main and Hard sample. We list column descriptions for the Main sample. In the few cases where there are differences in the column descriptions, we report in brackets the corresponding description for the Hard sample.  The last column of the catalog (ID\_MAIN) is present only in the Hard catalog. 

\begin{enumerate}
\item {\bf ERO\_Name}: eROSITA official source Name (see Brunner et al.)
\item {\bf ERO\_ID}: ID of eROSITA source in the Main Sample (from Brunner et al.) [ID of eROSITA source in the Hard sample (from Brunner et al.)]
\item {\bf ERO\_RA\_CORR};	J2000 Right Ascension of the eROSITA source (corrected) in degrees in the Main Sample (from Brunner et al.) [J2000 Right Ascension of the eROSITA source (corrected) in degrees in the Hard Sample (from Brunner et al.)] 
\item {\bf ERO\_Dec\_CORR}:J2000 Declination of the eROSITA source (corrected) in degrees in the Main Sample (from Brunner et al.) [J2000 Declination of the eROSITA source (corrected) in degrees in the Hard Sample (from Brunner et al.)] 
\item {\bf ERO\_RADEC\_ERR\_CORR}:	eROSITA positional uncertainty (corrected) in arcsec from the Main Sample (from Brunner et al.) [eROSITA positional uncertainty (corrected) in arcsec from the Hard Sample (from Brunner et al.)]
\item {\bf ERO\_ML\_FLUX}: 0.2-2.3 keV source flux in $\rm{erg~cm^{-2}~s^{-1}}$, converted from count rate assuming ECF=1.074e+12 (Gamma=2.0). [In the Hard catalog the column is called {\bf ERO\_ML\_FLUX\_3}: 2.3-5 keV source flux in $\rm{erg~cm^{-2}~s^{-1}}$, converted from count rate assuming ECF=1.147e+11 (Gamma=2.0)]. See Brunner et al.
\item {\bf ERO\_ML\_FLUX\_ERR}:	0.2-2.3 keV source flux error (1 sigma) in $\rm{erg~cm^{-2}~s^{-1}}$. [In the Hard catalog the column is called {\bf ERO\_ML\_FLUX\_ERR\_3}: 2.3-5 keV error (1 sigma) in $\rm{erg~cm^{-2}~s^{-1}}$]. See Brunner et al.
\item {\bf ERO\_DET\_LIKE}:	0.2-2.3 keV detection likelihood measured by PSF-fitting. [In the Hard catalog the column is called {\bf ERO\_DET\_LIKE\_3}: 2.3-5 keV detection likelihood measured by PSF-fitting]. See Brunner et al. 
\item {\bf ERO\_inArea90}: true if in the 0.2-2.3keV exp>500s region, which comprises 90\% area (from Brunner et al.). 
\item {\bf CTP\_LS8\_UNIQUE\_OBJID:} LS8 unique identifier for the proposed counterpart to the eROSITA source (Expression: toString(LS8\_BRICKID)+"\_"+toString(LS8\_OBJID))
\item {\bf CTP\_LS8\_RA}: J2000 Right Ascension of the LS8 counterpart in degrees 
\item  {\bf CTP\_LS8\_Dec}: J2000 Declination of the LS8 counterpart in degrees
\item  {\bf Dist\_CTP\_LS8\_ERO}: Separation between selected counterpart and eROSITA (corrected) position in arcsec.
\item {\bf CTP\_NWAY\_LS8\_UNIQUE\_OBJID}: LS8 unique identifier of the best LS8 countepart from NWAY (Expression: toString(LS8\_BRICKID)+"\_"+toString(LS8\_OBJID))
\item  {\bf CTP\_NWAY\_LS8\_RA}: J2000 Right Ascension of the LS8 counterpart from NWAY in degrees. 
\item {\bf CTP\_NWAY\_LS8\_Dec}: J2000 Declination of the LS8 counterpart from NWAY in degrees.
\item {\bf CTP\_NWAY\_dist\_bayesfactor}: Logarithm of ratio between prior and posterior, from separation, positional error and number density (see Appx. in Salvato et al 2018 for clarifications)			
\item {\bf CTP\_NWAY\_dist\_post}: Distance posterior probability comparing this association vs. no association (see Appx. in Salvato et al 2018 for clarifications)			
\item {\bf  CTP\_NWAY\_p\_single}: Same as dist\_post, but weighted by the prior (see Appx. in Salvato et al 2018 for clarifications)
\item {\bf CTP\_NWAY\_p\_any}: For each entry in the X-ray catalogue, the probability that there is a counterpart (see Appx. in Salvato et al 2018 for clarifications)	
\item {\bf CTP\_NWAY\_p\_i}:  Relative probability of the eROSITA/LS8 match (see Appx. in Salvato et al 2018 for clarifications)			
\item {\bf Dist\_NWAY\_LS8\_ERO}: Separation between the eROSITA position and the LS8 counterpart from NWAY in arcsec
\item {\bf CTP\_MLR\_LS8\_UNIQUE\_OBJID}: LS8 unique identifier of the LS8 counterpart from {\sc astromatch}	(Expression: toString(LS8\_BRICKID)+"\_"+toString(LS8\_OBJID))
\item {\bf CTP\_MLR\_LS8\_RA }:	J2000 Right Ascension of LS8 counterpart from {\sc astromatch} in degrees
\item {\bf CTP\_MLR\_LS8\_Dec}:	J2000 Declination of LS8 counterpart from {\sc astromatch} in degrees		
\item {\bf CTP\_MLR\_LR\_BEST}:	Likelihood Ratio value from {\sc astromatch}
\item {\bf CTP\_MLR\_REL\_BEST}: Reliability of the identification from {\sc astromatch}
\item {\bf Dist\_MLR\_LS8\_ERO}: Separation between the eROSITA position and the LS8 counterpart from {\sc astromatch} in arcsec	
\item {\bf CTP\_SAME}:	Comparison NWAY/MLR: true if the counterpart selected by the two method is the same 
\item {\bf CTP\_MLR}:	Comparison NWAY/MLR: true if the counterpart from NWAY(MLR) has p\_any(LR\_BEST) below(above) threshold			
\item {\bf CTP\_Hamstar}: Match to Hamstar: 1=same counterpart, 0=different counterpart, -99=no Hamstar (Schneider et al)			
\item {\bf CTP\_Hamstar\_p\_stellar}: Probability of association from Hamstar.
\item {\bf Dist\_CTP\_Hamstar}:	Separation between the counterpart proposed by Hamstar and the counterpart selected in this work.			
\item {\bf CTP\_quality}: counterpart quality: 4=best, 3=good, 2=with secondary, 1/0=unreliable (see paper)			
\item {\bf GaiaEDR3\_ID}:		ID in Gaia EDR3 source catalog 			
\item {\bf GaiaEDR3\_parallax}:	Parallax from Gaia EDR3 in mas			
\item {\bf GaiaEDR3\_parallax\_error}:	Parallax error from Gaia EDR3 in mas
\item {\bf GaiaEDR3\_parallax\_over\_error}:	Parallax/Parallax error. (a ratio $>$5 define a SECURE GALACTIC counterpart)			
\item {\bf GaiaEDR3\_pmra}:	Proper motion in RA from Gaia EDR3			
\item {\bf GaiaEDR3\_pmra\_error}:		Error on Proper motion in RA from Gaia EDR3			
\item {\bf GaiaEDR3\_pmdec}:	Proper motion in Dec from Gaia EDR3			
\item {\bf GaiaEDR3\_pmdec\_error}:	Error on Proper motion in Dec from Gaia EDR3			
\item {\bf GaiaEDR3\_phot\_g\_mean\_mag}: g band magnitude (VEGA) from Gaia EDR3
\item {\bf GaiaEDR3\_phot\_g\_mean\_mag\_error}: Error g band magnitude (VEGA) from Gaia EDR3	
\item {\bf GaiaEDR3\_phot\_bp\_mean\_mag}:	bp band magnitude (VEGA) from Gaia EDR3
\item {\bf GaiaEDR3\_phot\_bp\_mean\_mag\_error}: Error bp band magnitude (VEGA) from Gaia EDR3
\item {\bf GaiaEDR3\_phot\_rp\_mean\_mag}: rp band magnitude (VEGA) from Gaia EDR3
\item {\bf GaiaEDR3\_phot\_rp\_mean\_mag\_error}: Error rp band magnitude (VEGA) from Gaia EDR3	
\item {\bf FUV}:		Galex Far UV magnitude (AB magnitude)	
\item {\bf FUV\_ERR}:		Galex Far UV magnitude error (AB magnitude)			
\item {\bf NUV}:	Galex Near UV magnitude	(AB magnitude)		
\item {\bf NUV\_ERR}:		Galex Near UV magnitude error (AB magnitude)			
\item {\bf KiDS\_u	}:	KiDS u-band magnitude (AB magnitude)		
\item {\bf KiDS\_u\_ERR}:		KiDS u-band magnitude error	(AB magnitude)		
\item {\bf KiDS\_g}:	KiDS g-band magnitude	(AB magnitude)		
\item {\bf KiDS\_g\_ERR}:	KiDS g-band magnitude error	(AB magnitude)		
\item {\bf KiDS\_r}:	KiDS r-band magnitude	(AB magnitude)		
\item {\bf KiDS\_r\_ERR}:	KiDS r-band magnitude error	(AB magnitude)		
\item {\bf KiDS\_i}:	KiDS i-band magnitude (AB magnitude)			
\item {\bf KiDS\_i\_ERR}:	KiDS i-band magnitude error	(AB magnitude)		
\item {\bf omegac\_z}:	OmegaCAM z-band magnitude (AB magnitude)			
\item {\bf omegac\_z\_ERR}:		OmegaCAM z-band magnitude error	(AB magnitude)		
\item {\bf HSC\_g}:		HSC g-band magnitude (AB magnitude)			
\item {\bf HSC\_g\_ERR}:	HSC g-band magnitude error	(AB magnitude)		
\item {\bf HSC\_r}:		HSC r-band magnitude (AB magnitude)			
\item {\bf HSC\_r\_ERR}:	HSC r-band magnitude error	(AB magnitude)		
\item {\bf HSC\_r2}:	HSC r2-band magnitude (AB magnitude)			
\item {\bf HSC\_r2\_ERR}:	HSC r2-band magnitude error	(AB magnitude)		
\item {\bf HSC\_i}:		HSC i-band magnitude (AB magnitude)			
\item {\bf HSC\_i\_ERR}:	HSC i-band magnitude error	(AB magnitude)		
\item {\bf HSC\_i2}:	HSC i2-band magnitude (AB magnitude)			
\item {\bf HSC\_i2\_ERR}:	HSC i2-band magnitude error	(AB magnitude)		
\item {\bf HSC\_z}:		HSC z-band magnitude (AB magnitude)			
\item {\bf HSC\_z\_ERR}:	HSC z-band magnitude error	(AB magnitude)		
\item {\bf HSC\_Y}:		HSC Y-band magnitude (AB magnitude)		
\item {\bf HSC\_Y\_ERR}:	HSC Y-band magnitude error	(AB magnitude)		
\item {\bf VIKING\_z}:	VISTA/VIKING  z-band magnitude (AB magnitude)			
\item {\bf VIKING\_z\_ERR}:	VISTA/VIKING z-band magnitude error	(AB magnitude)		
\item {\bf VIKING\_Y}:	VISTA/VIKING Y-band magnitude	(AB magnitude)		
\item {\bf VIKING\_Y\_ERR}:	VISTA/VIKING Y-band magnitude error	(AB magnitude)		
\item {\bf VIKING\_J}:	VISTA/VIKING J-band magnitude	(AB magnitude)		
\item {\bf VIKING\_J\_ERR}:		VISTA/VIKING J-band magnitude error (AB magnitude)			
\item {\bf VIKING\_H}:	VISTA/VIKING H-band magnitude (AB magnitude)			
\item {\bf VIKING\_H\_ERR}:		VISTA/VIKING H-band magnitude error (AB magnitude)			
\item {\bf VIKING\_Ks}:		VISTA/VIKING Ks-band magnitude (AB magnitude)			
\item {\bf VIKING\_Ks\_ERR}: VISTA/VIKING Ks-band magnitude error (AB magnitude)			
\item {\bf W1}:	LS8/Wise W1 magnitude (AB magnitude)			
\item {\bf W1\_ERR}: LS8/Wise W1 magnitude error (AB magnitude)			
\item {\bf W2}:		LS8/Wise W2 magnitude	(AB magnitude)		
\item {\bf W2\_ERR}: LS8/Wise W2 magnitude error (AB magnitude)			
\item {\bf W3}:		LS8/Wise W3 magnitude	(AB magnitude)		
\item {\bf W3\_ERR}: LS8/Wise W3 magnitude error (AB magnitude)			
\item {\bf W4}:	LS8/Wise W4 magnitude (AB magnitude)			
\item {\bf W4\_ERR}: LS8/Wise W4 magnitude error (AB magnitude)		
\item {\bf LS8\_g}:		LS8 g-band magnitude (AB magnitude)			
\item {\bf LS8\_g\_ERR}:	LS8 g-band magnitude error	(AB magnitude)		
\item {\bf LS8\_r}:		LS8 r-band magnitude (AB magnitude)			
\item {\bf LS8\_r\_ERR}:	LS8 r-band magnitude error	(AB magnitude)		
\item {\bf LS8\_z}:		LS8 z-band magnitude (AB magnitude)			
\item {\bf LS8\_z\_ERR}:	LS8 z-band magnitude error (AB magnitude)			
\item {\bf VHS\_Y}:		VISTA/VHS Y-band magnitude	(AB magnitude)		
\item {\bf VHS\_Y\_ERR}:	VISTA/VHS Y-band magnitude error (AB magnitude)			
\item {\bf VHS\_H}:		VISTA/VHS H-band magnitude	(AB magnitude)		
\item {\bf VHS\_H\_ERR}:	VISTA/VHS H-band magnitude error (AB magnitude)			
\item {\bf VHS\_Ks}:	VISTA/VHS Ks-band magnitude	(AB magnitude)		
\item {\bf VHS\_Ks\_ERR}:	VISTA/VHS Ks-band magnitude error (AB magnitude)			
\item {\bf HCS\_g\_diff}:	Difference between psf and Kron magnitude in HSC g-band (AB magnitude)			
\item {\bf HCS\_r\_diff}:	Difference between psf and Kron magnitude in HSC r-band	(AB magnitude)		
\item {\bf HCS\_i\_diff}:	Difference between psf and Kron magnitude in HSC i-band	(AB magnitude)		
\item {\bf HCS\_z\_diff}:	Difference between psf and Kron magnitude in HSC z-band	(AB magnitude)		
\item {\bf HCS\_opt\_extended}:		Extension in HSC griz bands. 1=extended; -99=data missing 0=other from Aihara et al 2018		
\item {\bf CTP\_LS8\_phot\_flag}: Flag for LS8 photometry: true when  the source has simultaneously  g,r,z,w1 photometry in LS8 
\item {\bf CTP\_LS8\_Type}: Morphological model from LS8
\item {\bf in\_KiDS\_flag}:		Flag for KiDS coverage: 1: Source is in KiDS area; 0: otherwise	
\item {\bf in\_HSC\_flag}:		Flag for HSC coverage: 1: Source is in HSC area as from Aihara et al 2018; 0: otherwise						
\item {\bf SPECZ\_RA}:	Right Ascension (degrees) of the spectroscopic redshift entry in the original catalogue from which it was taken
\item {\bf SPECZ\_Dec}:	Declination (degrees) of the spectroscopic redshift entry in the original catalogue from which it was taken		
\item {\bf SPECZ\_Redshift}:		Spectroscopic redshift from original catalog 		
\item {\bf SPECZ\_NORMQ}:	Normalised quality of spectroscopic redshift: 3=secure, 2=not secure, 1=unreliable redshift/bad spectrum, -1= Blazar candidate		
\item {\bf SPECZ\_Origin}:	Catalogue which provided this spectroscopic redshift		
\item {\bf SPECZ\_Original\_ID}: Identifier of this spectroscopic redshift entry in the original catalogue from which it was taken			
\item {\bf SPEC\_Gal\_flag }:		true when the CTP has a reliable redshift above 0.002 (boolean)
\item {\bf SPEC\_Star\_flag	}:	true when the CTP has a reliable redshift below 0.002 (boolean)				
\item {\bf CTP\_Classification }:	counterpart classification: 	SECURE/LIKELY GALACTIC/EXTRAGALACTI, as from flowchart (see paper)		
\item {\bf PHZ\_LePHARE\_zphot }:		Photo-z from Le PHARE, but set to 0 for GALACTIC sources			
\item {\bf PHZ\_LePHARE\_zl68 }:		Le PHARE zphot min at 1 sigma			
\item {\bf PHZ\_LeePHARE\_zu68 }:		Le PHARE zphot max at 1 sigma			
\item {\bf PHZ\_LePHARE\_zl90 }:		Le PHARE zphot min at 2 sigma			
\item {\bf PHZ\_LePHARE\_zu90 }:		Le PHARE zphot max at 2 sigma			
\item {\bf PHZ\_LePHARE\_zl99 }:		Le PHARE zphot min at 3 sigma			
\item {\bf PHZ\_LePHARE\_zu99 }:		Le PHARE zphot max at 3 sigma			
\item {\bf PHZ\_LePHARE\_chi }:		Le PHARE chi2 value for best fitting galaxy/AGN template			
\item {\bf PHZ\_LePHARE\_ModelAGN }:		Le PHARE best template fitting the data			
\item {\bf PHZ\_LePHARE\_extlaw }:		Le PHARE Extinction Law applied to the template: Prevot (1) or none (0)			
\item {\bf PHZ\_LePHARE\_ebv }:		Le PHARE  E(B-V) applied to the template			
\item {\bf PHZ\_LePHARE\_pdz }:		Le Phare probability distribution. Photo-z more reliable when value is high				
\item {\bf PHZ\_LePHARE\_nband	}:	Le Phare number of bands used for the computation of photo-z			
\item {\bf PHZ\_LePHARE\_zp2 }:		Le Phare second best photo-z from {\sc {LePhare}}, if existing			
\item {\bf PHZ\_LePHARE\_chi2\_2 }:		Le Phare chi2 value for second best fitting template, if existing			
\item {\bf PHZ\_LePHARE\_ModelAGN\_2 }:	Le Phare second best template fitting the data, if existing			
\item {\bf PHZ\_LePHARE\_pdz2 }:		Le Phare probability distribution for secondary solution, if existing				
\item {\bf PHZ\_DNNz\_zphot}:		Photo-z from DNNZ (from Nishizawa et al.), but set to 0 for GALACTIC sources			
\item {\bf PHZ\_DNNz\_zl68 }:		DNNZ 1sigma min error on  photo-z			
\item {\bf PHZ\_DNNz\_zu68 }:	DNNZ 1 sigma max error on photo-z			
\item {\bf PHZ\_DNNz\_zl95 }:		DNNZ 2 sigma min error on  photo-z			
\item {\bf PHZ\_DNNz\_zu95 }:		DNNZ 2 sigma max error on  photo-z			
\item {\bf CTP\_REDSHIFT }:		Final redshift: zspec (NORMQ=3) when available, else photo-z from {\sc {Le Phare}} 			
\item {\bf CTP\_REDSHIFT\_GRADE }:	In a range from 5 (spectroscopy) to 0 (unreliable photo-z) (see text for details)
\item {\bf CLUSTER\_CLASS}: In range from 5 to 1: 5=most likely a cluster; 1= not a cluster (see text for details)
\item {\bf CTP\_CLASS }:		same as CTP\_Classification, but with numbers: 3: SECURE EXTRAGALACTIC; 2: LIKELY EXTRAGALACTIC; 1: SECURE GALACTIC; 0: LIKELY GALACTIC 
\item {\bf ID\_MAIN}: [column present only in the Hard sample: The source ID in the Main catalog for the sources in common.]
\end{enumerate}
\end{appendix}
\end{document}